\documentclass[journal]{IEEEtran}
\usepackage{cite}
\usepackage{amsmath,amssymb,amsfonts}
\usepackage{algorithmic}
\usepackage{graphicx}
\usepackage{textcomp}

\usepackage{booktabs}
\usepackage{tabularx}
\usepackage{pdflscape}
\usepackage{longtable}

\usepackage[hidelinks]{hyperref} 

\usepackage[T1]{fontenc}

\usepackage{caption}

\def\BibTeX{{\rm B\kern-.05em{\sc i\kern-.025em b}\kern-.08em
    T\kern-.1667em\lower.7ex\hbox{E}\kern-.125emX}}

\begin{document}

\title{A Systematic Literature Review of Retrieval-Augmented Generation: Techniques, Metrics, and Challenges}

\author{Andrew Brown, Muhammad Roman, and Barry Devereux}



\maketitle

\begin{abstract} 
    This systematic review of the research literature on retrieval-augmented generation (RAG) provides a focused analysis of the most highly cited studies published between 2020 and May 2025. A total of 128 articles met our inclusion criteria. The records were retrieved from ACM Digital Library, IEEE Xplore, Scopus, ScienceDirect, and the Digital Bibliography and Library Project (DBLP). RAG couples a neural retriever with a generative language model, grounding output in up-to-date, non-parametric memory while retaining the semantic generalisation stored in model weights. Guided by the PRISMA 2020 framework, we (i) specify explicit inclusion and exclusion criteria based on citation count and research questions, (ii) catalogue datasets, architectures, and evaluation practices, and (iii) synthesise empirical evidence on the effectiveness and limitations of RAG. To mitigate citation-lag bias, we applied a lower citation-count threshold to papers published in 2025 so that emerging breakthroughs with naturally fewer citations were still captured. This review clarifies the current research landscape, highlights methodological gaps, and charts priority directions for future research.
\end{abstract}

\section{Introduction} 

Large Language Models (LLMs) have, over the past five years, transformed the way researchers and practitioners process text. Retrieval-Augmented Generation (RAG) addresses key shortcomings of these models, such as hallucinated facts, stale world knowledge, and the challenges posed by knowledge-intensive and domain-specific queries, by allowing a generative model to query an external corpus at inference time, combining \emph{parametric} memory learnt during pre-training with \emph{non-parametric} evidence retrieved on demand \cite{RN1}.

Traditional retrieval systems locate relevant passages but cannot compose new text; purely generative models produce fluent language, yet risk factual errors when outside knowledge is required. RAG integrates both paradigms, offering factual grounding without sacrificing fluency.

Since Meta AI introduced RAG in 2020 \cite{RN1}, the field has diversified rapidly, incorporating hybrid retrievers, iterative retrieval loops, graph-based retrieval, and domain-specific pipelines have been proposed. However, the results are fragmented and the evaluation protocols are still evolving. Therefore, a transparent, protocol-driven synthesis of RAG is required.
Consequently, we follow the PRISMA 2020 statement (Preferred Reporting Items for Systematic Reviews and Meta-Analyses) \cite{PRISMA} to ensure transparency and reproducibility in the development of a systematic review of the state-of-the-art of RAG research. Each published paper selected for this review progressed through the four PRISMA flow stages --- identification, selection, eligibility, and inclusion --- while the reviewers verified that the study addressed at least one of our research questions \emph{and} met the predefined inclusion/exclusion criteria. To focus on work that has demonstrably shaped the field, we place special emphasis on the most frequently cited RAG articles, including only the most highly cited studies published between 2020 and May 2025.\footnote{Citation statistics were collected from Semantic Scholar on 13 May 2025.}  This citation-based filter serves as the primary gate in our PRISMA workflow, ensuring that the review concentrates on influential contributions while maintaining reproducibility.

The present study addresses these gaps by offering a citation-weighted, PRISMA-compliant systematic synthesis of 128 influential RAG studies that maps datasets, architectures, evaluation metrics, and open research challenges, thus advancing the field toward more aligned, robust, and scalable retrieval-augmented systems.

This review is aimed at both NLP researchers, who can use it to identify gaps and promising directions, and NLP engineers seeking practical guidance on applying RAG techniques. Our review catalogues datasets, novel methods, evaluation metrics, and RAG deployment challenges. In doing so, it delivers a cohesive overview of RAG architectures and offers actionable insights to inspire future innovations.

Based on this aim, we formulate four research questions (Table~\ref{tab:research_questions_label}).
Throughout the review, we treat the original RAG architecture of Lewis \emph{et al.} \cite{RN1} - Dense Passage Retriever plus sequence-to-sequence generator - as the \emph{standard baseline}. Variants are characterised relative to this reference point.

\begin{table}
  \caption{Summary of the research questions that guide this systematic review.}
  \label{tab:research_questions_label}
  \begin{tabular}
  {p{0.06\linewidth}p{0.35\linewidth}p{0.45\linewidth}}
    \toprule
    \textbf{Index} & \textbf{Research Question} & \textbf{Goal} \\
    \midrule
    RQ1 & What thematic topics have already been addressed by highly cited RAG studies? & Summarises the main topics in the field, outlining the current state of knowledge and identifying gaps in the literature.\\
    RQ2 & What are the innovative methods and approaches compared to the standard retrieval-augmented generation? & Provides a thorough overview of current research on RAG, assisting researchers and engineers in identifying common methodologies, existing studies, and exploring novel approaches in the field. \\
    RQ3 & What are the most frequently used metrics for evaluating the effectiveness of retrieval-augmented generation systems? & By identifying relevant metrics, researchers can conduct meaningful comparative analyses of systems, essential for benchmarking and advancing the field. \\
    RQ4 & What are the key challenges and limitations associated with retrieval-augmented generation techniques? & Identifies research gaps, enabling researchers to propose solutions or suggest areas for further exploration. \\ 
  \end{tabular}
\end{table}

The remainder of this paper is organised as follows. Section \ref{sec:methodology} details the methodology employed in this review, including search strategies and inclusion criteria. Section \ref{sec:results} presents the results, categorising the studies according to key themes and findings. Section \ref{sec:discussion} discusses the implications of these findings, addressing both the strengths and challenges of RAG. Finally, Section \ref{sec:conclusion} concludes the paper, summarising the key insights with concrete recommendations for researchers and engineers building the next wave of knowledge-aware language models.

\section{Methodology} 
\label{sec:methodology}

This systematic literature review adhered to the strategy and reporting guidelines of the Preferred Reporting Items for Systematic Reviews and Meta-Analyses (PRISMA 2020) \cite{PRISMA}. The PRISMA framework is recognised across various research domains for its robust approach to literature reviews. The review process was structured into three main phases: Identification, Screening, and Inclusion. 

\begin{figure} 
  \centering
  \includegraphics[width=\linewidth]{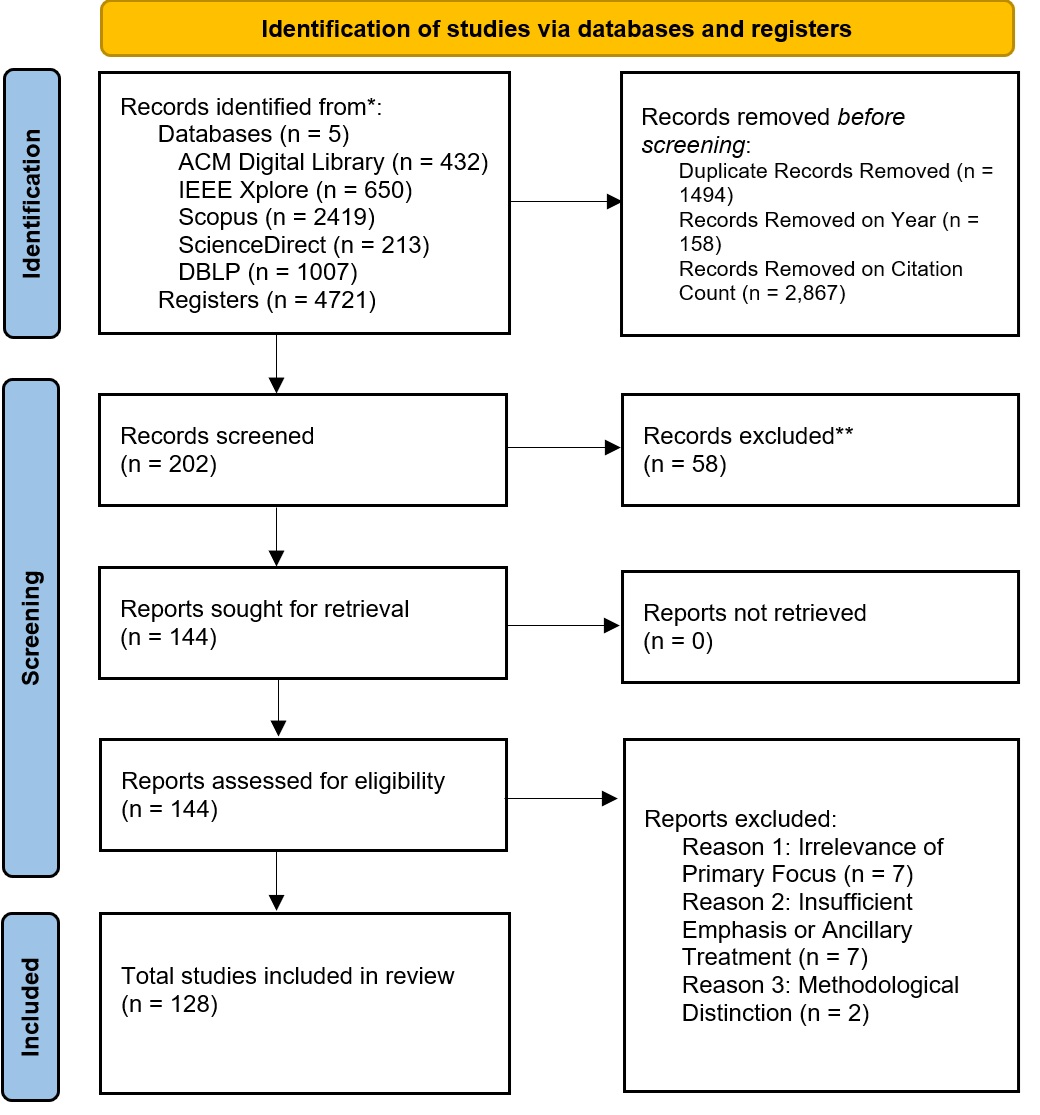}
  \caption{PRISMA 2020 flow diagram showing the stages of article selection in this systematic review}
  \label{fig:PRISMA-flow-diagram}
\end{figure}

We used the standard PRISMA flow diagram, as shown in Figure \ref{fig:PRISMA-flow-diagram}, which encompasses searches exclusively in specific databases and registers, although these are not detailed here. This section elaborates on our selection of the systematic review framework, detailing the search strategy, inclusion and exclusion criteria, data extraction processes, and quality assessment procedures to underscore our commitment to a transparent and reproducible methodology.

\begin{table*}
  \caption{Search queries used with each database.}
  \label{tab:research_queries_table}
  \centering
  \begin{tabular}{p{0.15\linewidth} p{0.8\linewidth}}
    \toprule
    \textbf{Database} & \textbf{Query} \\
    \midrule
    ACM Digital Library & Title:(retrieval AND augmented AND generation) OR Abstract:(retrieval AND augmented AND generation) \\
    \midrule
    IEEE Xplore & ("Document Title": retrieval augmented generation) OR ("Publication Title": retrieval augmented generation) OR ("Abstract": retrieval augmented generation) \\
    \midrule
    Scopus & TITLE-ABS-KEY ( retrieval AND augmented AND generation ) \\
    \midrule
    ScienceDirect & Title, abstract, keywords: retrieval AND augmented AND generation \\
    \midrule
    DBLP & retrieval augmented generation \\
    \bottomrule
  \end{tabular}
\end{table*}

\subsection{Systematic Review Framework Selection} 

The PRISMA 2020 guidelines provide an extensive framework for systematic reviews, especially suitable for multidisciplinary fields such as RAG. These guidelines emphasise updated methodological standards, including the synthesis of findings, the assessment of study biases, and the inclusion of various study designs. In contrast, Kitchenham's guidelines \cite{RN40}, designed specifically for software engineering literature reviews, do not offer the necessary interdisciplinary breadth required for RAG research. Similarly, Evidence-Based Software Engineering (EBSE) \cite{RN40} focuses primarily on the application of evidence-based principles to software engineering and does not adequately address the broader theoretical and application-based questions relevant to RAG. Therefore, PRISMA 2020 is valuable for facilitating the synthesis of various study methodologies and goals, which aligns well with the evolving and interdisciplinary nature of RAG research.

\subsection{Database Selection} 

We used four digital databases and the DBLP bibliographic index to improve coverage and deduplication. We targeted five key electronic resources, chosen for their extensive repositories and relevance to our research topics:

\begin{enumerate}
    \item ACM Digital Library: \href{https://dl.acm.org}{https://dl.acm.org}
    \item IEEE Xplore: \href{https://ieeexplore.ieee.org/}{https://ieeexplore.ieee.org/}
    \item Scopus: \href{https://www.scopus.com/}{https://www.scopus.com/}
    \item ScienceDirect: \href{https://www.sciencedirect.com/}{https://www.sciencedirect.com/}
    \item Digital Bibliography and Library Project (DBLP; bibliographic index):\\ \href{https://dblp.org/}{https://dblp.org/}
\end{enumerate}

\subsection{Inclusion and Exclusion Criteria}\label{sec:InclusionAndExclusionCriteria} 

In this section, we define the eligibility criteria for selecting studies for our systematic review. We focus on articles published between 2020 and 2025; this time frame coincides with the significant introduction of the RAG framework by Meta AI \cite{RN1}, a key milestone in natural language processing (NLP) research. Our selection includes studies that explicitly address the RAG framework or explore systems with similar functionalities, ensuring that our review comprehensively captures the latest innovations in this area.

\paragraph{Inclusion Criteria:}
\begin{enumerate}
    \item \textit{Focus:} Studies must address RAG or similar systems that rely on retrieval to support text output.
    
    \item \textit{Publication Date and Citations:} Only works from January 2020 to May 2025 are accepted. For 2025 publications, a minimum of 15 citations is required; for those from 2024 or earlier, at least 30 citations are needed.
    
    \item \textit{Original Contributions:} Only works that present new experimental data or fresh ideas are considered.
    
    \item \textit{Input and Output:} Studies may use various input types (e.g., text, images, audio) if retrieval is central, but the final output must be text.

\end{enumerate}

\paragraph{Exclusion Criteria:}
\begin{enumerate}
    \item \textit{Relevance:} Works that do not pertain to the topic are removed. 
    
    \item \textit{Language:} Studies not published in English are excluded. 
    
    \item \textit{Duplicates and Access:} Duplicate works or those with unavailable full text are omitted.
    
\end{enumerate}

\subsection{Search Strategy and Search Terms} 

We based our search terms on the core concept of the RAG framework by breaking down ``retrieval augmented generation'' into three parts: ``retrieval'', ``augmented'', and ``generation''. These parts became the basis for our search terms used in titles, abstracts, and keywords. Table \ref{tab:research_queries_table} lists the detailed queries for each database. Our systematic approach, combining the main keywords with related phrases such as "retrieval augmented text generation", gathered a wide range of relevant literature on RAG.

\subsection{Search Process} 
We queried five well-established digital databases and a bibliographic inde (ACM Digital Library, IEEE Xplore, Scopus, ScienceDirect, and DBLP) to collect relevant articles. The results were exported in BibTeX, CSV, or Excel formats as provided by the source. A Python script converted BibTeX files into Excel format, gathering key details such as titles, abstracts, publication years, authors, author counts, and journal names into one data table. Duplicate entries were first automatically removed by the script, followed by a manual check to verify accuracy.

\subsection{Screening Process} 

Articles were screened against a set inclusion and exclusion criteria linked to our research questions. Missing abstracts were retrieved from the original databases and manually added. Following PRISMA guidelines, a reviewer handled initial screening, full text review, and data extraction, while a second reviewer independently checked the results to reduce bias. This dual-review method strengthens the review's reliability. The process, illustrated in Figure \ref{fig:PRISMA-flow-diagram}, consisted of an initial screening and a review of the full text.


\subsubsection{Initial Screening} 

After removing duplicates and applying date and citation filters, two of the present authors (R$_1$, R$_2$) independently screened \emph{all} titles and abstracts ($n = 202$).
Each record was labelled \texttt{1}~(include) or \texttt{0}~(exclude) against the predefined eligibility criteria (\S\ref{sec:InclusionAndExclusionCriteria}). To aid, but not replace, human judgement, we provided both reviewers with LLM-generated suggestions from \textit{deepseek-ai/DeepSeek-R1-Distill-Llama-70B}; final decisions remained entirely with the reviewers.

\textbf{LLM-assisted suggestions.} To support decision-making, not replace human judgement, we provided both reviewers with five independent generations from \textit{deepseek-ai/DeepSeek-R1-Distill-Llama-70B} for each record. The LLM was prompted with our research questions and inclusion/exclusion criteria; its five binary recommendations were then collapsed into a single suggestion by majority vote. The final selection decisions remained exclusively with the human reviewers.

\subsubsection{Full Text Screening} 

Full texts were retrieved from the original sources indexed by our selected databases and the DBLP bibliographic index. During full-text screening, we applied a quality assurance protocol assessing soundness, validity, reliability, and statistical rigour to ensure the inclusion of only high-quality studies. During the screening, each article was evaluated against predefined criteria for inclusion and exclusion, which encompassed the scope and methodological robustness of the study, and was categorised with a '0' for exclusion or '1' for inclusion. Moreover, we encountered challenges concerning the interchangeable use of terms such as RAG, retriever+reader models, and retrieval-augmented LLMs. To address these challenges, we concentrated on clearly differentiating the retriever and generator components, thereby streamlining the analysis while ensuring a comprehensive comprehension of the fundamental elements.

\subsection{Data Extraction} 

Data extraction and management were handled using Google Sheets for organising data and EndNote for managing references. The data extracted from the articles were compiled into a structured database designed for easy access during subsequent analysis, synthesis, and reporting. Each entry was verified against the original articles to identify and correct any discrepancies, such as mismatched values or missing information.

\paragraph{Data-extraction workbook.} 
All coded variables, their operational definitions, and the raw study-level entries are available in a publicly accessible Google Sheets workbook \footnote{\href{https://docs.google.com/spreadsheets/d/1HDIwZnigCfd92GyYtAwA2Vn4z1QkmTnA}{\texttt{RAG\_Data\_Extraction.xlsx}}}

After verification, the data were synthesised to address the research questions of the systematic review. The synthesis used methods suited to the nature of the data and the review objectives, primarily through a descriptive approach that summarised and explained the data patterns by identifying trends, differences and similarities between studies. This method enabled us to draw meaningful conclusions from the diverse data collected during the review.

\subsubsection{Data Extraction Methodology: Domains, Specific Tasks, Technique and Results} 

The data extraction process followed our research question and eligibility criteria, focusing on topics, methods, and evaluation metrics. It recorded details such as Domain Area, which defines the field addressed by each study. For datasets, both public and private sets were included. The framework and components of the RAG system were documented, listing the "Retrieval Mechanism", "Chunking Mechanism", "Vector Space Encoder", and "Generation Model" while excluding any components not mentioned in the paper. All data were organised in a workbook under clear headings for easy access and analysis, providing complete coverage for detailed review. 

A single reviewer, using a RAG framework, independently extracted the data to confirm accuracy and reliability. The framework treated each article as a separate knowledge source, queried by the specific data required. This approach simplified the review process and offered a method of verifying the details. Using this framework confirmed that the data collection was complete and consistent with the research criteria and objectives.

However, the RAG framework poses two major challenges. The first is the risk of hallucination, where the system may generate information that does not exist. The second is that key data might be absent from the retrieved passages. Despite the framework's benefits in improving speed and precision, these issues call for careful cross-checking of the extracted data to maintain its authenticity and reliability. Addressing these challenges is essential to preserve the integrity of the data extraction process.

\subsubsection{Dataset Identification Methodology} 

We systematically examined all studies in this review and used citation tracking to identify and extract relevant datasets. The extracted information was organised in Google Sheets to form a structured, navigable database, ensuring the inclusion of the most impactful and widely used resources. This organisation supported more effective analysis and comparison, making sure that the most relevant and impactful datasets were included.

Each entry lists its source reference, full official name and common abbreviation, content overview, intended use, and frequency of citations, allowing researchers to assess scope and suitability. The relevance and popularity of each dataset are highlighted by the number of papers that have used it, indicating its significant impact and widespread adoption in the field. 

As shown in Table~\ref{tab:datasetsTable} (see Appendix~\ref{sec:appendix}, Table~\ref{tab:datasetsTable}), each dataset with the extracted fields: dataset name; content description, which includes details such as the number of questions; intended use, which may be described as designed to or as a high overview; citation frequency, which indicates the number of times the dataset has been mentioned in the reviewed academic papers.

\section{Results} 
\label{sec:results}

\newcommand{\dateOfSearch}{13/05/2025}
\newcommand{\numberOfSearchedDatabases}{5}
\newcommand{\listMajorDatabases}{ACM Digital Library, IEEE Xplore, Scopus, ScienceDirect, DBLP}
\newcommand{\searchStartYear}{2020}
\newcommand{\searchEndYear}{2025}

\newcommand{\numOfInitialRecords}{4721}
\newcommand{\numDuplicatesRemoved}{1494}
\newcommand{\numYearRemoved}{158}
\newcommand{\numCitationRemoved}{2867}
\newcommand{\numRecordsScreened}{202}
\newcommand{\numRecordsExcluded}{58}
\newcommand{\numReportsRetrieved}{144}
\newcommand{\numNotRetrieved}{0}
\newcommand{\numAssessedEligibility}{144}
\newcommand{\numExcludedIrrelevance}{7}
\newcommand{\numExcludedAncillary}{7}
\newcommand{\numExcludedMethodological}{2}
\newcommand{\numMeetingInclusionCriteria}{128}

We identified \numOfInitialRecords\ records; after removing duplicates (\numDuplicatesRemoved), out-of-range (\numYearRemoved) and below-threshold items (\numCitationRemoved), \numRecordsScreened\ were screened; \numReportsRetrieved\ full texts were assessed; \numMeetingInclusionCriteria\ studies were included (reasons in Fig.~\ref{fig:PRISMA-flow-diagram}).

\subsection{Excluded Studies} 

Following the screening of the title and abstract, 144 candidate records were recovered in full and assessed against the predefined inclusion criteria. Sixteen of these were excluded during the full text screening for the reasons summarised below. The reasons for exclusion were categorised as follows:

\begin{itemize}
    \item \textbf{Irrelevance of Primary Focus (n = 7):} Papers whose primary contributions lay outside the augmented generation of retrieval, e.g. robustness of dense search, long-context benchmarks, general GenIR evaluation or system-level optimisations, where RAG appeared only as a peripheral baseline or illustrative example \cite{RN85, RN1817, RN1819, RN1820, RN1821, RN1702, RN1824}.
    
    \item \textbf{Insufficient Emphasis or Ancillary Treatment (n = 7):} Studies that incorporated RAG merely as an auxiliary component within broader investigations---such as LLM-human hybrids for marketing research, domain-specific LLM development, knowledge graph construction workflows, multimodal agent toolkits, healthcare task automation, cost-effective classification or materials modelling pipelines---without substantive and dedicated analysis of RAG itself \cite{RN1816, RN334, RN1818, RN1822, RN1823, RN53, RN632}.
    
    \item \textbf{Methodological Distinction (n = 2):} Works focused on conceptually distinct paradigms from RAG, specifically generative retrieval or generation-augmented retrieval, which invert the standard RAG pipeline by predicting document identifiers rather than conditioning the generation on the retrieved content \cite{RN59, RN947}.
    
\end{itemize}

All exclusion decisions were systematically documented to ensure methodological rigour, transparency, and reproducibility.

\subsection{Yearly Distribution of Identified Articles} 

\begin{figure}
    \centering
    \includegraphics[width=\linewidth]{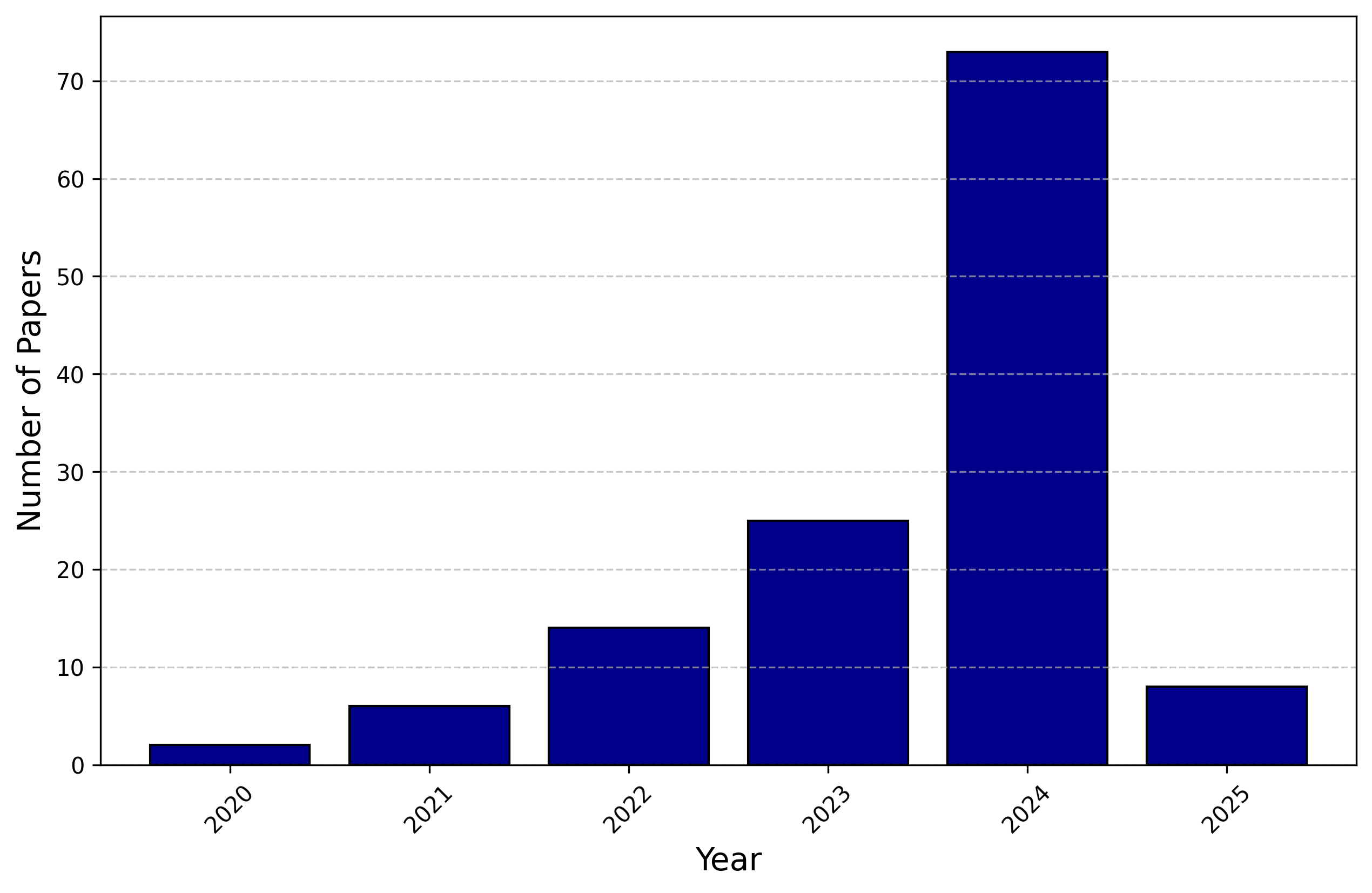}
    \caption{Yearly distribution of identified articles from 2020 to 2025}
    \label{fig:yearly_distribution}
\end{figure}

Across \searchStartYear–\searchEndYear, the number of identified articles increased year on year from 2020 to 2023, with a pronounced increase in 2024. As of \dateOfSearch, the count for 2025 is lower because the year is incomplete. Figure~\ref{fig:yearly_distribution} visualises the annual distribution; year-specific totals are listed in Table~\ref{tab:extractedData}.

These counts reflect the records that remained after deduplication and application of the eligibility criteria (Section~\ref{sec:InclusionAndExclusionCriteria}), including the citation thresholds ($\geq 30$ for publications up to 2024; $\geq 15$ for 2025). Consequently, year-to-year comparisons should be interpreted in light of (i) the staged indexing of databases and (ii) the partial coverage of 2025 at the time of the last search.

\subsection{Domain Characteristics of Included Studies} 

\begin{figure}
    \centering
    \includegraphics[width=\linewidth]{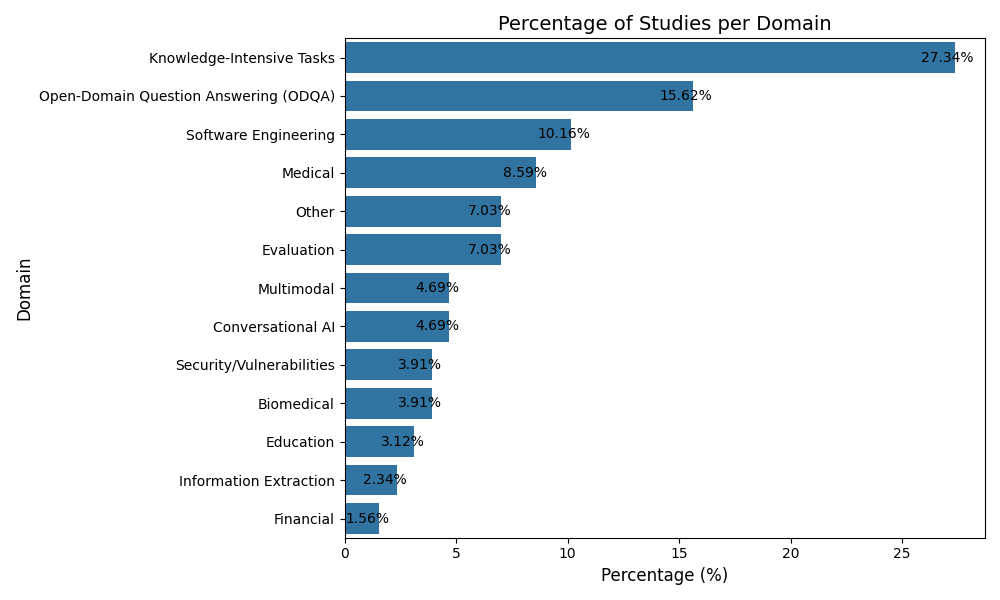}
    \caption{Distribution of Studies by Domain: This bar chart shows the percentage of studies conducted in various areas.}
    \label{fig:studiesPerDomainFigure}
\end{figure}

Studies were coded to a single \emph{primary} domain for proportional reporting; secondary tags (e.g., multimodal, conversational) were retained for analysis but are not double-counted in the primary distribution. Coding rules and examples appear in Table~\ref{tab:extractedData}. Proportions below refer to the included studies (Fig.~\ref{fig:studiesPerDomainFigure}).

Knowledge-intensive tasks accounted for 27.34\%, followed by open-domain question answering (ODQA) at 15.62\%, software engineering 10.16\% and medical 8.59\%. Evaluation comprised 7.03\%. The “Other” category (7.03\%) covers nine single-study niches: networking, counterfactual augmentation, content creation, personalisation, legal QA, recommender systems, chemistry, disaster response and personalised search. Multimodal and conversational AI each represented 4.69\%; security/vulnerabilities and biomedical 3.91\% each; education 3.12\%; information extraction 2.34\%; and finance 1.56\%.

These distributions indicate a concentration of work in knowledge-intensive and ODQA settings, with substantial activity in software engineering and medical applications and a long tail of niche areas. Full per-study domain labels and secondary tags are provided in Table~\ref{tab:extractedData}; percentages may not sum exactly due to rounding.

\setlength{\LTcapwidth}{1.2\textwidth}

\onecolumn
\begin{landscape}

\begin{longtable}{ 
  p{0.21\linewidth}  
  p{0.17\linewidth}  
  p{0.17\linewidth}  
  p{0.17\linewidth}  
  p{0.2\linewidth}  
}

  \caption[Study characteristics of included RAG papers]{Study characteristics of 128 included RAG papers by domain: datasets, chunking mechanisms, retrieval mechanisms, vector-space encoders, and generation models.}
  \label{tab:extractedData} \\
  \toprule
    \multicolumn{1}{p{0.21\linewidth}}{\textbf{Datasets}} &
    \multicolumn{1}{p{0.17\linewidth}}{\textbf{Chunking Mechanism}} &
    \multicolumn{1}{p{0.17\linewidth}}{\textbf{Retrieval Mechanism}} &
    \multicolumn{1}{p{0.17\linewidth}}{\textbf{Vector Space Encoder}} &
    \multicolumn{1}{p{0.2\linewidth}}{\textbf{Generation Model}} \\
  \midrule
  \endfirsthead

  \multicolumn{5}{c}{{\bfseries Table \thetable\ continued from previous page}} \\
  \toprule
  \textbf{Datasets} & \textbf{Chunking Mechanism} & \textbf{Retrieval Mechanism} & \textbf{Vector Space Encoder} & \textbf{Generation Model} \\
  \midrule
  \endhead

  \midrule \multicolumn{5}{c}{{(Continued on next page)}} \\
  \bottomrule
  \endfoot

  \bottomrule
  \endlastfoot
\multicolumn{5}{l}{{\textbf{Domain: Knowledge-Intensive Tasks}}} \\
\end{longtable}

\end{landscape}
\twocolumn

\section{Discussion}
\label{sec:discussion}

\subsection{What are the key topics that are already addressed in RAG?} \label{sec:basic_rag}

\subsubsection{Retrieval mechanism}\label{sec:retrieval} 

Retrieval-augmented generation systems depend uniformly on an external retriever to select a relevant context for a language model. In general, the mechanisms surveyed fall into five interrelated categories.

\textbf{Sparse term-based methods} (e.g., BM25) remain vital for their efficiency and interpretability, yet they struggle with semantic recall and gaps \cite{RN347}. Dense retrievers, built on dual encoder networks such as DPR, map queries and documents into continuous vector spaces and leverage the maximum inner-product search for semantic matching \cite{RN1}. Hybrid approaches combine sparse pruning of candidates with dense re-ranking to balance recall and precision across domains \cite{RN1637}.

\textbf{Encoder–decoder query generators reformulate inputs}, especially conversational or multihop questions, into standalone search queries, improving recall at the cost of added latency \cite{RN349}. Reclassification modules (e.g., CRAG) apply lightweight evaluators or preference-aligned models to reorder initial top k results, mitigating noisy retrievals, and aligning outputs with downstream generation needs \cite{RN1395}.

By organising passages or entities into \textbf{knowledge graphs}, graph retrieval methods extract sub-graphs or paths most relevant to a query.  Steiner tree formulations collecting prizes yield coherent multi-hop contexts with explicit reasoning chains, albeit at significant computational cost for large graphs \cite{RN672, RN1626}.

\textbf{Iterative frameworks interleave retrieval and generation}: LLM outputs refine subsequent queries, progressively bridging semantic gaps in complex tasks \cite{RN379}. Although this feedback loop improves multistep reasoning, it incurs increased latency and requires careful stopping criteria to prevent error propagation \cite{RN825}.

\textbf{Specialised retrievers adapt core architectures to different data modalities or domains}, such as code snippet retrieval by edit distance scoring \cite{RN363}, multimodal CLIP-based retrieval for image captioning \cite{RN367}, or clinical report retrieval using vision language embeddings \cite{RN913}. These systems achieve high task relevance but demand bespoke engineering and corpus maintenance.

Together, these mechanisms illustrate a landscape where advances in semantic embeddings, input optimisation, structural reasoning, adaptive feedback, and domain adaptation coalesce to enrich the context of LLM. Each category presents distinct trade-off between efficiency, scalability, interpretability, and domain generality, highlighting open avenues for unified, explainable, and resource-efficient retrieval in future RAG research.

\subsubsection{Vector Database} 

The vector database is fundamental to RAG, enabling fast similarity searches over dense embeddings through approximate nearest neighbor (ANN) techniques such as hierarchical navigable small world (HNSW) graphs and FAISS-based flat or inverted indices, which achieve sub-millisecond Maximum Inner Product Search (MIPS) performance in production settings but must negotiate accuracy–latency trade-offs and memory footprint constraints \cite{RN1,RN349,RN948}. Research has extended these core indexing methods to distributed and dynamic environments, employing GPU-sharded indices and cloud-native services like Pinecone to ingest and serve millions of vectors across training and inference pipelines; however, synchronization latency, update throughput, and cost-efficiency remain pressing concerns \cite{RN341,RN1617}. Concurrently, domain-specific vector stores have emerged—tailored for code retrieval (e.g., RepoCoder), biomedical concept embeddings (Chroma), financial knowledge bases, and multimodal memory systems (MuRAG, Re-ViLM)—to address the unique representational, alignment, and privacy demands of specialized data \cite{RN825,RN1569,RN144}. Finally, managed vector database offerings integrated via frameworks such as LangChain, LlamaIndex, Weaviate, and Qdrant have streamlined deployment in commercial RAG pipelines, albeit at the expense of potential vendor lock-in, hybrid architecture complexity, and unpredictable operational costs \cite{RN1711,RN1571}.

Despite the maturity of these infrastructures, several cross-cutting research gaps persist. Notably, adaptive indexing algorithms capable of real-time inserts and deletes without degrading search performance are under-explored, while cost-aware scaling strategies that balance query latency against infrastructure expenditure remain scarce. Moreover, ensuring seamless interoperability across heterogeneous vector database services and embedding formats presents an ongoing challenge and a fertile avenue for future RAG innovation.

\subsubsection{Document chunking}\label{sec:chunking} 

Document chunking is the decomposition of large inputs into smaller, retrievable units. It is a critical preprocessing step in RAG. Highly cited studies have converged on four principal approaches:

\textbf{Static fixed-length segmentation.} Early RAG architectures adopt uniform, size-bounded splits to simplify indexing and conform to transformer context limits. Common configurations include 100-word segments \cite{RN1,RN846}, fixed-size 64-token chunks (with optional 32-token flexible intervals) \cite{RN1633}, and approximately 600-character spans \cite{RN1642}. These static splits require minimal linguistic preprocessing and integrate readily with vector stores (e.g., FAISS), but frequently bisect semantic units, resulting in context loss and a trade-off between index growth (for smaller chunks) and retrieval precision (for larger ones).

\textbf{Semantic boundary–aware splitting.} To preserve discourse coherence, the subsequent work aligns the chunk boundaries with the inherent text structure. Techniques include sentence-level chunking, where each sentence becomes a chunk \cite{RN389}, and paragraph-level segmentation, merging short paragraphs and truncating overly long ones \cite{RN948}. More advanced methods leverage hierarchical section markers (e.g., PDF sub-sections) to define semantically coherent units \cite{RN1613,RN1744}. These approaches mitigate fragmentation and often improve retrieval relevance, at the cost of additional preprocessing complexity and the absence of standardised coherence metrics.

\textbf{Domain and modality specific chunking.} Recognising that different types of data exhibit unique structures, specialised chunking strategies have been developed:
\begin{itemize}
  \item \emph{Source code}: partitioning by function or Code Property Graph nodes to capture logical code blocks \cite{RN206,RN363}.
  \item \emph{Knowledge graphs}: aggregating graph triples into textual statements for embedding \cite{RN805}.
  \item \emph{Legal documents}: breaking cases into ($question$, $snippet$, $entity$, $answer$) tuples \cite{RN1615}.
  \item \emph{Biomedical texts}: micro-chunking into fixed five-token units to capture fine-grained concepts \cite{RN1643}.
  \item \emph{Multimodal inputs}: splitting image–text pairs into aligned patches or entries for vision–language RAG \cite{RN144}.
\end{itemize}
These tailored pipelines yield superior performance within their target domains, but require manual configuration and do not generalise easily across new data types.

\textbf{Adaptive dynamic chunking.} The most recent research line seeks to automate chunk-size and overlap selection based on query characteristics or retrieval performance. Representative techniques include sliding windows (for example, 1000-token windows with 200-token overlaps in LangChain \cite{RN1629}, fixed-size 1200-token chunks with dynamic overlap \cite{RN1756}), automated parameter search for domain-specific corpora (e.g., clinical notes \cite{RN1617}), and half-stride overlapping to balance novelty and context continuity \cite{RN1746}. Adaptive methods aim to integrate the benefits of static, semantic, and domain-specific approaches, yet remain largely experimental, facing challenges in hyperparameter optimisation, runtime overhead, and cross-domain robustness.

Over time, RAG document chunking has evolved from simple one-size-fits-all splits to sophisticated, context and domain-aware pipelines. Static segmentation offers scalability but suffers semantic fragmentation; semantic boundary methods enhance coherence but add preprocessing costs; domain-specific chunkers exploit structural priors at the expense of generality; and adaptive strategies promise end-to-end automation but require further validation. Future work should establish standardised coherence benchmarks, develop unified frameworks that dynamically leverage linguistic and domain signals, and evaluate scalability in large-scale RAG deployments.

\subsubsection{Vector encoders}\label{sec:encoders} 

In RAG systems, the \emph{vector space encoder} projects both user queries and document chunks into a shared high-dimensional embedding space, enabling efficient similarity-based retrieval.  Influential RAG studies fall into three principal paradigms:

\textbf{Sparse encoders.} Traditional IR techniques convert text into high-dimensional sparse vectors of term weights (e.g. TF-IDF, BM25), scoring relevance via inverted indices. BM25 remains a robust baseline, often combined with dense methods to increase recall in open-domain QA and hybrid pipelines \cite{RN846,RN672}. In specialised settings such as code and graph retrieval, additional sparse schemes are common. For example, one can measure the overlap between the query and a language-specific token inventory (e.g., Java identifiers, keywords, and API names), or employ weighted n-gram counts to capture local lexical structure. These approaches deliver low latency and scale efficiently but provide limited deep semantic modelling; in RAG, they are therefore used primarily as recall-orientated components that are complemented by dense retrieval and/or learnt re-ranking \cite{RN1746,RN825}.

\textbf{Dense encoders.} Deep learning–based dense encoders map inputs to continuous embeddings that capture contextual and semantic nuances:
\begin{itemize}
  \item \textbf{Transformer-based bi-encoders.}  Frameworks such as DPR, ANCE, REALM, ORQA, and dual encoder BERT variants embed queries and passages separately, optimising retrieval metrics (Recall@\(k\), MRR) through end-to-end fine-tuning \cite{RN1758,RN349}.  
  \item \textbf{Sentence and paragraph embeddings.} Models such as Sentence-BERT, MPNet, paraphrase-mpnet-base-v2 and Contriever produce fixed-length vectors for larger text spans, improving semantic similarity on standard benchmarks \cite{RN370,RN952,RN1516}.  
  \item \textbf{Foundation \& specialised models.} API-driven encoders (e.g., text-embedding-ada-002, text-embedding-3-small / large) and proprietary systems (Dragon, E5, BGE) deliver broad coverage with minimal tuning \cite{RN1758,RN1627,RN1636}.  Domain-adapted variants, MedLLaMA-13B for biomedicine \cite{RN1643}, PubMedBERT for clinical language \cite{RN1569}, CodeBERT / CodeT5 for source code, demonstrate versatility in specialised vocabularies \cite{RN206,RN209}.
\end{itemize}

\textbf{Hybrid \& multi-modal encoders.}
To retrieve across heterogeneous sources, modern RAG systems fuse sparse and dense signals or jointly encode multiple modalities.
\begin{itemize}
  \item \textbf{Sparse–dense hybrids.}  Elastic Learnt Sparse Encoder (ELSER) integrates learnt sparse representations with dense sentence embeddings, balancing latency and recall \cite{RN1614}.  
  \item \textbf{Vision–language models.}  CLIP (text + image), LXMERT, ALBEF, and temporal deformable convolutional encoders support multimodal retrieval for visual QA and image-based generation \cite{RN365,RN888,RN71}.  
  \item \textbf{Graph and sequence models.}  Graph Transformers and Graph Attention Networks embed structured data (knowledge graphs, ASTs) into vector spaces for retrieval-augmented reasoning \cite{RN1626,RN681}.  
\end{itemize}

The selection of encoders in RAG reflects a trade-off among retrieval accuracy, computational efficiency, and domain adaptability.  Future work should target out-of-domain robustness, real-time index updates, and unified frameworks that seamlessly integrate sparse, dense, and multimodal representations.

\subsubsection{Training} 

Training of Retrieval-Augmented Generation (RAG) models has coalesced into five interrelated paradigms, each addressing distinct trade-offs in performance, efficiency, and domain applicability. 

\textbf{Joint end-to-end} training optimises retriever and generator components simultaneously by minimising a combined negative marginal log-likelihood loss, often through expectation-maximisation loops that alternate reader and retriever updates \cite{RN1,RN341}. Although this yields cohesive alignment of retrieval-generation and can leverage implicit retrieval supervision, it incurs a high computational cost due to frequent document-encoder refreshes and requires careful weighting between retrieval versus generation gradients to avoid collapse of one component \cite{RN362,RN379}.

\textbf{Modular two-stage} approaches decouple training—pre-fine-tuning dense retrievers (e.g., DPR) before generator tuning—trading end-to-end optimality for pipeline stability and simplified hyperparameter search \cite{RN49,RN230}. Although this separation can ease convergence and allow retrieval-specific objective design, it may lead to suboptimal global coordination and requires additional engineering to integrate retrieval scores during generation.

\textbf{Parameter-efficient fine-tuning} (PEFT) and \emph{instruction tuning} techniques update only a small subset of model parameters, via low-rank adapters (LoRA), prefix-tuning, or lightweight mapper modules, dramatically reducing GPU memory requirements while preserving downstream performance \cite{RN1609,RN681}. These methods have been successfully applied in financial forecasting (e.g., StockGPT) and clinical QA, yet remain sensitive to adapter rank, learning-rate schedules, and the diversity of instruction data used \cite{RN1741,RN1751}.

\textbf{Specialized training objectives} increase standard cross entropy with contrastive losses (to distinguish relevant from irrelevant documents), self-critical sequence training (SCST) for sequence-level rewards and analogy or style-aware losses to capture higher-order relations or lexical emphasis \cite{RN144,RN79}. Such multiobjective schemes can yield significant gains in task-specific metrics (BLEU, CIDEr, accuracy) but introduce additional hyperparameter tuning complexity and obscure training dynamics.

\textbf{Domain and modality specific adaptation} tailors RAG pipelines to code (e.g., RepoCoder in large codebases \cite{RN825}), vision-language (e.g., ReViLM's gated cross-attention for radiology reports \cite{RN888}), and specialised legal or biomedical corpora \cite{RN1615,RN1643}. Although these systems achieve state-of-the-art benchmark results, they face challenges in data scarcity, overfitting, modality alignment, and cross-domain generalisation.

Collectively, these training paradigms illustrate the field’s evolution from monolithic joint optimisation to modular, resource-aware and domain-focused strategies, each of which presents open problems in objective balance, compute efficiency, and transferability that continue to drive RAG research forward.

\subsubsection{Generation Model} 

Since its introduction, retrieval-augmented generation (RAG) has evolved from a proof-of-concept dual-encoder retriever paired with an encoder–decoder backbone to a rich landscape of end-to-end retrieval–generation pipelines. The original RAG framework demonstrated that the addition of a T5-style encoder-decoder with an open-domain retriever markedly improved question answers over purely generative baselines \cite{RN128}. Fusion-in-Decoder subsequently refined this approach by fusing multiple retrieved passages via late-stage cross-attention, yielding more coherent multidocument summaries \cite{RN521}. In parallel, models with decoder only, such as RETRO, showed that fragment level retrieval could be interleaved within autoregressive decoding, laying the groundwork for lightweight, scalable RAG in conversational settings \cite{RN180}. More recent work like Self-RAG has pushed toward self-supervised alignment of latent retrieval signals, bypassing external supervision, and underscoring a trajectory from loosely coupled retriever–generator pairs to fully integrated systems \cite{RN9}.

Beyond open-domain QA and summarization, highly cited studies have extended RAG to specialised and multimodal tasks. In biomedicine, BioGPT applied retrieval-augmented generation to clinical question answering, demonstrating improved factuality on medical benchmarks \cite{RN1807}. Legal research platforms such as Lexis+ AI and Ask Practical Law AI have tailored retrievers to statutory and case law corpora, helping practitioners with contextually grounded legal drafting \cite{RN1759}. Code-centric work, using models like CodeT5 and Codex, has recovered API documentation at generation time, enhancing code synthesis and reducing syntactic errors \cite{RN537,RN1657}. More recent multimodal RAG approaches (e.g. MiniGPT-4, LLaVA, Qwen2-VL) incorporate image retrieval to support visually grounded question answering, pointing to an expanding modality scope within the RAG paradigm \cite{RN1678,RN1558,RN1653}.

\subsubsection{Generative Model Families}

Since the original RAG paper, model families have proliferated, each contributing distinct architectures, scale points, and fine-tuning strategies that shape retrieval integration:

\textbf{Anthropic (decoder-only).} Claude-3-Opus and Claude-3.5-Sonnet (2024) interleave retrieved context with safety-orientated controls to mitigate hallucinations in conversational QA \cite{RN1656}.

\textbf{BigScience (encoder–decoder).} Bloom (2022) provided a multilingual, multisize foundation; RAG adapters later enabled domain-agnostic retrieval experiments atop this family \cite{RN1655}.

\textbf{DeepSeek (decoder-only).} DeepSeek-V2-Chat (2024) embeds a lightweight retriever within a proprietary autoregressive backbone, optimising low-latency RAG for chatbots \cite{RN1667}.

\textbf{EleutherAI (decoder-only).} GPT-J (2021) and GPT-Neo variants served as open alternatives to evaluate the impact of retrieval on QA without instruction tuning \cite{RN526}.

\textbf{Google (encoder–decoder \& decoder-only).} Flan-T5 (base to XXL, 2022) set the standard for cross-attention fusion in summarization and QA \cite{RN522,RN521}, while PaLM-2 (XXS to 540B, Text-Bison) and the Gemini/Gemma chat series (2023–24) explore retrieval adapters in massive decoder-only contexts \cite{RN1674}.

\textbf{Meta AI (encoder–decoder \& decoder-only).} BART (2020) pioneered the integration of retrieval through cross-attention \cite{RN518}. The Llama family (2023-2025)---Llama-1/2/3 in sizes 7B to 70B, with LoRA and quantised variants---illustrates how scale and parameter-efficient fine-tuning affect RAG on conversational and QA tasks \cite{RN122,RN524,RN1666,RN1682,RN1683}.

\textbf{Mistral AI (decoder-only).} Mistral-7B (2023) and its Instruct, quantised, and Mixtral-8×7B ensemble (2024) probe the trade-offs between open source accessibility, instruction alignment, and retrieval fluency \cite{RN121,RN558}.

\textbf{Nomic AI \& NVIDIA.} GPT4All (2025) offers on-device prototyping for lightweight RAG \cite{RN1670}. NVIDIA’s NeMo GPT-43B (2023) and Llama3-ChatQA (8B/70B, 2024) combine large-scale proprietary pre-training with retrieval-aware objectives for enterprise applications \cite{RN1637,RN1754}.

\textbf{OpenAI (decoder-only).} From GPT-2 (2019) through GPT-3/3.5 (2020), ChatGPT/3.5-turbo (2022), to GPT-4/GPT-4o (2023–24), OpenAI has incrementally embedded retrieval: early work pre-generated snippets to GPT-2 input \cite{RN516}, while GPT-4-turbo (2024) dynamically issues retrieval calls via system prompts \cite{RN531,RN529,RN1644}.

\textbf{Qwen-1.5 (decoder-only).} The Qwen-1.5 lineup (0.5B to 72B, chat variant) explores multilingual retrieval for both text and code generation \cite{RN539}.

Despite this rich diversity, two broad patterns emerge. Encoder–decoder models (Flan-T5, BART, Bloom) excel at multipassage fusion via cross attention, making them well suited for tasks demanding precise grounding (e.g., summarization, QA). Decoder-only families (GPT-J to GPT-4, Mistral, Claude) leverage token-insertion or adapter-based retrieval, trading architectural simplicity, and inference speed for conversational flexibility. Open challenges persist: the absence of a unified, modality-spanning RAG benchmark suite; systematic evaluation of retrieval noise versus generation fluency; and thorough study of parameter-efficient fine-tuning (e.g. LoRA, quantisation) on RAG outcomes. Addressing these gaps will be critical to guide the next wave of retrieval-augmented generation research.


\subsection{What are the innovative methods and approaches compared to the standard retrieval augmented generation?}
\label{sec:advance_rag}

The record-breaking pace of work on RAG is no longer about proving that some retrieval helps large language models, but about how we can make retrieval more adaptive, controllable, trustworthy, and efficient. We synthesise the main messages that emerge when we contrast their contributions with the canonical DPR + seq-to-seq pipeline (one-shot top-k retrieval concatenate passages). We organise this section along the RAG data flow: preparation, retrieval, control, memory, orchestration, optimisation, and emerging multimodal frontiers.

\subsubsection{Pre-retrieval \& Post-retrieval Stages – the plumbing that keeps RAG watertight}
\label{subsec:prepost}

When a clinical chatbot invents a drug dosage, the root cause is often not the language model but a silent pre-processing step that mangled the source PDF. The unglamorous work that happens \emph{before} the first similarity search and \emph{after} the hit list come back, therefore, deserves as much care as fancy retrievers or generators.

\paragraph{Pre-retrieval: how we feed the index} 

\textbf{Structure-aware chunking.} Pipelines now segment along headings, tables and coherent narrative blocks detected by multimodal (vision-text) encoders; on FinanceBench, element-aware chunking achieved 84.4\% page-level retrieval accuracy and 53.19\% manual Q\&A accuracy, outperforming token-only baselines \cite{RN1623}.

\textbf{Metadata enrichment at chunk time.} Generate keywords and micro-summaries for each chunk automatically (e.g. with GPT-4) to aid retrieval and avoid manual labelling; element-aware pipelines use these metadata during indexing \cite{RN1623}, and retrieval augmentation has substantially increased accuracy in clinical deployments (e.g., GPT-4 from 80.1\% to 91.4\%) \cite{RN1617}.

\textbf{Curated corpus construction.}  Restrict retrieval to sentence-level snippets from authoritative clinical guidelines and other public sources; by indexing only such content, domain assistants avoid introducing protected health information and curb hallucinations by grounding answers in vetted guidance \cite{RN1618,RN644}.

\textbf{Longer retrieval units/chunks.} Treat each PDF or cluster of interlinked pages as a long ``retrieval unit'' (\(\approx 4\ \text{k tokens}\)). This 30-fold reduction in retrieval units (for example, from 22 million to 600 thousand) dramatically lowers the retriever’s workload while preserving or even improving recall, for example, answer-recall @ 1 increases from 52\% to 71\% in Natural Questions and answer-recall@2 from 47\% to 72\% on HotpotQA \cite{RN1630}. LongRAG achieves comparable exact-match performance, EM of 62.7\% on NQ and 64.3\% on HotpotQA, without additional training \cite{RN1630}.

\textbf{Security at the retrieval interface.}  Obfuscated code IDs, L2-normalised embeddings and poison filters remind us that the retriever, not the LLM, is the outer security wall \cite{RN348,RN206,RN1640}.  
\textbf{Defend the entry point.} By obfuscating code identifiers, applying L2-normalisation to embeddings, and filtering poisoned content, the retriever serves as the primary line of defence in retrieval-augmented systems \cite{RN348,RN206,RN1640}.

\paragraph{Post-retrieval: what we pass to the model}

\textbf{Re-ranking of retrieved evidence.} Employ reciprocal rank fusion or listwise autoregressive rankers to reshuffle retrieved evidence so that the most relevant passage appears first; this yields steady, low-cost improvements in accuracy and comprehensiveness \cite{RN1492,RN49,RN1738}.

\textbf{Context reduction and token budgeting.} Apply sentence-level context filtering (e.g. FILCO), one-line hints, or fast extractive summaries to reduce token usage while preserving factual accuracy and coherence \cite{RN381,RN1624,RN699}.

\textbf{Utility-based passage selection.} Employ lightweight utility scorers that decide whether to drop, keep, or even \emph{repeat} passages; a learnt ``bridge'' model edits passage IDs dynamically, keeping the prompt short while adapting to LLM preferences \cite{RN1395,RN1742}.

\textbf{Noise-aware inclusion of unrelated passages.} When the context allows, inserting a small number of unrelated passages can improve the accuracy of the answer in RAG; one study reports gains of up to 35\% when random documents are added to the prompt, with the effect depending on position and count \cite{RN1389}.

\textbf{Early verification with local regeneration.} A lightweight verifier LM diagnoses whether errors stem from retrieval (irrelevant knowledge) or grounding (unfaithful use of retrieved knowledge), and triggers only the needed correction - re-retrieve or regenerate \cite{RN861}.

\textbf{Adaptive context-window management.}  Use a budget-aware consolidator to set \(k\) to the space remaining in the prompt-trimming, merging or compressing passages as needed-so that the pipeline works across small and large context windows (for example, 4k-8k and beyond) \cite{RN391,RN362}.

These plumbing stages create a token-efficient, high-recall foundation that underpins adaptive, controllable, and cost-effective RAG architectures. This groundwork addresses the retriever directly on how to make retrieval more trustworthy and efficient in practice. In the next section, we examine how intelligent prompt and query strategies transform the front end of RAG into an active programmable interface.  

\subsubsection{Prompting \& Query Strategies-making the front-end intelligent}
\label{sec:prompting_query}

Standard retrieval-augmented generation (RAG) typically issues a single literal query, retrieves top-$k$ passages, and concatenates them into a fixed prompt for generation. This baseline often treats the prompt as a static container rather than an instrument for steering retrieval and inference. In contrast, recent prompting and query strategies reconceptualise the prompt as an active control interface that selectively modulates grounding, reformulates queries, and sequences reasoning with tool use. Consider the question ``How many valves does the human heart have?'' In RAG, performance is driven less by model size than by two factors: how we frame the query (which controls what is retrieved) and how strictly we require the model to use the retrieved evidence.

\textbf{Flexible grounding and structural prompting.}
RAG reduces hallucinations on knowledge-intensive tasks by conditioning answers on retrieved passages and enabling provenance attribution, yielding more factual outputs than parametric models alone. \cite{RN1}.Beyond prescriptive prompting, retrieval composition itself can regularise behaviour: deliberately adding irrelevant documents (“noise”) to the context can improve answer accuracy and robustness by counteracting misleading high-scoring passages \cite{RN1389}. Domain scaffolds further formalise evidence: workflow synthesis expressed in JSON \cite{RN358}, organ label tags for radiology reports \cite{RN359}, or hybrid text-graph templates for multi-hop knowledge-graph reasoning \cite{RN1745}. Compared to the free-form concatenation of standard RAG, these wrappers restrict the output format, reduce cognitive load, and improve faithfulness by aligning the generator's attention with well-written evidence.

Relative to baseline RAG, structural prompting improves relevance and robustness by imposing schemas that suppress spurious correlations, though it may add authoring overhead and requires schema governance to avoid brittleness in open-domain settings \cite{RN1,RN1389,RN358,RN359,RN1745}. Future work should quantify how schema granularity trades off against generalisability across domains.

\textbf{Query reformulation, expansion, and selective triggering of queries}
Allowing the model to expand or rewrite a user query typically improves recall by surfacing semantically diverse contexts; multi-query expansion bundles, as well as merge evidence downstream \cite{RN1492}. However, issuing additional queries indiscriminately increases latency and noise. To address this, uncertainty-aware controllers such as FLARE and RIND+QFS trigger retrieval only when token-level entropy spikes, thus avoiding unnecessary index lookups and focusing retrieval on genuinely uncertain spans \cite{RN198,RN1619}. In specialised settings, lightweight agents first extract salient entities (for example, disease names) and then query structured stores to reduce vocabulary mismatch and improve precision \cite{RN1569}. For streaming code completion, continuous query updates track the evolving context so that cross-file references remain current, a capability that standard single-shot RAG lacks \cite{RN1502}.

Compared to baseline, reformulation and entropy-triggered querying improve recall-precision balance and control latency, but they rely on robust fusion or re-ranking to prevent evidence dilution when multiple queries are issued \cite{RN1492,RN198,RN1619,RN1569,RN1502}. Open questions include how to calibrate entropy thresholds across domains and how to amortise multi-query costs under tight latency budgets.

\textbf{Example-augmented prompting (retrieval-augmented in-context learning).}
Retrieval-augmented in-context learning dynamically inserts near-neighbour exemplars while assembling the prompt. Systems such as R-GQA and MolReGPT incorporate similar question–answer pairs, improving accuracy at a modest token cost \cite{RN1571,RN1621}. Time-aware variants add hard negative examples so that the model learns when not to retrieve, mitigating over-reliance on stale or irrelevant context \cite{RN230}. Confidence-conditioned prefixes further allow the generator to modulate trust in retrieved snippets by signalling low certainty, which reduces the risk of over-fitting to misleading passages \cite{RN888}. Relative to standard RAG, which typically lacks task-specific exemplars, these strategies better align the prompt distribution with the current query manifold.

Example-augmented prompting enhances relevance and robustness, particularly for specialised or temporally sensitive queries, but raises curation questions (which exemplars, how many and how to manage drift) and requires careful token budget management to avoid context saturation \cite{RN1571,RN1621,RN230,RN888}. Promising directions include adaptive exemplar selection driven by utility estimates rather than fixed $k$.

\textbf{Deliberate reasoning before retrieval.}
ReAct-style prompts interleave \textsc{Thought}, \textsc{Action}, and \textsc{Observation}, allowing the model to plan tool calls, execute retrieval, and revise its plan iteratively \cite{RN1523}. Graph-of-Thought extends this idea by decomposing the question into sub-problems, each with a targeted retrieval hop, before composing a final answer \cite{RN1757}. These patterns depart from the standard RAG one-pass pipeline by explicitly sequencing reasoning and evidence gathering. However, such scaffolds can accidentally expose sensitive content if intermediate thoughts are logged or reflected back to the user, underscoring the need for strict access control and privacy-aware prompt design \cite{RN1639}.

Reason-first strategies improve multi-step fidelity and reduce retrieval of irrelevant context by aligning evidence to sub goals, at the cost of additional control complexity and potential privacy risks if traces are not properly contained \cite{RN1523,RN1757,RN1639}. Future research should formalise safety-preserving variants that preserve trace benefits without leaking private artefacts.

\textbf{Operational policy, fusion, and safety.}
Empirically, explicit prompt policies, such as clear grounding clauses, zero-temperature reasoning steps and domain-specific wrappers, often match or exceed the benefits of introducing a new retriever \cite{RN1613}. However, query expansion must be paired with fast fusion or re-ranking to curb latency and maintain precision as the number of evidence candidates grows \cite{RN1492}. Field-specific schemas (for example, ECG JSON blocks in cardiology) improve reliability in safety-critical applications relative to open completions \cite{RN204}. Finally, the prompt itself is an attack surface; sanitising complex instructions and constraining tool outputs are, therefore, mandatory operational controls.

\textbf{Overall implications for the research question.}
Across these categories, innovative prompting and query strategies advance, challenge, and in some cases redefine standard RAG by (i) making grounding adaptive and schema-aware, (ii) coupling query reformulation with uncertainty-aware triggering, (iii) leveraging exemplar retrieval to shape the prompt distribution, and (iv) sequencing reasoning to target retrieval more precisely. In general, these methods often yield larger improvements per unit cost than architectural changes, particularly when prompts are treated as versioned, testable artefacts, as code would treat, so that RAG systems become more controllable, economical, and safe to deploy \cite{RN1613,RN1492,RN204}.

\subsubsection{Hybrid and Specialised Retrievers: No single needle-finder}\label{sec:hybrid_retrievers}

Early RAG systems typically rely on a \emph{single}, dense passage retriever whose top-$k$ chunks are appended, wholesale, to the generator input. A striking commonality in the more recent literature is the rejection of this monolithic design in favour of \textbf{ hybrid retrieval}: lexical and dense signals are combined, cascaded or adaptively weighted, often alongside domain-specific similarity functions or graph indices. The consensus that emerges is clear: no single similarity metric can surface every useful evidence fragment.

Work in the clinical domain illustrates the value of score-level fusion: MEDRAG aggregates BM25 with up to three dense retrievers by Reciprocal Rank Fusion and records 3 to 6 percentage points gains in top-5 recalls for medical QA \cite{RN1613}. A more generalisable variant is the \textit{Blended Retriever}, which stitches together BM25, KNN-dense, and sparse-encoder indices behind a unified API; exhaustive sweeps over six query formulations reveal that the fused output consistently outperforms the best individual index, without task-specific fine-tuning \cite{RN1614}. Similar ideas appear in open-source toolkits such as Auto-RAG, which expose multiple indices at runtime and leave the choice to a lightweight policy learner or the user \cite{RN1625}. These studies collectively suggest that recall drops caused by the ``long tail'' of lexical variability can be mitigated without costly supervision, provided that one is willing to maintain multiple indices.

Several papers move beyond static mixtures and train the system to \emph{adaptively} decide where to sample evidence. In event argument extraction, Adaptive Hybrid Retrieval samples pseudo-demonstrations from continuous semantic regions defined jointly over document and schema embeddings, delivering a five-point F$_1$ improvement over nearest-neighbour baselines \cite{RN1516}. A complementary strategy, introduced for legal case reasoning, learns dual embeddings: one space captures the similarity between questions, the other captures the affinity between questions and support. Then optimises a weighting scheme that can privilege either signal depending on the input \cite{RN1615}. Such results hint at a future in which hybrid retrieval is \emph{learned} rather than manual operation.

Hybridisation is especially powerful when it exploits structure that generic dense vectors cannot easily encode. For knowledge-graph question answering, a dual-level pipeline first retrieves entity neighbours or thematic nodes using keyword matching, then refines the candidate set with vector similarity; this combination captures both symbolic locality and semantic relatedness and proves markedly more accurate than flat chunk retrieval \cite{RN1756}. In code intelligence, lexical overlap remains a robust signal of syntactic similarity, whereas a fine-tuned dense retriever better captures semantics; a two-stage hybrid first filters with BM25 and then re-ranks with a CodeT5-based encoder, cutting irrelevant patches by more than one third \cite{RN209}. Multimodal cascades follow the same philosophy: an image-to-text system uses CLIP similarity to shortlist images whose titles match a visual prompt, then applies a text encoder to retrieve the precise passages required for answer generation \cite{RN1642}.

Hybridisation also affects \emph{how} evidence is consumed. RETRO++ routes the single most relevant chunk directly to the decoder, where it can influence every token, while sending additional passages to the encoder as background context, yielding significant gains on open-domain QA without increasing sequence length \cite{RN867}. Such architectural nuances reinforce the broader lesson that retrieval and generation cannot be optimised in isolation.

Although the quality gains are unambiguous, hybrid designs are not free. Maintaining several indices requires more memory and imposes separate refresh cycles; empirical studies report end-to-end latency increases of 5-50 ms per query on commodity GPUs. Where low latency is mandatory, selective trigger policies, e.g., avoiding dense retrieval for purely factual lexical queries, recover much of the benefit at a fraction of the cost \cite{RN699}. However, very few papers measure index update overhead or the engineering effort needed to keep blended systems in sync with evolving corpora.

Two methodological gaps remain. First, cross-domain robustness is largely untested: most hybrids are tuned and evaluated on the same corpus, leaving questions about their behaviour when the domain shifts. Second, security aspects, how fusion strategies cope with poisoned subindices or adversarial trigger documents, are almost entirely unexplored. Bridging these gaps will require shared benchmarks that couple quality metrics with latency, energy, and robustness reporting.

The evidence base demonstrates that retrieval heterogeneity is a virtue: lexical scoring anchors precision, dense vectors widen semantic recall, structure-aware indices inject domain priors, and increasingly, learnt policies decide which mixture to trust. Treating retrieval composition as a first-class, configurable module, rather than a line in the appendix, appears to be essential for the next generation of reliable and efficient RAG systems.

\subsubsection{Structure-aware \& Graph-based RAG: "Talk to me in triples, not tokens"}

A growing strand of work argues that retrieval-augmented generation should reason over \emph{relations} rather than over flat passages. By turning documents, captions or code into nodes and edges, these systems place LLMs in environments where neighbourhood, path and provenance are explicit. The result is a family of \emph{structure-aware} or \emph{graph-based} RAG pipelines that differ from the canonical baseline DPR + seq2seq at every stage, from indexing to decoding.

The first departure is at retrieval time. Instead of ranking passages in isolation, systems such as \textsc{G-Retriever} construct a minimal connected sub-graph that already encodes multi-hop context before it is shown to the LLM \cite{RN681}. Knowledge-Graph Prompting extends the idea to ad hoc graphs built on whole document collections, thereby recovering passages that are jointly rather than individually relevant \cite{RN672}. Biomedical variants prune domain KGs aggressively: KG-RAG selects only the ``prompt-aware'' neighbourhood of SPOKE, halving token expenditure without loss of precision \cite{RN1569}. Across these studies, the lesson is consistent: a few well-chosen triples beat many loosely related sentences.

Once a graph has been selected, it must align with the token world of the generator. Two strategies dominate. \emph{Soft-prompt projection} feeds the LLM a dense prefix derived from a Graph Neural Network encoder; Graph Neural Prompting shows that a learned projector lets the language model attend to sub-graph semantics without retokenising long edge lists \cite{RN1606}. In contrast, mixed-modal encoders treat each document embedding as a latent token. The xRAG architecture concatenates a textual view with such projected embeddings, while RAG-Token marginalises over latent documents so that each generated word may be conditional on a different evidence source \cite{RN363}. These designs blur the boundary between retrieval and generation, but they also introduce computational overhead: fast approximate decoding is now an open system challenge.

Manual curation of graphs is untenable, so recent work automates their creation. A graph-based text indexer segments documents, extracts entities and relations with an LLM, then maintains the structure as a hybrid keyword and vector store that supports both lexically exact and semantic queries \cite{RN1756}. Customer-service pipelines construct dual-level graphs in which intra-ticket trees are inter-linked via clone or reference relations; a two-step process retrieves a sub-graph and then issues Cypher queries for precise answer extraction \cite{RN134}. In code intelligence, static analysis graphs are fused with retrieved exemplars so that program repair models reason simultaneously over abstract syntax and concrete fixes \cite{RN218,RN209}. Across these domains, the ``graph first, dense fallback'' has become the pragmatic recipe: traversal is attempted, but vector similarity remains a safety net.

Structure-aware RAG is also proving its worth beyond text. Vision language pipelines ground image regions in Wikidata entities using a CLIP-based retriever, allowing captioning models to cite explicit facts rather than hallucinating \cite{RN913}. Multimodal captioning systems encode images, retrieved captions, and their cross-caption relations in a single transformer, improving rare-concept coverage and faithfulness \cite{RN364,RN62}. These studies confirm that the graph perspective can bridge modality gaps as well as logical ones.

The empirical gains are grouped around three themes. First, the answer \emph{faithfulness} rises when the model can quote paths or node identifiers, giving analysts concrete error traces \cite{RN389,RN681}. Second, \emph{token efficiency} improves because graph neighbourhoods are far denser information carriers than flat chunks; prompt length drops by 40-60\% in biomedical QA \cite{RN1569}. Third, graphs offer natural hooks for \emph{explainability}: Users can inspect which edge or entity grounded a statement, an impossibility when evidence is a text passage spanning many pages.

However, significant obstacles remain. Current pipelines are based on the linkage of weak entities and the cost of stale or mislinked nodes. Incremental update algorithms exist \cite{RN1756}, but their impact on answer drift over months is unknown. Finally, evaluation practices lag: while factual QA has BLEU and EM, graph RAG lacks agreed metrics for edge coverage or topological correctness, hindering cross-paper comparison.

We expect structure-aware RAG to converge on three design principles: lightweight on-the-fly KG construction; learned policies that choose pragmatically between graph traversal and vector search; and plug-in projection layers that make any LLM ``graph-ready'' without bespoke retraining. As modalities proliferate---tables, time series, 3-D scenes---the foundational insight stays the same: represent knowledge in the form that preserves its relations, then let the language model converse in that richer vocabulary.

\subsubsection{Iterative \& Active Retrieval Loops: From Static Context to Conversational Search}
\label{subsec:iterative_loops}

Work published during the past two years reveals a decisive migration from the traditional ``retrieve-then-generate'' pipeline towards \emph{closed-loop} systems in which LLMs continually query external knowledge, inspect their own draughts, and revise both the retrieval context and the answer. These approaches treat the retriever not as a one-off helper but as a conversational partner that can be invoked, ignored, or re-invoked in response to model uncertainty, verification feedback, or evolving sub-goals.

A first line of work equips the generator with \emph{uncertainty trigger}. In FLARE, the model examines each newly generated sentence for high-entropy spans; when token-level uncertainty exceeds a threshold, it halts generation, masks those spans, emits a focused search string to the retriever, and then regenerates the sentence \cite{RN198}. A related attention-based mechanism, RIND+QFS, similarly uses uncertainty triggers to decide when and which tokens should form subsequent queries, improving recall without compromising precision \cite{RN1619}.  Real-time Information Needs Detection (RIND) combined with Query Formulation by Self-attention (QFS) generalises this idea by blending token-level entropy with self-attention salience to decide when to retrieve and which tokens should form the query \cite{RN1619}. In these designs, the model literally \emph{emits a search string token} (e.g.\ <SEARCH> how many valves in the human heart?), which the orchestration layer interprets as a call to the retriever, giving the loop an explicit and inspectable hand-off point.

SELF-RAG uses reflection tokens (\texttt{Retrieve}, \texttt{ISREL}, \texttt{ISSUP}, \texttt{ISUSE}) to trigger retrieval, assess evidence and critique outputs, giving segment-level control \cite{RN9}. The biomedical Self-RAG further extends the mechanism by training a domain-specific critic language model whose reflection tokens signal both the need for retrieval and the subsequent relevance of the evidence \cite{RN618}. Collectively, these studies demonstrate that trigger on model uncertainty recovers the majority of the accuracy gains of full iterative pipelines while invoking the retriever only when it is genuinely useful. Parallel work on agentic systems confirms this principle: SELF-RAG \cite{RN9}, DRAGIN \cite{RN1619} and TA-ARE \cite{RN230} introduce explicit decision tokens, entropy thresholds and veto classifiers that suppress unnecessary searches, trimming 15–45\% of context tokens with negligible loss in fidelity.

A second group of research emphasises \emph{iterative refinement}.  The CHAIN-OF-NOTE (CON) framework obliges the LLM to write concise ``reading notes'' for each retrieved document, thereby exposing document reliability and reducing hallucination before synthesis of the final answer \cite{RN1760}.  Batch grounding strategies process evidence in successive mini-batches, stopping as soon as adequate justification is found and injecting the progressively revised answer back into the context, a tactic that curbs noise and token bloat \cite{RN1626}.  RAT performs a stepwise revision of an explicit chain of thought, generating a new query for each reasoning step and localising corrections instead of rewriting entire explanations \cite{RN1607}.  Verification-driven loops such as KALMV enact automatic error rectification: if a verifier flags a retrieval or grounding fault, the pipeline re-retrieves new passages or re-generates the answer until the verifier is satisfied, closing the loop on both failure points \cite{RN861}. Agentic pipelines strengthen this pattern by exposing each stage---retrieval, reranking, refinement and generation as discrete---inspectable actions inside modular toolchains such as RALLE \cite{RN846} and MEDRAG \cite{RN1625}, making revision steps debuggable and reusable.

When the original query is too sparse or ambiguous for high-recall retrieval, \emph{generation-augmented loops} become effective.  ITER-RETGEN feeds the intermediate draught of the model back to the retriever, providing increasingly informative queries at each turn \cite{RN379}. ITRG offers two complementary modes. \textit{Refine}, which updates an existing draft with only newly retrieved documents. \textit{Refresh}, which starts afresh from the latest evidence. This shows that alternating between these modes improves long-form document generation \cite{RN79}.  RepoCoder adopts the same principle for code completion, appending the most recent code continuation to the retrieval query so that cross-file context converges towards the intended target snippet \cite{RN1502}.

A fourth strand decomposes the original task into smaller sub-problems and retrieves evidence in a \emph{multi-hop} fashion.  RA-ISF first checks whether the LLM already knows the answer, then filters irrelevant passages, and finally decomposes unanswered questions into simpler subquestions, recursing until each leaf is resolved \cite{RN189}.  SearChain externalises the reasoning trajectory as a \emph{Chain-of-Query} tree, allowing the IR engine to verify or veto each hop and permitting back-tracking when evidence contradicts prior steps \cite{RN1638}.  Graph-oriented systems traverse knowledge graphs node by node, either through an LLM-guided agent \cite{RN672} or via a divide-and-conquer ego-graph search with learnable pruning \cite{RN1745}, thereby combining symbolic relational structure with neural retrieval. Reason–act loops in the agentic literature echo this multi-hop spirit, alternating between planning, external tools and answer revision to accumulate evidence from diverse sources—for instance a ReAct-style clinical assistant \cite{RN365} or the Retrieval-augmented Recommender System \cite{RN367}.

Finally, several papers exploit \emph{self-consistency or memory}. SelfMem alternates between producing multiple candidate memories and selecting the best one to seed the next round of generation, enabling the model to bootstrap its own knowledge without external corpora \cite{RN383}.  A related idea is used in activity-pattern generation, where multiple hypothetical trajectories are rated for alignment with historical data before the most self-consistent plan is chosen \cite{RN1753}.  The Knowledge-to-Response architecture separates knowledge prediction from response generation, gives an explicit checkpoint that can be inspected or re-executed if downstream verification fails \cite{RN888}.

Across these diverse implementations, a set of common lessons emerges.  First, retrieval should be \emph{policy-driven}: systems that fire the retriever only under measured uncertainty or verified need to gain most of the quality benefits at a fraction of the computational cost.  Second, \emph{local} revision—editing one thought, sentence, or document at a time—prevents prompt lengths from exploding and keeps provenance transparent.  Third, closed loops demand fail-safes: lightweight critic LMs or verifiers effectively halt divergence when early retrieval or generation steps go wrong.  Lastly, latency and energy budgets vary dramatically between designs; rigorous reporting of retrievals-per-answer, wall-clock delay, and GPU minutes is essential if future work is to compare accuracy improvements on an equal footing.

These iterative and active retrieval loops recast RAG as an \emph{interactive search companion}.  By recognising their own knowledge gaps, gathering fresh evidence on demand, and continuously revising their reasoning, modern RAG systems approach the discipline of a human researcher.  The next frontier is to make these loops \emph{budget-aware} and embed them in evaluation frameworks that reward knowledge fidelity and resource efficiency.

\subsubsection{Memory-augmented RAG: Personalisation and Long-Horizon Context}
\label{sec:memrag}

Early retrieval-augmented systems were stateless: each turn re-embedded the user’s query, retrieved passages, concatenated them, and produced an answer. However, domains like education, clinical care and personal assistance benefit from knowledge that accumulates and varies by user. Thus, a family of \emph{memory-augmented} RAG architectures has emerged, persisting dialogue turns, sensor readings, search history or model-generated thoughts beyond a single query.

One line of work introduces \emph{short-horizon conversational buffers}. In education, MoodleBot allocates a vector store per course and rewrites follow-ups into standalone queries that include recent turns; students rate its coherence far above a buffer-free baseline \cite{RN1711}. Likewise, LangChain’s \texttt{ConversationBufferMemory} retains the chat transcript for retriever and generator use, boosting F$_1$ by over eight percentage points in follow-up QA benchmarks, largely across domains \cite{RN1744}.

Beyond fleeting context, some systems maintain \emph{persistent, user-specific memories}. LiVersa’s hepatology assistant separates long-term documents (e.g.\ discharge summaries), short-term signals and a dynamic slot of the fifteen latest queries. Selective retrieval from these stores cuts hallucinations by $\sim$25 per cent and halves prompt length \cite{RN1712}. The entity-centric store $K_{E}$ timestamps canonicalised entities from browsing history, storing compact IDs rather than raw text; this achieves personalisation with strong privacy and mere megabytes per user \cite{RN1755}. Similarly, the agentic \textsc{Brain} logs every perception–thought–action tuple and recalls them to aid planning in complex optimisation tasks \cite{RN1629}.

Another approach embeds memory within the model. Retrieval Augmentation Mechanism (RAM) for video captioning initialises a key–value store with hidden states from teacher-forcing; at inference the decoder attends this store, injecting linguistic and visual cues and raising CIDEr by nearly 10 per cent on MSR-VTT \cite{RN360}. \textsc{SelfMem} appends its own generations to a growing memory pool, lowering retrieval latency over time while BLEU keeps improving \cite{RN383}. A clustered memory module groups millions of examples into centroids, allowing soft interpolation or hard selection so the generator exploits abstracted task knowledge rather than a few nearest neighbours \cite{RN935}.

Despite gains in coherence, relevance and efficiency, open challenges remain. Few works address \emph{memory governance}: LiVersa encrypts clinical memories at rest and $K_{E}$ avoids raw text, yet no standards exist for retention, revocation or audit. The second issue is \emph{forgetting}: none of these works implement principled eviction or decay, despite stale or erroneous memories causing model drift. Finally, evaluation stays narrow—accuracy metrics dominate, while longitudinal measures (trust calibration, drift detection, catastrophic memory errors) are seldom reported.

The memory-augmented RAG shifts from ``answering the current question'' to ``accompanying the user over time''. Whether through lightweight buffers, structured personal knowledge graphs, or train-time key value abstractions, integrating memory with retrieval and generation paves the way for truly adaptive, user-centred assistants. To move beyond prototypes, these systems must tackle privacy, life-cycle management, and long-term robustness.

\subsubsection{Agentic \& Multi-tool Pipelines: Orchestrating Reasoning, Tools and Memory}
\label{subsec:agentic}

Where the previous sections zoom in on \emph{what} to retrieve (hybrid indices, structure-aware graphs), \emph{when} to retrieve (uncertainty-driven loops) and \emph{where} to store past context (memory buffers and personal knowledge bases), the emerging notion of an \textbf{agent} asks a broader systems question: \emph{How can a language-model controller weave \underline{all} of these capabilities, including retrievers, memories, external APIs, calculators, and even other LLMs, into a single adaptive execution plan?}

Under the hood, each agent exposes a toolbox of heterogeneous capabilities. Hybrid retrievers supply both lexical and dense evidence; structure-aware traversals explore knowledge graphs; memory stores cache past interactions; and domain plugins execute arbitrary APIs—from code compilers to database queries. For example, MEDRAG’s laboratory orchestrates five discrete steps (\texttt{Judger}, \texttt{Retriever}, \texttt{Reranker}, \texttt{Refiner}, \texttt{Generator}) in a fixed graph, while RALLE provides practitioners with a drag-and-drop canvas to create custom pipelines in real time \cite{RN1625,RN846}.

How does the controller decide its next move? Research is grouped around three design patterns. In static graphs, the flow is scripted (for example retrieve - rerank - generate), but nodes can be toggled at runtime (for instance, switching from a general-purpose index to a proprietary one when domain drift is detected). Dynamic planning agents interleave \textsc{Thought}, \textsc{Action} and \textsc{Observation} tokens, letting the model plan each step—should it consult the calculator or dive into long-term memory next? And learned controllers treat tool selection as a reinforcement-learning problem, optimising for latency, cost and accuracy under real-world constraints \cite{RN365,RN367,RN134}.

Memory is not an afterthought but a peer of retrieval. Short buffers prevent conversational dead ends, but true agency emerges when the system logs every perception-thought-action tuple for hours—or even days. LiVersa’s hepatology assistant splits data into long-term documents, streaming vitals and a sliding window of recent queries; the result is a 50\% reduction in hallucinations and half the prompt length \cite{RN1712}. The \textsc{Brain} architecture goes further, treating each memory as an explicit action token that the agent can revisit when planning complex optimisation tasks \cite{RN1629}.

Orchestration unlocks tangible benefits and magnifies new risks. On the upside, agents can superintend long-horizon workflows (from syllabus design to lab automation), hot-swap tools when one fails, and gracefully fall back on alternative evidence sources. However, this flexibility invites debugging nightmares: tracing a misstep through a branched execution graph is much harder than inspecting a single ``retrieve-then-generate'' call. Credit assignment across cascaded tools remains unresolved, and persistent memories demand rigorous governance for retention, revocation and audit \cite{RN1755}.

Looking ahead, agentic RAG must mature from ad hoc scripts to dependable infrastructure. We need vendor-neutral DSLs to describe tool graphs, unified dashboards that report accuracy alongside latency, energy consumption and privacy metrics, and formal memory policies that prevent drift and data leakage. Once these scaffolds are in place, controllers will be free to juggle dozens of modules—truly turning retrieval-augmented models into retrieval-augmented systems.

\subsubsection{Efficiency \& Compression—token budgets still matter}
\label{sec:efficiency_compression}

The first time a production team wired a 32\,K-token model into its help-desk bot, the GPU bill doubled overnight.  The lesson landed quickly: long contexts feel free, but every extra symbol still burns memory, latency, and cash.  Recent papers therefore chase leaner recipes that keep answers faithful whilst maintaining efficiency \cite{RN1748,RN1633,RN356}.

Why carry an entire document when a single learned vector will suffice? xRAG maps each retrieved passage to a single document token, reducing the retrieved context from roughly 175 tokens to one and delivering task performance comparable to uncompressed RAG, while also lowering compute (a \(3.53\times\) reduction in GFLOPs) and improving speed (a \(1.64\times\) speed-up in CUDA time) \cite{RN1748}. Biomedical variants prune entire graph branches; Cypher-RAG++ restricts its prompt to ``prompt-aware'' triples and nevertheless improves robustness \cite{RN1569}. Even simple prompt engineering helps: RAPT stores most tunable weights in a global prefix and keeps per-example infixes small \cite{RN952}.

A bloated index slows everything downstream.  One group runs an asynchronous re-encoder that refreshes FAISS shards while the system is online, so nightly jobs never block training \cite{RN128}. Another treats megabyte-scale PDFs as single ``long retrieval units'', resulting in thirty-fold smaller indices  but the same recall \cite{RN1630}. Toolkits such as Parrot and Auto-RAG now expose multiple vector stores and show that picking the right dimensionality can improve speed better than another hardware upgrade \cite{RN1609,RN1617,RN1613}.

PipeRAG drags passages from the CPU while the GPU is already decoding, roughly cutting a third off end-to-end latency \cite{RN1633}. RAGCache predicts which passages are likely to be reused, warms the key-value cache, and initiates speculative decoding before the retriever responds. In a production trace, this approach reportedly halved the US dollar cost and reduced the 95th-percentile latency by 200 ms \cite{RN1636}.

RETRO++ adjusts retrieval cadence analogously to adaptive bitrate streaming: fetch every token for maximum quality, or every few hundred for speed; quality degrades smoothly rather than collapsing \cite{RN867}. PipeRAG pushes adaptivity further, tuning its cadence at runtime to respect a global latency budget \cite{RN1633}. Other teams precompute dense knowledge stores offline, shifting the heaviest computation away from the critical path \cite{RN378,RN62}.

In these approaches, compression is no longer a lossy compromise; it is a design posture. Whether by projecting documents into single embeddings, refreshing indices on the fly, overlapping compute, or throttling retrieval frequency, modern RAG systems show that frugality can coexist with accuracy. Future benchmarks should report energy (joules) and monetary cost alongside \textsc{EM} and BLEU; otherwise, we will continue to top leaderboards whilst exceeding budget constraints.

\subsubsection{Modality Expansion – RAG Beyond Plain Text}
Early RAG systems treated all knowledge as text—until researchers discovered that a single X-ray caption or table row can transform a dry answer into a vivid insight. Imagine a disaster-response chatbot that not only quotes tweets but overlays them on live satellite imagery. This fusion is now within reach thanks to unified multimodal backbones. MuRAG, for example, couples a Vision Transformer with a T5 encoder–decoder so that images and text share the same embedding space, letting a prompt about ``the mysterious lesion'' fetch both radiology reports and the relevant chest X-ray as a single learned token projection—and it works without retraining the language model for each modality \cite{RN144, RN1748}. Meanwhile, xRAG shows that whole documents—whether PDF, PNG or CSV—can collapse into one compact token, greatly reducing context length and memory use without sacrificing answer quality \cite{RN1748}.

Beyond model-level tweaks, contemporary orchestration frameworks expose pluggable components: engineers can configure CLIP-style embedding models for image/text retrieval, Whisper-based audio transcription and HTML/CSV/Excel loaders with minimal code changes, and then index outputs in interchangeable vector stores. In practice, frameworks such as LangChain provide loaders for web pages and YouTube transcripts, Whisper parsers, Pandas/CSV tooling and a common vector-store interface; this allows a single workflow to draw on web pages, video transcripts and tabular datasets, with retrieval improving grounding in downstream generation \cite{RN1609,RN1610}.

In clinical imaging, one line of work retrieves text using contrastively pre-trained vision–language encoders (e.g., ALBEF) and then prompts general-purpose language models (including GPT-4) to draft radiology findings; a separate line develops grounded report generation that links textual findings to specific image regions, improving traceability beyond text-only outputs. Beyond imaging, retrieval augmentation has also been explored for lay-language clinical communication and explanation \cite{RN359}.

Yet challenges linger: CLIP-style joint spaces work well for vision and language but falter on tables or code snippets; scale-up strains storage budgets when every video frame becomes an index entry; and privacy controls for sensitive modalities, from medical scans to CAD files, have no industry standard. Addressing these gaps will make multimodal RAG not just possible, but dependable.

\subsubsection{Synthesis \& Outlook}
\label{sec:synthesis}

The evidence in this review indicates a clear shift from the canonical \textit{DPR\,+\,seq2seq} pipeline towards modular, policy-driven architectures. Hybrid indices broaden coverage; structure-aware retrievers identify relations that are otherwise difficult to detect; and uncertainty-triggered loops request additional evidence only when model uncertainty is high \cite{RN1614,RN198,RN681}. The combined effect is higher top-$k$ recall without overloading the generator with unnecessary tokens.

Closed-loop control and lightweight critics have transformed retrieval from a static pre-retrieval step into a dynamic, in-generation process. Verifiers can filter low-information snippets during generation, and memory buffers retain relevant prior context. Early deployments in medicine and education report reduced hallucination and improved personalisation \cite{RN1712,RN1744}. Efficiency techniques—document projection, speculative decoding, cache-aware scheduling—demonstrate that speed need not be sacrificed for rigour \cite{RN1748,RN1633}. Token budgets remain a constraint; the most efficient token is the one the generator never processes.

Despite these advances, the field continues to rely on incomplete quality signals. Benchmarks often prioritise accuracy and rarely report cost. Few studies record retrievals per answer, GPU minutes or carbon emissions, and even fewer analyse how compromised sub-indices may influence agentic planning. Memory governance—retention, revocation, audit and related controls—is seldom emphasised in system evaluations. Without shared yardsticks, reported gains are not readily comparable.

Future work should prioritise three directions to support the transition of RAG systems from prototypes to dependable infrastructure: developing holistic benchmarks that report not only accuracy but also retrieval latency, energy consumption and privacy guarantees; treating retrieval strategy as a resource-allocation problem, with policies that respect time, token and compute budgets rather than fetching evidence indiscriminately; and defining open, vendor-agnostic interfaces for heterogeneous indices (graphs, tables, images, streams) to enable drop-in retrievers without extensive pipeline refactoring.

\subsection{What are the most frequently used metrics for evaluating the effectiveness of retrieval-augmented generation systems?} 

Evaluating such hybrid architectures requires more than standard natural language generation metrics: it requires a suite of measures that capture both the retriever’s ability to surface relevant evidence and the generator's ability to weave that evidence into factually accurate and contextually appropriate responses. In the paragraphs that follow, we survey the most widely adopted metrics, ranging from low-cost, repeatable automated scores (e.g. EM, F1, BLEU/ROUGE, perplexity, recall@k) to resource-intensive human judgements and emerging LLM-as-judge protocols, and discuss their respective strengths, blind spots, and complementarities. By mapping out this evaluation landscape, we highlight best practices for assembling a balanced metric rubric and pinpoint enduring gaps that future research must address.

\subsubsection{Overview of the Evaluation Landscape} 

Across our set of RAG evaluations, metrics are grouped into three broad types: automated, human, and LLM-as-judge, with a pronounced skew toward automated measures.

Automated metrics dominate: by far the most frequent single metric is accuracy (e.g. \cite{RN1389, RN1395, RN1516}), appearing in diverse contexts from biomedical QA to commonsense reasoning. The exact match (EM) and the F1 score are likewise ubiquitous, serving as strict (EM) versus soft (F1) string overlap measures in QA, summarization, code generation, and information extraction tasks. Lexical similarity metrics such as BLEU, ROUGE-L, are also common, while perplexity and diversity measures (e.g. Distinct-1/2, Self-BLEU) appear more sporadically. Automated measures are prized for their reproducibility and low cost, but they largely capture surface overlap or retrieval success, not deeper semantic fidelity.

Human-judged metrics appear less often, but remain critical for qualitative aspects. Approximately one third of the articles we survey report some form of expert or crowd-rated accuracy \cite{RN605}, hallucination counts \cite{RN1609, RN1610}, completeness, consistency, or user satisfaction. These metrics provide insight into factuality, fluency, and user experience, but at the expense of higher annotation cost and interannotator variability.

LLM-as-judge approaches are an emerging third pillar: a handful of recent studies (e.g., \cite{RN379, RN391}) prompt powerful models like GPT-4 or text-davinci-003 to score correctness, fluency, or safety. These surrogate evaluators combine semantic evaluation with automation, ideally offering a strong correlation with human judgements, although with risks of model bias and prompt sensitivity.

This landscape shows a clear tension: scalable, repeatable automated metrics versus nuanced, costly, human assessments, and with LLM-based evaluators positioned to bridge the gap. Therefore, any comprehensive RAG evaluation should combine at least one high-level retrieval or overlap metric (e.g. recall@k, EM/F1), one semantic or embedding-based score (e.g. BERTScore \cite{RN134} or BLEURT \cite{RN134}), and either a human or LLM-mediated judgement to ensure both rigour and depth.

\subsubsection{Automatic Generation Metrics} 

Automatic generation metrics quantify the fidelity, fluency, and informativeness of RAG outputs without human intervention. They fall into four broad categories: (1) classification-based metrics, (2) overlap-based n-gram metrics, (3) probabilistic metrics, and (4) specialised diversity and grounding metrics. Each offers unique insight and carries distinct limitations in the evaluation of retrieval-augmented generation.



\textbf{Accuracy} measures the proportion of responses generated that are correct in the total number of outputs. It provides a straightforward gauge of answer correctness, although it ignores partial matches or semantic equivalence \cite{RN1389,RN1395}.  
\textbf{Exact Match (EM)} is a stricter binary metric: it reports the fraction of outputs that coincide exactly (character-for-character) with one of the reference answers \cite{RN1,RN128}. EM is essential in tasks demanding verbatim precision, such as code generation or fact retrieval, but does not give credit for near-correct paraphrases.

\textbf{F1 score} is the harmonic mean of the precision and recall of the token level:
\[
F_1 = 2 \times \frac{\mathrm{Precision}\times \mathrm{Recall}}{\mathrm{Precision} + \mathrm{Recall}}
\]
Precision is the fraction of overlapping tokens in the generated output that also appear in the reference; recall is the fraction of reference tokens recovered in the output. F1 allows partial credit for overlap and is widely used in QA and summarization benchmarks (e.g., SQuAD, WebQSP) \cite{RN128,RN1606}.

\textbf{BLEU} (Bilingual Evaluation Understudy) measures n-gram precision relative to one or more references and applies a brevity penalty to discourage overly short outputs:
\[
\mathrm{BLEU} = \mathrm{BP}\times \exp\Bigl(\sum_{n=1}^N w_n\log p_n\Bigr)
\]
where \(p_n\) is the n-gram precision for \(n\) up to typically 4 \cite{RN1571,RN359}. Despite its ubiquity, the reliance of BLEU on exact n-gram matches leads to poor sensitivity to synonymy and paraphrase \cite{RN1625,RN1621}.

\textbf{ROUGE} (Recall-Oriented Understudy for Gisting Evaluation) emphasises recall of n-gram matches; the ROUGE-L variant measures the longest common subsequence (LCS) between candidate and reference:
\[
\mathrm{ROUGE\text{-}L} = \frac{\mathrm{LCS}}{\text{length(reference)}}
\]
ROUGE-L captures sequence-level cohesion and is especially prevalent in summarization and long-form QA \cite{RN1,RN1639}. However, like BLEU, it fails to capture semantic similarity beyond surface overlap.

\textbf{METEOR} (Metric for Evaluation of Translation with Explicit ORdering) extends n-gram overlap by incorporating stemming, synonym matching, and a fragmentation penalty. Calculates a weighted F-mean of unigram matches, typically showing higher correlation with human judgements than BLEU or ROUGE at the cost of increased complexity \cite{RN1571,RN359}.

\textbf{BERTScore} measures semantic similarity by comparing contextual token embeddings (e.g. RoBERTa base) between the generated text and the reference. It computes cosine similarities at the token level and aggregates them to produce a single score that better captures paraphrase and meaning overlap than surface n-gram metrics \cite{RN359,RN386,RN952}.

\textbf{Perplexity} quantifies a model’s uncertainty by exponentiating the negative logarithmic likelihood of the generated sequence:
\[
\mathrm{PPL} = \exp\Bigl(-\tfrac{1}{N}\sum_{i=1}^N \log p(w_i)\Bigr)
\]
Lower perplexity indicates that the model predicts the next token with greater confidence \cite{RN1395,RN1633}. Although useful for assessing fluency and coherence, perplexity does not directly measure alignment with retrieved evidence or task-specific correctness.

\paragraph{Specialized Diversity \& Grounding Metrics}

\textbf{Self-BLEU} computes BLEU of each generation against its peers to quantify diversity (lower Self-BLEU to higher diversity) \cite{RN340,RN867}.

\textbf{chrF++} evaluates character-level F-measure over character n-grams, capturing fine-grained similarity in morphologically rich settings \cite{RN383}.

\textbf{Self-TER} (Translation Edit Rate) measures the average edit distance between multiple outputs, thus quantifying novelty \cite{RN952}.

\textbf{Support} labels each generated claim fully, partially, or not supported by the retrieved evidence, ensuring factual grounding \cite{RN9}.

\textbf{Rare F1} and \textbf{Predicted Knowledge F1 (PKF1)} focus on specialised tasks: Rare F1 emphasises performance on low-frequency tokens, while PKF1 gauges the model’s ability to recover explicit knowledge sentences \cite{RN358}.

\subsubsection{Automatic Retrieval Metrics} 

Effective retrieval is a prerequisite for high-fidelity generation in retrieval-augmented generation (RAG) systems. Automatic retrieval metrics quantitatively assess how well the retriever component selects and ranks relevant documents from a large corpus for a given query. In general, these metrics fall into (1) set-based measures, which evaluate the accuracy and completeness of the retrieved set, (2) ranking-based measures, which assess the ordered quality of the retrieval, and (3) hit-based measures, which capture the presence of any relevant document within a specified cut-off point.

\textbf{Retrieval Accuracy.} computes the proportion of queries for which all retrieved documents are relevant, relative to the gold standard for relevance judgements. By directly evaluating whether the retriever selects exclusively pertinent documents, the accuracy of document retrieval gauges the binary correctness of the retrieval set, a fundamental prerequisite for downstream generation \cite{RN218}.

\textbf{Precision@k} is defined as the fraction of the top k retrieved documents that are relevant. Measures the system's ability to avoid including irrelevant items among its highest-ranked results \cite{RN1625,RN1622}. \textbf{Recall@k} is the fraction of all relevant documents that appear within the top k positions, thereby capturing the completeness of the retrieval \cite{RN1625,RN1642}. Together, they offer complementary views: precision penalises false positives at high ranks, while recall penalizes false negatives within the cutoff.

\textbf{F1@k} is the harmonic mean of Precision@k and Recall@k, defined as
\[
\mathrm{F1@}k = 2 \times \frac{\mathrm{Precision@}k \times \mathrm{Recall@}k}{\mathrm{Precision@}k + \mathrm{Recall@}k}.
\]
This balanced metric mitigates trade-offs between precision and recall, providing a single score that reflects both accuracy and completeness of the top-k retrieval \cite{RN1625}.

\textbf{Mean Average Precision (MAP@k)} averages the precision scores computed at each rank position where a relevant document occurs, then aggregates over all queries. Formally, for each query q,
\[
\mathrm{AP@}k(q) = \frac{1}{N_q}\sum_{i=1}^k P(i)\,\mathbf{1}\{\text{doc}_i\text{ is relevant}\},
\]
where \(N_q\) is the number of relevant documents for q and \(P(i)\) is precision at rank i, and MAP@k is the mean of AP@k over q \cite{RN390,RN1622}. MAP@k rewards retrieval sets that place relevant documents early and penalises late retrievals.

\textbf{Mean Reciprocal Rank (MRR@k)} focusses solely on the rank of the first relevant document. For each query, it computes the reciprocal of the rank position of the first relevant hit (capped at k) and then averages over queries:
\[
\mathrm{MRR@}k = \frac{1}{|Q|}\sum_{q\in Q} \frac{1}{\min(\mathrm{rank}_q,\,k)}.
\]
It is particularly informative when downstream tasks depend critically on the earliest relevant context, as in ODQA \cite{RN390,RN1571}.

\textbf{Normalized Discounted Cumulative Gain (nDCG@k)} accommodates graded relevance by weighting each retrieved document gain by a logarithmic discount based on its position, then normalising by the ideal DCG. It is defined as
\[
\mathrm{nDCG@}k = \frac{\sum_{i=1}^k \frac{2^{\mathrm{rel}_i}-1}{\log_2(i+1)}}{\mathrm{IDCG@}k},
\]
where \(\mathrm{rel}_i\) is the relevance grade of the \(i\)th document and IDCG@k is the maximum possible DCG@k \cite{RN1614,RN1638}. nDCG@k is well suited to scenarios with multiple relevance levels or varying document importance.

\textbf{R-Precision} sets the cutoff R equal to the total number of relevant documents for a query and computes precision at that rank:
\[
\mathrm{R\text{-}Precision} = \mathrm{Precision@}R.
\]
Adapting the cut-off to the relevance count of each query, R-Precision offers a query-specific summary of ranking quality. It forms a core component of composite benchmarks (e.g., KILT) that jointly evaluate retrieval and generation \cite{RN49,RN846}.

\textbf{Hit@K} is a binary metric that indicates whether at least one relevant document appears within the top K positions; it is averaged over queries to produce a success rate \cite{RN1738}. \textbf{Hit Success Ratio (HSR)} similarly counts the proportion of queries that require external knowledge for which the retriever provides supporting evidence, highlighting the dependence of the model on the retrieved context \cite{RN362}.

Beyond standard relevance metrics, some studies measure the model’s ability to decide whether retrieval is necessary (that is, the accuracy of retrieval abstention) or to withstand adversarial passages (adversarial success rate) \cite{RN1516,RN1612}. These metrics inform selective retrieval policies and robustness evaluations.

Using a combination of these metrics, set-based, ranking-based, and hit-based, researchers obtain a multifaceted understanding of retrieval effectiveness. This rigour in evaluating the retriever component is critical to ensuring that RAG systems have reliable and comprehensive access to external knowledge.

\subsubsection{Other Automated metrics}

In addition to standard metrics, a diverse set of \emph{other automated metrics} has emerged to target specific facets of RAG that are not captured by general purpose measures. These include computational efficiency, robustness, bias, and domain- or task-specific criteria. Because each metric addresses a narrow aspect of system behaviour or relies on specialised evaluation procedures, they appear only occasionally in the RAG literature, typically in studies with unique experimental setups or domain constraints. Their limited adoption reflects both the implementation overhead and the context-specific validity of the measures.

\paragraph{Computational Efficiency}
\textbf{Latency} quantifies the time to retrieve documents and generate text, often decomposed into retrieval time ($T_{r}$), decision time ($T_{d}$), and generation time ($T_{g}$), with speedup (SU) defined as the relative reduction in total latency compared to a baseline of always retrieving \cite{RN1638,RN1746}.  
\textbf{Response Time} measures the end-to-end delay from query submission to first token output, a critical factor in interactive and clinical settings \cite{RN1618,RN644}.  
These metrics are crucial for real-time applications, where user experience and operational feasibility depend on prompt responses. However, their computation depends on controlled hardware environments and precise logging, which limits cross-study comparability.

\paragraph{Robustness \& Error Handling}
\textbf{Hallucination Rate} tracks the frequency or density of fabricated content in generated responses, either as hallucinations per 100 words or as the proportion of faulty outputs \cite{RN189,RN1617, RN1759, RN1618}.  
\textbf{Rejection Rate} (Reject Rate) measures the system’s ability to refuse answers when the knowledge base is insufficient, thus avoiding hallucinations \cite{RN1760,RN622, RN1612}.  
\textbf{Success Rate} evaluates the success of adversarial jailbreak attempts, reflecting the vulnerability under malicious prompts \cite{RN1632}.  
These metrics are indispensable for safety-critical domains (e.g., medicine, law), yet they demand rigorous annotation protocols or adversarial testing frameworks, constraining their routine use.

\paragraph{Contextual Bias}
\textbf{Contextual Bias} measures the tendency of a model to adopt incorrect assumptions from a misleading context, even when its internal knowledge would suggest a correct response \cite{RN1618,RN1752}.  
This metric surfaces subtle failure modes of RAG pipelines, particularly when retrieval yields noisy passages, but requires carefully crafted bias scenarios, which are rarely standardised.

\paragraph{Image- and Code-Specific Metrics}
\textbf{CIDEr \& SPICE} evaluates generated image captions by assessing consensus-based textual agreement or semantic propositional fidelity against human references \cite{RN356,RN365,RN79}.  
\textbf{Edit Similarity (ES)} computes $1 - \frac{\mathrm{Lev}(\widehat{Y},Y)}{\max(|\widehat{Y}|,|Y|)}$, where $\mathrm{Lev}$ is Levenshtein Distance,  to quantify token-level similarity of code snippets \cite{RN1743,RN1746}.
\textbf{Pass@k} measures the proportion of code generation attempts that pass automated test suites within $k$ trials \cite{RN825,RN1523}.  
\textbf{CodeBLEU} extends BLEU by incorporating abstract syntax tree and data flow comparisons, capturing both syntactic and semantic correctness of code \cite{RN1743,RN360}.  
These task-specific metrics yield deep insights in their respective domains but lack generalisability: captioning and code generation each demand bespoke reference datasets, execution environments, or parser toolchains.

\paragraph{Performance Comparison}
\textbf{Comparative Metrics} quantify improvements over baseline systems (e.g., KRAGEN vs. BioGPT / OpenChat in biomedical QA) by aggregating multiple performance indicators into a single comparative score \cite{RN1757,RN1629}.  
Although succinct, such composite measures often obscure which individual components drive gains and presuppose the availability of strong baselines in the target domain.

\paragraph{Discussion \& Recommendations}
These \emph{other automated metrics}, while rarely applied in general RAG research, play a pivotal role in specialised studies by illuminating efficiency, safety and domain-specific quality attributes. Their sporadic use stems from (1) the high cost of bespoke dataset creation or annotation; (2) dependencies on hardware and execution environments; and (3) the lack of universally accepted standards for task-specific evaluation. To enhance comparability and encourage broader adoption, we recommend the following.
\begin{itemize}
  \item \textbf{Modular Reporting:} Package each specialised metric within containerised pipelines to facilitate deployment.
  \item \textbf{Benchmark Extensions:} Propose extensions to popular RAG benchmarks (e.g., adding hallucination annotations to QA datasets).
  \item \textbf{Open-Source Toolkits:} Contribute wrappers for less common metrics, such as ES and contextual bias, to public evaluation libraries.
\end{itemize}
By situating these metrics alongside standard automated measures in future studies, researchers can achieve a more holistic assessment of RAG systems without imposing prohibitive setup costs.

\subsubsection{Human Evaluation Metrics}\label{sec:human} 

Human evaluation remains indispensable for assessing aspects of RAG that escape purely automatic measures. By soliciting judgments on dimensions such as correctness, relevance, fluency, and factuality, researchers gain insight into real-world performance and user impact \cite{RN359, RN1635}.


\paragraph{Correctness \& Accuracy.}  
Accuracy gauges the degree to which the generated outputs match the expert-validated answers. In the clinical settings of RAG, evaluators verify whether the responses of the model reflect consensus recommendations \cite{RN605}. Legal RAG evaluations similarly require that each response be both factually correct and properly grounded in authoritative sources \cite{RN1759}. Educational chatbots assess 'correctness' using multipoint rating scales applied by subject matter experts \cite{RN1744, RN1745}.

\paragraph{Relevance.}  
Relevance measures how well the context retrieved or the generated text aligns with the user’s query. Human raters typically score summaries or answers on a binary or Likert scale for topical pertinence, grammatical coherence, and external information appropriateness \cite{RN359}. In personalised RAG frameworks, relevance judgements of retrieved passages ensure that augmentation truly addresses user intent \cite{RN1755}.

\paragraph{Hallucination \& Groundedness.}  
Hallucination metrics capture instances of fabricated or misattributed content. Annotators label responses as 'Extrinsic' (not supported by any input), 'Intrinsic' (incorrectly synthesised from input) or 'Misgrounded' (false citation) \cite{RN1610, RN1759}. Human evaluation thus directly quantifies the tendency of the model to invent facts, a critical safety concern in high-stakes domains such as healthcare care and law \cite{RN358, RN1609}.

\paragraph{Factual Correctness \& Consistency.}  
Beyond binary correctness, human judges assess whether a response maintains internal consistency, avoids contradictions, and remains factually accurate throughout longer interactions \cite{RN349, RN358}. This qualitative lens captures subtle semantic errors that are not detected by overlap metrics.

\paragraph{Comprehensiveness \& Quality.}  
Comprehensiveness evaluates depth of coverage: whether the generated text addresses all aspects of a query \cite{RN1492}. General quality scales (for example, 1 to 5 points) combine relevance, coherence, and absence of typos, resulting in a single interpretable score \cite{RN935, RN994}.

\paragraph{User-Centric Metrics}  

\textbf{User Satisfaction} is measured via post-interaction surveys; satisfaction scores reflect perceived usefulness and clarity \cite{RN1618, RN644}.

\textbf{System Usability (SUS)}: a standardised 5-point questionnaire assesses accuracy, clarity, relevance, and ease of understanding \cite{RN61}. 

\textbf{Technology Acceptance (TAM):} structures such as perceived usefulness and ease of use are quantified through validated survey instruments, offering insight into the likelihood of adoption \cite{RN1711}.

\paragraph{Annotation Protocols \& Reliability}

Most studies use three to five human annotators to rate system outputs against predefined criteria. Common protocols include Likert scales (3--5 points) to assess relevance, fluency, and factuality \cite{RN359,RN1635}; binary judgements (yes/no), particularly for retrieval relevance or groundedness \cite{RN1755,RN1759}; comparative judgements (win/tie/loss) for head-to-head model comparisons \cite{RN1754}; and error classification, in which incorrect outputs are sampled and error types are categorised (e.g.\ reasoning versus retrieval failures) \cite{RN805}. To support reliable annotation, studies typically provide clear guidelines with worked examples for each rating level, pilot the scheme on a small subset and refine the instructions, and report inter-annotator agreement (e.g.\ Cohen's $\kappa$), including both raw agreement and chance-corrected statistics \cite{RN1759}.

\paragraph{Strengths, Limitations \& Recommendations}

Human-judged metrics capture nuanced aspects of RAG output, such as hallucination, conversational coherence, and user trust, that automated measures often miss. However, they are time-intensive, costly, and susceptible to annotator bias, with inter-annotator agreement frequently below $\kappa=0.7$, reflecting subjectivity in complex judgements \cite{RN1759}. 

To maximise rigour and reproducibility, evaluations should combine measures spanning core dimensions (e.g., accuracy, relevance, hallucination, comprehensiveness, and satisfaction), report annotation scales, rater qualifications, and agreement statistics transparently, and consider hybrid designs that supplement expert judgements with carefully prompted LLM-as-judge procedures to increase scale while retaining depth. Making annotation guidelines and code openly available further facilitates external replication and community benchmarking. 

When protocols are defined \emph{a priori}, each metric is grounded in previous work and reliability is reported, the human evaluation section can more convincingly demonstrate both the real-world viability and the limitations of an RAG system.

\subsubsection{LLM-as-Judge Metrics} 

Recent advances in evaluation methodologies have shifted toward the use of LLMs themselves as automated judges of generated content. Rather than relying solely on surface-level overlap or costly human annotation, LLM-as-judge approaches prompt a high-capacity model, such as GPT-4, to assess outputs along dimensions such as correctness, relevance, coherence, and safety.


\paragraph{Accuracy via Advanced LLM Verification}  
One common formulation applies an LLM (e.g. text-davinci-003) to re-evaluate model outputs against ground-truth answers, flagging semantically correct yet lexically divergent generations as accurate \cite{RN379}. This ``LLM-verified accuracy'' provides a more robust correctness estimate than exact-match metrics, particularly in question-answering settings where paraphrase is common.

\paragraph{GPT-Based Correctness and Quality Ratings}  
A suite of studies instruct ChatGPT or GPT-4 to assign binary or scalar judgements to outputs:

\textbf{Binary correctness:} ChatGPT classifies each response as correct or incorrect, yielding a proportion-correct score \cite{RN1612}. 

\textbf{Quality scales:} Responses are rated on a 1–10 scale for overall quality—including helpfulness, relevance, and depth—by ChatGPT \cite{RN1612}, and similarly by GPT-4 across multiple facets (relevance, clarity, depth) in fully automated scoring systems \cite{RN1744}.  

\textbf{Sentiment assessment:} ChatGPT assesses the polarity of model outputs (positive vs.\ negative) to gauge tone and user experience \cite{RN1612}.

\paragraph{Benchmarking Against GPT-4 judgements}  
To validate internal model evaluations, some works compare their own LLM’s judgements with those of GPT-4. For example, GPT-4 is used as a reference judge for self-knowledge, passage relevance, and question-decomposition tasks, establishing a reliability benchmark \cite{RN1635}.

\paragraph{Harmfulness and Safety Classification}  
Ensuring ethical outputs, researchers prompt GPT-4 to detect and classify harmful or toxic content, computing the proportion of harmful responses or the worst-case toxicity over multiple samples \cite{RN1640}. This approach complements traditional toxicity metrics by leveraging the LLM’s contextual understanding of offensiveness.

\paragraph{LLM-Fact-Checker Chains}  
Leveraging frameworks such as LangChain, an LLM (e.g., gpt-3.5-turbo) is embedded in a fact-checking pipeline: it cross-verifies chatbot responses against course content or reference materials and generates confusion-matrix statistics (accuracy, precision, sensitivity, specificity) to automate what was formerly manual evaluation \cite{RN1711}.

\paragraph{G-EVAL: Comprehensive LLM-Judged Evaluation}  
G-EVAL uses GPT-4 to score generated text on coherence, consistency and fluency using a 1–5 rubric, outperforming traditional overlap metrics in correlating with human judgements \cite{RN1747}. It has been used to evaluate the generation of domain-specific reports, such as flood incident summaries, demonstrating superior alignment with expert evaluators \cite{RN1747}.

\paragraph{Semantic Accuracy via LLM Instruction Models}  
By prompting gpt-3.5-turbo-instruct to compare generated answers semantically against ground truths, ``semantic accuracy'' metrics capture meaning preservation beyond exact tokens, addressing limitations of classical exact-match scores \cite{RN1745}.

\paragraph{Discussion \& Recommendations}  
LLM-as-judge metrics offer scalable, semantically rich evaluation but inherit potential biases and prompt-sensitivity from their host models. To mitigate these issues, we recommend calibrating LLM prompts against a small human-annotated validation set, reporting multiple perspectives (e.g. combining binary correctness with a scalar quality score) and disclosing prompt templates and model versions to ensure reproducibility. Adopting these practices can harness the efficiency of LLM-judged evaluation while maintaining rigorous, transparent assessment standards.

\subsubsection{\textbf{Automated Frameworks}} 
 
Automated evaluation frameworks are pivotal for assessing RAG systems by mitigating subjectivity, scalability issues, and bias. Two notable systems, ARES \cite{RN385} and RAGAS \cite{RN147}, concentrate on three core metrics: context relevance, answer faithfulness, and answer relevance.

ARES adopts a quantitative approach, using fine-tuned language models and Kendall’s $\tau$ to align automated scores with human judgements \cite{RN385}. This method delivers high precision and nuanced insights into response fidelity; however, its reliance on extensive annotated data may restrict scalability. In contrast, RAGAS employs a reference-free strategy that uses cosine similarity to measure semantic relationships between queries, retrieved contexts, and generated responses \cite{RN147}. Although this technique improves objectivity and accelerates evaluations, it is more sensitive to prompt variations, which can reduce consistency.

ARES and RAGAS thus represent two contrasting yet complementary approaches to RAG system evaluation. ARES offers detailed, human-aligned assessment but can be hampered by scalability issues due to its dependency on annotated data. Conversely, RAGAS provides operational efficiency through automated semantic similarity measurements, albeit with potential variability due to prompt sensitivity. This juxtaposition highlights the trade-off between detailed, qualitative insights and streamlined quantitative evaluation, prompting critical questions about whether future frameworks might integrate the strengths of both methods to achieve a balanced, robust evaluation strategy.

It is important to note that automated evaluation frameworks relying on large language models are not immune to inherent biases, which can subtly skew outcomes and misrepresent true system performance. To mitigate these issues, future evaluation strategies could benefit from hybrid approaches that integrate LLM-based assessments with calibrated human oversight, balancing the objectivity and scalability of automated methods with the nuanced insights of human evaluators.

Practical implications of these frameworks include guiding the design of adaptable RAG systems that lower annotation costs while upholding rigorous evaluation standards. Future research may further integrate qualitative elements and refine metrics to address emerging concerns, as discussed in Section~\ref{sec:human}. Ultimately, merging ARES’s detailed human insight with RAGAS’s operational efficiency may offer the most balanced strategy to advance the evaluations of the RAG system.

\subsubsection{Holistic Evaluation of RAG Benchmarks} 

RAG benchmarks, although diverse in their origins and target domains, collectively map out a multidimensional landscape of model performance. At the core, each benchmark isolates particular capabilities—be it resilience to noise, financial forecasting acuity, medical question precision, multi-hop reasoning, or CRUD-style text operations—and in doing so, they both complement and challenge one another.

\paragraph{Connecting the Four Pillars of RGB to Broader RAG Metrics}
The Retrieval-Augmented Generation Benchmark (RGB) explicitly dissects RAG capability into noise robustness, negative rejection, information integration, and counterfactual robustness \cite{RN622}. These four axes are not arbitrary: they represent the basic dilemma of ``when and how to trust retrieved context.'' Noise robustness measures whether a model can sift signal from distractors—a requirement shared by nearly all other RAG tasks, since any retrieval pipeline may surface irrelevant documents \cite{RN622}. Negative rejection, on the other hand, examines the model’s restraint: ability to say ``I don’t know'' rather than hallucinate. This restraint is critical in high-stakes domains such as medicine, where wrong answers can mislead practitioners \cite{RN1613}. Information integration overlaps naturally with multi-hop retrieval and summarization: it probes the model’s capacity to aggregate evidence from multiple sources, akin to what MultiHop-RAG quantifies through MAP@K and answer accuracy \cite{RN390}. Finally, counterfactual robustness examines error detection and correction—an echo of CRUD-RAG’s ``Update'' task, which tests factual revision in generated text \cite{RN392}.

\paragraph{Quantitative Meets Qualitative: Trade-offs in evaluation}
While RGB relies primarily on exact match accuracy and rejection / error rates, AlphaFin extends evaluation to financial performance metrics: annualized rate of return (ARR), Sharpe ratio, drawdowns, along with traditional language metrics like ROUGE and human preferences \cite{RN1741}. This duality highlights a fundamental trade-off: quantitative metrics (ARR, MAP@K, accuracy) offer objectivity and comparability, yet may miss subtleties of fluency, coherence, or interpretability that qualitative human studies and chain-of-thought evaluations capture. For example, a model that achieves high ARR by blindly following market trends may still produce explanations that fail regulatory standards or mislead users; here, GPT-4 preference judgements in financial Q\&A illuminate whether the model’s reasoning is human-aligned \cite{RN1741}. In contrast, purely qualitative assessments can be subjective and difficult to standardise in large test beds such as the 7,663 medical questions of MIRAGE \cite{RN1613}.

\paragraph{Domain-Specific Demands and Broader Trends}
The emergence of domain-tailored benchmarks—AlphaFin in finance, MIRAGE in medicine—reflects a broader shift in RAG research: One-size-fits-all evaluation is giving way to specialised suites that capture domain nuances. In medicine, zero-shot versus retrieval-augmented evaluations in MIRAGE reveal that RAG can increase accuracy by up to 18\%, but also surface 'lost in the middle' issues when too much context overwhelms the model \cite{RN1613}. MultiHop-RAG similarly shows that retrieval itself remains a bottleneck: even GPT-4 reaches only 56\% accuracy with real retrieval versus 89\% with ground-truth contexts \cite{RN390}. These findings spark questions: \emph{How might improvements in retriever architectures reorder the current performance hierarchy?} And \emph{can domain-agnostic LLMs ever match domain-specific ones once retrieval pipelines are fully optimised?}

\paragraph{Methodological Reflections: Why These Metrics?}
Each benchmark's metric choices go back to its core use case. The rejection rate metric of RGB emerges from the need to induce hallucinations in open domain QA, while the AlphaFin ARR and Sharpe ratio base the evaluation on financial risk-reward trade-offs \cite{RN622,RN1741}. The reliance of MIRAGE on established medical QA datasets (MMLU-Med, MedQA-US, BioASQ) ensures comparability with previous work, but by layering the retrieval into zero-shot and multichoice settings, it exposes where medical LLMs overuse or underuse external evidence \cite{RN1613}. The combination of retrieval metrics (MAP@K, MRR@K) and generation (accuracy) of MultiHop-RAG mirrors the two-stage reality of the RAG pipelines, allowing separate diagnostics for the retriever and the generator \cite{RN390}. The taxonomy of CRUD-RAG in the Create, Read, Update, Delete tasks underscores the need for full-lifecycle assessment of text operations, not just answering questions \cite{RN392}.

\paragraph{Practical Implications and Future Directions}
In practice, these benchmarks guide system design: a retrieval pipeline optimised for MAP@10 may not yield the best error correction performance in counterfactual settings; a model fine-tuned for ROUGE in financial summaries could underperform in drawdown mitigation metrics. Thus, practitioners face calibration challenges: \emph{Which trade-off between retrieval depth and generative precision aligns best with their application’s risk profile?}

Looking ahead, several avenues merit exploration. First, \emph{integrating qualitative fluency measures directly into quantitative benchmarks} could bridge the gap between human-centric evaluation and automated metrics. Second, \emph{extending benchmarks to multilingual or cross-modal contexts}---combining text with tables, charts, or code---would reflect real-world uses. Finally, as interactive RAG agents grow, \emph{dynamic benchmarks that simulate user feedback loops} will be critical to measure adaptability and continuous learning.

RGB, AlphaFin, MIRAGE, MultiHop-RAG, and CRUD-RAG form a tapestry of complementary benchmarks: each covers a slice of the performance spectrum: signal filtering, domain-specific reasoning, error detection, evidence synthesis, and text lifecycle operations. Their varied metrics—accuracy, rejection rates, ARR, ROUGE, MAP@K, Sharpe ratios—highlight that no single number suffices. A holistic evaluation demands a suite of metrics that reflect both quantitative rigour and qualitative nuance. As RAG systems advance, our benchmarks must evolve in tandem, posing ever more challenging questions: \emph{Can we craft unified metrics that capture trustworthiness, utility, and user alignment in one framework?} Only through such integrative efforts can the next generation of RAG applications realise their full potential.

\subsubsection{Datasets} 

\newcommand{\totalUniqueDatasets}{343 }

In our systematic survey, we find that researchers have used a large array of approximately \totalUniqueDatasets unique datasets to evaluate RAG systems, illustrating the multifaceted nature of performance assessment. Open-domain resources such as Wikipedia \cite{RN1}, Natural Questions \cite{RN468}, and MS MARCO \cite{RN455} provide a baseline, particularly for question-answer tasks. These datasets excel in benchmarking fluency and general comprehension, but may not fully represent specialised applications. In contrast, domain-specific collections, ranging from legal (e.g. ALQA \cite{RN1572}, LEDGAR \cite{RN1427}) to biomedical sources (e.g. CORD-19 \cite{RN428}, KGRAGQA \cite{RN1524}), offer in-depth evaluation in high-stakes contexts, although they often suffer from inconsistent preprocessing and versioning practices. Table \ref{tab:datasetsTable} summarises the content description and intended use of these datasets.

Multi-hop QA sets, including HotPotQA \cite{RN445} and 2WikiMultihopQA \cite{RN406}, challenge systems with complex reasoning tasks, highlighting strengths in multistep inference, while also revealing limitations in current methodologies. Similarly, multimodal and code-centric corpora, such as COCO \cite{RN420} for image-text pairs and CodeSearchNet \cite{RN418} for code-centric evaluations, extend performance evaluation beyond traditional text, addressing broader application domains, yet introducing variability due to differences in data segmentation and annotation standards.

This diversity reflects both advantages and trade-offs: Although open domain datasets support benchmark consistency, specialised datasets provide critical insight into domain-specific challenges \cite{RN1, RN468, RN455, RN1427}. The absence of standardised dataset preparation, ranging from segmentation to versioning, poses a significant methodological challenge and raises questions about the reproducibility and comparability of RAG evaluations. For example, how might emerging frameworks for dataset processing and standardised evaluation metrics improve consistency between studies?

The interplay among these datasets underscores a broader trend toward holistic, multidimensional evaluation strategies in the development of the RAG system. By integrating both quantitative benchmarks and qualitative assessments, researchers can better capture the strengths and limitations of current models, ultimately guiding future innovations and establishing more robust operational standards.

\subsection{What are the key challenges and limitations associated with retrieval-augmented generation techniques?}

As detailed in Section~\ref{sec:advance_rag}, recent RAG systems have been propelled by dynamic query generation, universal-scheme frameworks, multimodal fusion, and iterative refinement. These innovations revolutionise data transformation and context preservation \cite{RN349,RN389,RN362,RN363}. However, their very integration reveals a set of stubborn obstacles that constrain performance, scalability, and adaptability.  The remainder of this section therefore organises these obstacles into six thematic challenges, tracing how each one limits today’s retrieval-augmented generation pipelines.

\subsubsection{Computational and Resource Trade-offs}
\label{sec:comp_tradeoffs}

The first obstacle is the raw cost, both in time and in hardware.  Dynamic query rewriting, iterative retrieve-and-refine loops, and extended context attention improve relevance, but each extra pass increases the wall clock latency and memory footprint \cite{RN349,RN699,RN364,RN379,RN399}.  Universal-schema frameworks compress representations and trim indices, yet they handle novel patterns brittlely and still struggle to keep edge devices within real-time budgets \cite{RN389}.  Training compounds the strain: full end-to-end tuning of large retriever–generator stacks can consume days of multi-GPU time and hundreds of gigabytes of RAM \cite{RN1609,RN364,RN79}.  Lighter alternatives, such as adapter layers, retrieval-only updates, or prompt tuning, reduce cost but usually eliminate domain specificity \cite{RN362,RN948}.  

Even at inference, resources rarely align neatly.  CPU-bound vector search typically precedes GPU-bound decoding, so one processor idles while the other works; bespoke schedulers such as PipeRAG and RAGCache attempt to hide lookup time by overlapping ANN search with generation, yet they demand careful profiling and remain sensitive to corpus size \cite{RN1633,RN1636,RN846}.  Approximate-nearest-neighbour indices halve retrieval latency but lower recall, whereas exhaustive search inverts the trade-off.  Progress therefore hinges on adaptive scheduling policies that co-optimise ANN depth, speculative decoding, and device utilisation, plus resource-efficient joint objectives that align retriever and generator without prohibitive fine-tuning \cite{RN1643}.

\subsubsection{Noise, Heterogeneity, and Multimodal Alignment}
\label{sec:noise_hetero}

RAG pipelines are only as good as their inputs, yet most inputs are noisy and heterogeneous.  Vision-to-Language transformers compress complex scenes into terse captions, suppressing spatial clues such as gaze or depth \cite{RN362}.  Code-Property graphs balloon super-linearly with project size, so aggressive pruning saves space but can excise rare, security-critical constructs \cite{RN363}.  Selective densification, reinjecting previously filtered snippets when retrieval confidence dips, offers a middle ground, although it still inflates indices \cite{RN1623,RN389}.  

Noise also lurks in hybrid retrieval itself.  Dense vectors, sparse keywords and rule filters score on incompatible scales; naive normalisation swings between overrecall and underrecall, while cross-encoders that fix the problem add 2–5 times the latency \cite{RN340,RN353}.  Learnable weighting gates are promising but lack cross-domain evidence \cite{RN1606}.  Multimodal encoders introduce another layer of fragility: CLIP-style models often suffer ``semantic bleeding'', where irrelevant visual regions influence text similarity, a serious risk in radiology and surgical robot logs \cite{RN365,RN367,RN386,RN1623}.  Fine-grained alignment losses mitigate leakage but add both milliseconds and supervision cost. Lightweight validation schemes, such as attention entropy checks or cross-view consistency regularisers, offer a protection at marginal run-time cost and do not require dense pixel labels \cite{RN1623}.  

Finally, knowledge graph retrieval excels in multi-hop reasoning, yet depends on noisy entity linking and heuristic pruning; over-pruning deletes long-tail nodes, under-pruning explodes memory, and classic graph metrics correlate only weakly with downstream QA \cite{RN1569,RN1571,RN1606}.  What the field needs are learnable fusion frameworks that expose per-channel uncertainty and graph-aware benchmarks that reveal the real cost-benefit envelope of noise mitigation strategies.

\subsubsection{Domain Shift, Dataset Alignment, and Generalisation}
\label{sec:domain_shift}

Our focus now shifts to distributional robustness---RAG models that shine in one domain often stumble in another. Systems tuned to PubMed outperform BM25 on biomedical queries but falter in legal corpora without costly retraining \cite{RN128,RN381}. Hybrid pipelines that anchor language-agnostic schemas with thin domain-specific rules travel better, but add engineering overhead and still require careful calibration when knowledge is fragmented across disconnected sources \cite{RN1635}.

Repository freshness compounds the problem. Stale or erroneous material propagates directly into answers, a high-stakes liability in finance and medicine \cite{RN386,RN1607}. Index refreshes mitigate drift but demand labour-intensive validation pipelines and may introduce their own lag \cite{RN1629}. Worse, most evaluation sets lean heavily on English-language Wikipedia, masking specialist failure modes and inflating scores through train–test overlap \cite{RN353,RN1643}. Corpus choice is thus decisive: biomedical encoders dominate on PubMed but misfire elsewhere \cite{RN1613}, and multilingual retrieval remains hamstrung by scarce aligned data and inconsistent terminology \cite{RN1615}. Adaptive ``retrieval triggers'' that fire the retriever only when the generator signals high uncertainty appear attractive; yet, when they misfire, they either waste compute or omit indispensable evidence \cite{RN618}.

Seemingly mundane hyperparameters—chunk size, hierarchical fragmentation strategy, the number of documents $k$ to return, and undocumented caching policies—can shift accuracy–latency curves by double-digit margins: small windows fracture discourse; large ones bloat latency; and inconsistent choices of $k$ thwart reproducibility \cite{RN1608,RN1630,RN935,RN1606,RN1636}. Closing these gaps will require continuous validation pipelines and unified cross-domain, multilingual testbeds that expose real-world brittleness while tracking accuracy–latency trade-offs.

\subsubsection{Modular Pipelines and Error Cascades}
\label{sec:error_cascades}

Even when knowledge is fresh and well aligned, architectural glue can fail. In this section, we focus on interfaces that link retrieval to generation.
Splitting retrieval, reranking, and generation kerbs hallucination but creates brittle processing chains.  A misranked passage in the first stage can irreversibly bias the generator, and although deep cross-encoders lift ranking fidelity, their compute cost still forces approximate first-pass filters whose scores are tuned ad hoc \cite{RN358,RN342,RN383,RN49}.  

Iterative and memory-augmented pipelines add another wrinkle.  External memories curb repetition but introduce staleness and snowballing: cached errors are re-retrieved in later turns \cite{RN1629}.  Content-based decay, which weights cache entries by both recency and reuse, cuts latency by up to 40\% without hurting precision, yet evidence remains limited to small-scale experiments \cite{RN1608}.  Ultimately, combining interface patterns that expose calibrated model confidence with uncertainty-triggered safeguards, for example, probability/entropy thresholds that proactively invoke retrieval, verification, abstention, or rollback, can prevent error cascades from taking hold \cite{RN349,RN402}.

\subsubsection{Large-Language-Model Constraints and Safety Risks}
\label{sec:llm_limits}

Next, we examine the constraints of the generator (the LLM) that produces user-facing text. Commercial LLM APIs deliver strong performance but impose per-token fees, usage limits and a requirement for internet connectivity. Open-weight models avoid vendor lock-in and can run locally, however require substantial hardware and usually offer fewer tuning options \cite{RN379,RN340}.  Fixed context windows—often four thousand tokens or fewer—truncate multi-document evidence, forcing lossy chunking that undercuts retrieval depth; long-context variants help, but do not fully restore cross-passage reasoning \cite{RN356,RN378}.  

Bias, toxicity, and hallucinations remain endemic.  Encoding a user’s information need in only a few tokens is brittle, and attempts to map that intent into structured formats (e.g. JSON) often break under domain drift \cite{RN347,RN1619,RN1502}.  Automatically generated search strings show the same fragility: ill-formed queries invite off-topic retrieval and can launch an irrelevant evidence cascade \cite{RN349}.  Skewed pre-training corpora, meanwhile, inject demographic bias and toxic completions; retrieval softens but does not eliminate hallucination, and lapses are especially hazardous in medicine \cite{RN204,RN1609,RN1627}.  Prompt design does not offer a silver bullet: minor syntactic edits shift coherence and factuality, while adversarial prompts can bypass guardrails or surface-corrupted evidence \cite{RN846,RN1569,RN1632,RN134}.  Progress therefore hinges on bias- and hallucination-aware losses, adversarial-prompt test suites, and extended-context architectures that enlarge windows at sustainable cost.

Skewed pre-training corpora drive demographic bias and toxic completions; retrieval dampens but does not eliminate hallucination, which is especially problematic in medicine, where factuality lapses carry real harm \cite{RN204,RN1609,RN1627}.

\subsubsection{Security Threats in Retrieval-Augmented Generation}
\label{sec:security_rag_unified}

Even the best-engineered and safest pipelines remain vulnerable to deliberate attack, so finally we consider the RAG security landscape. The same external knowledge that makes RAG systems powerful also opens up a new attack surface: \emph{the retrieval corpus itself}. Because the language model is trained to trust whatever the retriever returns, even a \textit{single} poisoned document or a carefully crafted query can steer generation, violate safety policies, or leak private data. Recent work exposes three broad threat families—\textbf{(i) corpus-poisoning back-doors}, \textbf{(ii) data-exfiltration and privacy attacks}, and \textbf{(iii) jailbreak and policy-evasion triggers}—all of which exploit the loose coupling between the retriever and generator.

AGENTPOISON \cite{RN1758} and Phantom \cite{RN1749} show that an attacker needs to tamper with \emph{fewer than 0.1 \%} of corpus items—sometimes only one passage—to create a back-door that fires when a secret trigger word appears.
A constrained trigger optimisation maps those queries to a compact, unique region of the embedding space, guaranteeing retrieval while remaining stealthy (low perplexity, robust to paraphrase).
The result is alarming: across six dense retrievers and multiple LLMs, retrieval success exceeds 80\%, and end-to-end malicious action rates sit around 60\% with virtually no drop in benign accuracy.
These findings underline a systemic weakness: current RAG deployments rarely authenticate or provenance-stamp the documents they ingest, so ``sleeper'' passages can lie dormant until the attacker issues the right query.

BadRAG \cite{RN1612} extends the idea to \emph{content-only poisoning}: its COP / ACOP / MCOP techniques craft passages that are \emph{only} retrieved under specific trigger conditions and then bias the generated output (Alignment-as-an-Attack, Selective-Fact-as-an-Attack).
With as few as ten poisoned passages, the framework achieves a 98\% trigger-retrieval rate and slashes GPT-4 accuracy from 92\% to 19\%.
Crucially, attacks bypass naive defences such as perplexity filters or keyword blacklists and can even nudge sentiment or political stance without overtly toxic text, highlighting how difficult covert bias detection will be once adversaries understand retrieval scoring.

A different axis of vulnerability is privacy.
``Follow My Instruction and Spill the Beans'' \cite{RN1739} demonstrates that simply appending a malicious system or user prompt can coerce instruction-tuned models to copy verbatim from their private datastores.
Across nine open-source LLMs and 25 production GPTs, the leakage success hit 100\% in at most two queries; larger models leaked \textgreater 70 BLEU points of text.
Leakage worsens with coarse, semantically coherent chunks and when prompts are injected at the start or end of the context, painting a clear blueprint for would-be attackers.
Mitigations such as PINE (Position-bias Elimination) and safety-aware system prompts halve—yet do not eliminate—the reconstruction rate, signalling that stronger retrieval-side controls are required.

Pandora \cite{RN1632} and Phantom \cite{RN1749} move beyond bias or leakage to full \emph{policy evasion}.
By injecting adversarial content that the retriever dutifully surfaces, the attacker sidesteps the usual guard-rail prompt hierarchy; GPT-4, normally resilient to direct jailbreaks, yields prohibited outputs in 35\% of cases once the supporting evidence comes from a poisoned corpus.
Because the unsafe text reaches the generator as ``ground truth'', refusal classifiers often let it pass.
These results expose an uncomfortable asymmetry: alignment layers supervise \emph{prompts}, yet poisoned retrieval arrives as ``context'' and therefore inherits implicit trust.

In practice, commonly deployed defences provide only limited protection. Perplexity-based filtering and query rephrasing reduce AgentPoison’s end-to-end success by at most single-digit percentage points in some tasks (e.g., 9.6 percentage points and 6.8 percentage points in Agent-Driver), but do not produce any reduction - and sometimes an increase - in others (ReAct-StrategyQA). Moreover, AgentPoison’s triggers remain low-perplexity and thus difficult to flag \cite{RN1758,RN1612}. Query rephrasing or majority vote reranking is similarly ineffective because trigger optimisation tends to cluster poisoned queries tightly in embedding space; paraphrases remain within the backdoor region. Safety prompting and refusal classifiers cannot, in general, distinguish benign evidence from adversarially retrieved content, and therefore authorise harmful completions \cite{RN1632}. Blacklisting triggers is also brittle: Phantom shows that an unseen synonym can reactivate the attack, and adversaries can optimise entirely new token sequences that were not present at the defence time \cite{RN1749}.

These limitations motivate a set of complementary directions. Strengthening corpus provenance and attestation is a priority: practical mechanisms to sign, version, and audit documents in large-scale vector stores remain scarce, but append-only logs based on Merkle trees, together with proofs from trusted execution environments, could make retroactive poisoning detectable. Retrieval-time anomaly detection also merits attention; distance-based or density-ratio detectors in embedding space may identify outlier triggers, provided they operate at millisecond latency and resist adaptive manipulation. A further avenue is joint retriever--generator training: current pipelines typically ``freeze'' the retriever at deployment, coupling the retriever’s gradients to downstream safety losses may instead lead the system to \emph{unlearn} reliance on poisoned sources. In parallel, refusal mechanisms should assess the provenance of retrieved spans—not only the prompt—so that unsafe evidence is withheld before it reaches the LLM. Continued progress will depend on rigorous benchmarking, since most leaderboards emphasise hallucination and factuality rather than integrity; a standard suite that measures attack-specific metrics—retrieval attack success rate, end-to-end ASR under transfer (ASR-t), and drift in benign accuracy—would enable systematic evaluation.

Security threats in RAG are no longer theoretical. With a handful of poisoned passages or a single prompt injection, adversaries can bias, leak, or jailbreak state-of-the-art systems while evading current defences. The community must therefore treat the retrieval pipeline---and, by extension, the knowledge base---as a first-class security boundary, on a par with the language model itself.

\subsubsection{Synthesis and Outlook}
The six challenges examined above form an interlocking system rather than a menu of orthogonal pain points.
Compute budgets shape how much noise can be tolerated (\S\ref{sec:comp_tradeoffs}~\ensuremath{\leftrightarrow}~\S\ref{sec:noise_hetero});
data cleanliness conditions the severity of domain shift (\S\ref{sec:noise_hetero}~\ensuremath{\leftrightarrow}~\S\ref{sec:domain_shift});
imperfect domain coverage magnifies error cascades (\S\ref{sec:domain_shift}~\ensuremath{\leftrightarrow}~\S\ref{sec:error_cascades});
architectural fragility limits the safe operating range of large-language models (\S\ref{sec:error_cascades}~\ensuremath{\leftrightarrow}~\S\ref{sec:llm_limits});
and every residual weakness enlarges the attack surface (\S\ref{sec:llm_limits}~\ensuremath{\leftrightarrow}~\S\ref{sec:security_rag_unified}).

Progress hinges on \emph{co-design}: latency-aware scheduling across retrieval and generation; benchmarks that jointly score robustness to noise, distribution shift, and security; extended-context models balanced by adaptive retrieval depth; and probabilistic defences that propagate calibrated uncertainty end-to-end. Tackling these dependencies together will yield RAG systems that are efficient, reliable, and resilient amid rapidly evolving knowledge and threats.

\section{Limitations of the Systematic Review}

The methodological choices we made in selecting the literature for our systematic review were intended to maximise focus, transparency, and reproducibility, but they entail constraints that should be acknowledged. The citation thresholds (\S\ref{sec:InclusionAndExclusionCriteria}) foreground influential contributions and kept screening tractable, however, they also risk citation-lag bias, under-representing very recent breakthroughs and niche, domain-specific work that has not had time to accrue citations. Future updates could mitigate this by adopting time-normalised criteria (for instance, citations per month since publication), reporting a short sensitivity analysis with relaxed cut-offs, and optionally tracking an expert-curated ``emergent'' set alongside the main corpus.

Coverage was restricted to five major digital libraries plus DBLP, and to English-language publications. This scope improves deduplication and comparability, but probably undersamples grey literature, non-indexed preprints, and research disseminated in other languages. Terminological inconsistency in the field (``RAG'', ``retriever-reader'', ``retrieval-augmented LLMs'') further complicates study selection; a component-based inclusion rule was applied to reduce misclassification, although borderline cases may remain.

Screening and extraction procedures also introduce potential bias. Titles and abstracts were double-screened, while full-text data extraction was performed by a single reviewer with verification; In addition, suggestions assisted by LLM were used to support, not replace, human judgement. These steps accelerated the workflow but may still allow selection or extraction bias despite audit trails. Given the pace of the area and our search window ending in May 2025, temporal generalisability is limited. Periodic updates, a broader database and language coverage, preregistered protocols, and a brief sensitivity analysis in future iterations would strengthen robustness.

\section{Conclusion} 
\label{sec:conclusion}

In this systematic review, we synthesised a comprehensive picture of the RAG research landscape through the lens of 128 highly cited studies on retrieval-augmented generation (2020–May 2025) using a citation-weighted PRISMA protocol. Although DPR + seq-to-seq remains a strong baseline, recent progress centres on hybrid retrieval, structure- and graph-aware indexing, uncertainty-triggered and iterative retrieval loops, efficiency-orientated compression, and the integration of memory and multimodality. Evaluation practice is broad but uneven: overlap metrics still dominate, with growing use of retrieval quality measures (e.g., Recall@k, MAP), human judgements, and LLM-as-judge protocols.

However, important gaps remain. Reporting seldom couples accuracy with cost and latency; systems generalise brittlely under domain shift and evolving corpora; and defences against retrieval-side poisoning and prompt-in-context attacks are still immature. We therefore recommend holistic benchmarks that combine quality, efficiency, and safety; treating retrieval depth and tool use as budget-aware policy decisions; and provenance-aware retrieval pipelines that surface uncertainty and provide traceable evidence. Looking ahead, modular adaptive retrieval generation stacks that allocate compute based on uncertainty, unified multilingual and multimodal benchmarks, and end-to-end security and privacy frameworks will be key to moving RAG from promising prototypes to reliable and scalable systems.

\newpage

\onecolumn
\begin{landscape}
  \setlength\LTleft{0pt}
  \setlength\LTright{0pt}

\begin{footnotesize}
\appendix
\section{Appendix} 
\label{sec:appendix}

\begin{longtable}{ 
    p{0.3\linewidth}  
    p{0.3\linewidth}  
    p{0.3\linewidth}  
    p{0.05\linewidth}  
}
    \caption{Summary of datasets utilised in the studies included in this systematic literature review of RAG. It outlines the key characteristics and origins of each dataset, offering an overview that enhances understanding of the data employed across the reviewed research articles. This summary supports an analysis of the trends and methodologies specific to RAG, showcasing the variety and scope of datasets applied in this area of research.}
    \label{tab:datasetsTable} \\
    \toprule
    \textbf{Dataset Name} & \textbf{Content Description} & \textbf{Intended Use} & \textbf{Citation Frequency} \\
    \midrule
    \endfirsthead

    \toprule
    \textbf{Dataset Name} & \textbf{Content Description} & \textbf{Intended Use} & \textbf{Citation Frequency} \\
    \midrule
    \endhead

    \midrule \multicolumn{4}{r}{{Continued on next page}} \\
    \bottomrule
    \endfoot

    \bottomrule
    \endlastfoot
    Natural Questions (NQ) \cite{RN468} & 323,045 QA examples across train/dev/test splits. & Train and evaluate open-domain QA systems. & 27 \\ 
    HotPotQA \cite{RN445} & 113,000 multi-hop QA pairs. & Train/test QA with multi-hop reasoning and explanations. & 26 \\ 
    Wikipedia \cite{RN1} & 6 million articles of text and metadata. & General corpus of Wikipedia text for NLP tasks. & 19 \\ 
    TriviaQA (TQA) \cite{RN489} & ~96,000 QA pairs with six supporting documents each. & Develop comprehension models requiring complex inference. & 18 \\ 
    2WikiMultihopQA (2WikiMQA) \cite{RN406} & 192,606 multi-hop QA pairs from Wiki data. & Multi-hop QA using structured and unstructured sources. & 11 \\ 
    Multihop Questions via Single-hop Question Composition (MuSiQue) \cite{RN462} & ~25,000 2–4-hop questions (50,000 with contrast). & Multi-hop QA by composing single-hop questions. & 9 \\ 
    Fact Extraction and VERification (FEVER ) \cite{RN439} & 185,445 claims annotated with evidence. & Designed for verifying claims using Wikipedia as the textual source & 8 \\ 
    Microsoft MAchine Reading COmprehension (MS MARCO) \cite{RN455} & 100,000 questions and 1 M passages from web docs. & Reading comprehension and QA from real web data. & 8 \\ 
    StrategyQA \cite{RN485} & 2,780 yes/no questions with step-by-step reasoning. & Benchmark Boolean QA needing implicit multi-hop reasoning. & 8 \\ 
    Wizard of Wikipedia (WoW) \cite{RN499} & 22,311 dialogues (202K utterances) using Wiki info. & Dialogue with a ``wizard'' answering via Wikipedia. & 8 \\ 
    WebQuestions (WebQ) \cite{RN493} & 6,642 QA pairs from real user web queries. & Semantic parsers using Freebase KG. & 7 \\ 
    Arc-Challenge \cite{RN1414} & 2,590 science multiple-choice questions. & Benchmark deep-reasoning QA systems. & 5 \\ 
    Explain Like I'm Five (ELI5) \cite{RN437} & 72k QA pairs with supporting web documents. & Long-form QA understandable by five-year-olds. & 5 \\ 
    Massive Multitask Language Understanding (MMLU) \cite{RN1462} & 57 task-specific single-sentence summaries. & Benchmark broad knowledge and reasoning coverage. & 5 \\ 
    NarrativeQA \cite{RN1453} & 1,572 narratives, 46,765 QA pairs. & QA over long narratives and summaries. & 5 \\ 
    PopQA \cite{RN1415} & ~14,000 Wikipedia QA pairs across 16 relations. & QA focusing on Wikidata relationship types. & 5 \\ 
    WebQuestions Semantic Parses (WebQSP) \cite{RN1405} & SPARQL queries for 4,737 questions, 1,073 partial. & KB-QA research using Freebase semantic parses. & 5 \\ 
    Wikipedia English (December 2018) \cite{RN495} & 21 Million passages from December 2018 English Wikipedia. & Passage corpus for retrieval and QA tasks. & 5 \\ 
    Answer Summaries for Questions which are Ambiguous (ASQA) \cite{RN410} & 12,632 ambiguous QA annotations. & Long-form QA for ambiguous factoid questions. & 4 \\ 
    OpenbookQA (OBQA) \cite{RN1416} & 6k science MCQs with 1,326 core facts. & Multi-hop science QA using core facts. & 4 \\ 
    Stanford Question Answering Dataset (SQuAD) \cite{RN1425} & 23k passages, 108k questions (span answers). & Reading comprehension with span answers. & 4 \\ 
    Triple-based Relation Extraction (TREx) \cite{RN488} & 3.09 Million abstracts with 11 Million triples. & Relation extraction and KB population tasks. & 4 \\ 
    TruthfulQA \cite{RN1463} & 817 questions across 38 categories. & Evaluate factual consistency in QA. & 4 \\ 
    Zero Shot RE (zsRE) \cite{RN503} & Over 30 M QA examples for relation extraction. & Zero-shot relation extraction without examples. & 4 \\ 
    Conversational Question Answering (CoQA ) \cite{RN427} & 127k questions from 8k multi-turn dialogues. & Build conversational QA systems. & 3 \\ 
    MultifieldQA-en (MFQA) \cite{RN1480} & 150 docs, 150 cases, ~4.6k words each & Single-document long-context QA. & 3 \\ 
    Physical Interaction: Question Answering (PIQA) \cite{RN1459} & 16,000 physical commonsense MCQs. & Reason about everyday physical tasks. & 3 \\ 
    PubMedQA \cite{RN1524} & PubMed abstracts QA (yes/no/maybe) & Biomedical QA benchmarking. & 3 \\ 
    Qasper (QASP) \cite{RN1479} & 416 papers, 371 cases, 4.7k tokens per doc & Academic QA over research papers. & 3 \\ 
    Unified Medical Language System (UMLS) \cite{RN1424} & Integrated biomedical vocabularies. & Standardize medical terminologies. & 3 \\ 
    Wikipedia Aspect-based summarization (WikiAsp) \cite{RN494} & 320,272 docs with section-title aspects. & Aspect-based summarization of Wikipedia articles. & 3 \\ 
    WikiQA \cite{RN1454} & 3,047 questions with Wikipedia candidate sentences. & Evaluate answer-sentence selection in QA. & 3 \\ 
    Bamboogle \cite{RN412} & 125 handcrafted 2-hop reasoning questions. & Evaluate compositional reasoning capabilities. & 2 \\ 
    BioASQ \cite{RN1529} & 4k+ PDFs and 1k domain-specific questions & Biomedical retrieval \& QA tasks. & 2 \\ 
    BoolQ \cite{RN1458} & 16,000 yes/no questions with passages. & Boolean question answering & 2 \\ 
    C Code Summarization Dataset (CCSD) \cite{RN415} & 95k function–summary pairs. & Source code summarization. & 2 \\ 
    CNN/Daily Mail \cite{RN1493} & News articles paired with human-written summaries. & Summarization and hallucination benchmarking. & 2 \\ 
    Code mixed-language GLUE (General Language Understanding Evaluation) (CodeXGLUE) \cite{RN416} & Millions of code - NL pairs across tasks. & Code understanding and generation. & 2 \\ 
    CodeSearchNet (CSNet) \cite{RN418} & 6M functions, 2M docstring pairs in six langs. & Semantic code search evaluation. & 2 \\ 
    Colossal Clean Crawled Corpus (C4) \cite{RN1448} & Billions of English tokens from web. & Unsupervised pre-training for NLP models. & 2 \\ 
    Common Crawl dump of the internet (CCNet) \cite{RN419} & 1.5B documents, 532B tokens across 174 langs. & Pre-training large-scale language models. & 2 \\ 
    Common Objects in Context (COCO) \cite{RN420} & 330k images, 1.5M captions. & Object recognition and image captioning. & 2 \\ 
    CommonsenseQA \cite{RN1457} & 12,247 MCQs from ConceptNet subgraphs. & Evaluate commonsense question answering. & 2 \\ 
    Conceptual Caption (CC) \cite{RN421} & 3.3M image-text pairs. & Pretrain vision-language models. & 2 \\ 
    Dolly \cite{RN1528} & ~15k human-crafted instruction–response pairs & Instruction-following model training. & 2 \\ 
    Enron Email \cite{RN360} & 500k corporate emails for PII extraction tasks & Evaluate PII detection and removal & 2 \\ 
    ExplaGraphs \cite{RN1406} & 3,166 belief-argument-explanation graphs. & Commonsense reasoning via explanation graphs. & 2 \\ 
    Flickr30k \cite{RN579} & 30k images with five captions each. & Image captioning research. & 2 \\ 
    Google Search corpus (GSfull) \cite{RN444} & 280k sentences from Google Search snippets. & Visual QA (OK-VQA) supporting data. & 2 \\ 
    HellaSwag \cite{RN1464} & 70k multiple-choice from ActivityNet/WikiHow & Commonsense reasoning evaluation. & 2 \\ 
    Incomplete Information Reading Comprehension Questions (IIRC) \cite{RN1437} & 13,441 questions, 5,698 paragraphs. & Challenging reading comprehension. & 2 \\ 
    LAION \cite{RN449} & Billions of image-text pairs. & Train multi-modal language-vision models. & 2 \\ 
    MultimodalQA \cite{RN464} & ~30k questions, ~58k images, text, tables. & Multi-modal QA requiring joint reasoning. & 2 \\ 
    Outside-Knowledge Visual Question Answering (OKVQA) \cite{RN474} & 14k visual questions needing external knowledge. & Visual QA with outside knowledge. & 2 \\ 
    PubHealth \cite{RN1430} & True/false health-claim questions. & Health-claim verification. & 2 \\ 
    PubMed Clinical Papers \cite{RN507} & Millions of biomedical abstracts. & Biomedical literature retrieval. & 2 \\ 
    QMSum \cite{RN1508} & Meeting transcripts with query-based summaries. & Query-focused dialogue summarization. & 2 \\ 
    RealNews \cite{RN479} & 120 GB news articles from Common Crawl. & News summarization benchmark. & 2 \\ 
    RealTimeQA \cite{RN1516} & Weekly news quizzes on politics, business, entertainment. & Evaluate QA on current events requiring retrieval. & 2 \\ 
    RepoEval \cite{RN1503} & Curated GitHub repos for code completion benchmarks. & Evaluate repository-level code completion. & 2 \\ 
    WikiData \cite{RN509} & Structured knowledge graph for Wikipedia. & Knowledge-base for various QA tasks. & 2 \\ 
    Wikipedia (December 2021) \cite{RN510} & 37M passages, 78-word average. & Updated Wikipedia text corpus. & 2 \\ 
    Wikipedia Event (WikiEvent) \cite{RN496} & 246 docs, 6,132 sentences, 3,951 events. & Event extraction and coreference analysis. & 2 \\ 
    WikiText (WikiText) \cite{RN498} & 103M words (103); 2M words (2). & Evaluate long-context language modeling. & 2 \\ 
    1,000-User Benchmark Subset \cite{RN1793} & 1,000 user-session sample with 493 queries avg. & Train and evaluate personalized query prediction. & 1 \\ 
    14 De-identified Clinical Scenarios \cite{RN1435} & 14 anonymized patient scenarios with structured data. & Evaluate clinical query handling. & 1 \\ 
    2019 TREC Deep Learning track (TREC DL19) \cite{RN1525} & 2019 deep-learning track for passage ranking. & Benchmark passage ranking in IR. & 1 \\ 
    2020 TREC Deep Learning track (TREC DL20) \cite{RN1526} & 2020 deep-learning track for passage ranking. & Benchmark passage ranking in IR. & 1 \\ 
    35 Preoperative Guidelines \cite{RN1435} & 35 guidelines on preoperative assessment and care. & RAG knowledge for pre-op instructions. & 1 \\ 
    ACE04 \cite{RN407} & 300k words train, 50k words evaluation & Entity/relation extraction. & 1 \\ 
    ActivityNet Captions \cite{RN1507} & Consists of 20,000 YouTube videos with ~100k localized sentences. & Dense video event description modeling. & 1 \\ 
    ade-corpus-v2 \cite{RN1585} & Sentences labeled for adverse drug reactions. & Text classification focus on ADE detection in biomedical texts & 1 \\ 
    Adversarial Benchmark (AdvBench ) \cite{RN1552} & 520 harmful queries simulating jailbreak attacks. & Support defense against adversarial prompts. & 1 \\ 
    Adversarial NLI (ANLI) \cite{RN1533} & Adversarial inference examples. & Evaluating the inference and reasoning robustness of language models. & 1 \\ 
    Adverse Drug Effect (ADE) \cite{RN408} & 2,972 documents on adverse drug effects & Train ADE extraction models. & 1 \\ 
    Agent-Driver \cite{RN1796} & 23,000 driving episodes with states, objects, reasoning chains, actions. & Retrieval-based memory for safe driving planning. & 1 \\ 
    Aggregated flood event listings from EMSR, GDACS, and ReliefWeb \cite{RN1747} & Curated list of major global flood disasters. & Provide event codes for UI. & 1 \\ 
    AGNews \cite{RN1550} & ~496k news articles in four topics. & Topic classification in news. & 1 \\ 
    AI Tutor \cite{RN1530} & Course PDFs, HTML, and video transcripts. & Retrieve source-based answers for students. & 1 \\ 
    AIDA CoNLL-YAGO \cite{RN1466} & CoNLL03 news articles linked to YAGO entities. & Named entity disambiguation tasks. & 1 \\ 
    Alzheimer’s Disease Interventions (ADInt) \cite{RN1581} & Pharmaceutical interventions entries. & Advance AD intervention knowledge extraction. & 1 \\ 
    Alzheimer’s knowledge graph (AlzKB) \cite{RN1795} & Neo4j dump of genes, diseases, drugs with NL statements and embeddings. & Drive precise biomedical RAG for Alzheimer’s queries. & 1 \\ 
    Amazon Book Reviews \cite{RN1510} & Reviews with user, product IDs, ratings. & Analyze book recommendation and sentiment. & 1 \\ 
    Amazon Movie Reviews \cite{RN409} & 42M reviews, 10M users, 3M items. & Recommender-system and sentiment analysis & 1 \\ 
    AmbigQA \cite{RN1455} & 14,042 ambiguous open-domain questions with rewrites. & Benchmark QA systems’ disambiguation ability. & 1 \\ 
    American Association for the Study of Liver Diseases (AASLD) \cite{RN1436} & 30 liver disease clinical practice guidelines & Reference for hepatology QA tasks & 1 \\ 
    Apnea-ECG Dataset (Sleep Apnoea Detection) \cite{RN1565} & 70 long ECG recordings with minute-wise apnea labels. & Detect sleep apnoea via ECG variability. & 1 \\ 
    Arc-Easy \cite{RN1414} & 5,197 easy science multiple-choice questions & Benchmark simple science QA & 1 \\ 
    Australian Open Legal QA (ALQA) \cite{RN1572} & 232K legal docs, 69.5M lines, 1.47B tokens. & Legal AI research on Australian law. & 1 \\ 
    Automatic Content Extraction 2005 (ACE 2005) \cite{RN407} & 625k annotated words in English, Arabic, Chinese & Train entity, relation, event extraction. & 1 \\ 
    Avocado Research Email Collection \cite{RN1762} & Corporate email archive with threads and metadata. & Retrieval-augmented personalized email drafting. & 1 \\ 
    Bias Benchmark for Question Answering (BBQ) \cite{RN1551} & Multiple-choice QA testing nine social bias categories. & Diagnose representational harms in QA. & 1 \\ 
    BigPatent \cite{RN413} & 1.34M patent documents & Abstractive text summarization. & 1 \\ 
    Bing Search Logs \cite{RN1684} & Three months of anonymized Bing queries and clicks. & Build search-history memory for query suggestion. & 1 \\ 
    BioChatter Continuous-Monitoring Benchmark Suite \cite{RN1740} & Growing suite of biomedical LLM workflow tasks. & Track performance over evolving system features. & 1 \\ 
    BioChatter Knowledge-Graph Query-Generation Benchmark \cite{RN1740} & QA pairs with correct BioCypher graph queries. & Evaluate LLM-to-KG query translation accuracy. & 1 \\ 
    Biography \cite{RN1429} & Long-form biographical narratives of various entities. & Test biographical text generation. & 1 \\ 
    Biomedical Instructions \cite{RN618} & 18k generated biomedical and clinical instruction sets. & Fine-tune models on diverse biomedical tasks. & 1 \\ 
    Biomedical Multiple Choice Questions (MCQ) \cite{RN1570} & Biomedical MCQs with five answer options. & Evaluate biomedical multiple-choice QA. & 1 \\ 
    CaseHOLD \cite{RN1567} & 846K contract provisions with 12.6K refined labels. & Benchmark legal question-answering systems. & 1 \\ 
    Census/projection-disaggregated gridded population datasets \cite{RN1771} & 2020 global population grid disaggregated by census. & Quantify populations in flood zones. & 1 \\ 
    Chain-of-thought \cite{RN1544} & Explicit multi-step reasoning demonstrations & Foster coherent stepwise reasoning & 1 \\ 
    ChEBI-20 \cite{RN1447} & 33,010 molecule-caption pairs & Chemical image captioning models & 1 \\ 
    Chemical Protein Interaction Corpus (ChemProt) \cite{RN1584} & 2,432 PubMed abstracts annotated with interactions. & Chemical-protein relationships and advancing the performance of biomedical relation extraction algorithms & 1 \\ 
    ClashEval Drug Dosage \cite{RN1752} & 249 QA pairs on drug dosages with perturbed contexts. & Benchmark precise dosage retrieval from text. & 1 \\ 
    ClashEval Locations \cite{RN1752} & 200 QA pairs asking for place names from entries. & Test place-name retrieval under context errors. & 1 \\ 
    ClashEval Names \cite{RN1752} & 200 QA pairs querying two-word proper names. & Benchmark proper-noun retrieval against noise. & 1 \\ 
    ClashEval News \cite{RN1752} & 238 numeric QA pairs from AP headline excerpts. & Assess numerical answer extraction under noise. & 1 \\ 
    ClashEval Sports Records \cite{RN1752} & 191 QA pairs on Olympic-record tables with perturbations. & Evaluate correct sports record retrieval. & 1 \\ 
    ClashEval Wikipedia Dates \cite{RN1752} & 200 QA pairs asking for four-digit years from text. & Test year retrieval robustness under corruption. & 1 \\ 
    Clinical Practice Guidelines \cite{RN1679} & Curated guideline articles from MEDITRON. & Support clinical decision-making tasks. & 1 \\ 
    Code Refinement Dataset (CRD) \cite{RN417} & 2.3M bug-fix function pairs. & Code repair and refinement. & 1 \\ 
    CodeMatcher \cite{RN1763} & 10.5M Java methods paired with first doc sentence. & Retrieve exemplar code snippets for generation. & 1 \\ 
    codeparrot/github-jupyter \cite{RN1575} & 165k Jupyter notebooks with metadata & Train code exemplar retrieval & 1 \\ 
    Cognitive Reviewer \cite{RN1530} & Research PDFs analyzed and ranked for reviews. & Facilitate literature reviews via RAG. & 1 \\ 
    ConceptNet \cite{RN1815} & Multilingual commonsense KG with everyday concept triples. & Augment LLM QA with retrieved commonsense subgraphs. & 1 \\ 
    Conceptual 12M (CC12M) \cite{RN1498} & 12M image-text pairs from the web. & Pretrain vision-and-language models. & 1 \\ 
    Concode \cite{RN422} & 100k train, 2k val/test of NL-to-Java examples. & Generate code from natural language. & 1 \\ 
    Conference on Natural Language Learning 2003 (CoNLL03) \cite{RN423} & 301k English/German tokens for NER. & Named-entity recognition benchmark. & 1 \\ 
    Conference on Natural Language Learning 2004 (CoNLL04) \cite{RN424} & ~2k sentences for NER and SRL. & Joint NER and semantic-role labeling. & 1 \\ 
    Conversation QA (QAConv) \cite{RN425} & 10,259 conversations; 34,608 QA pairs. & QA from informative multi-turn conversations. & 1 \\ 
    ConvFinQA (CFQA) \cite{RN1790} & Financial QA grounded in tables and text, requiring math. & Table comprehension and arithmetic in dialogues. & 1 \\ 
    Corpus for Enhancement of Lay Language Synthesis (CELLS) \cite{RN359} & 62,886 abstract - lay summary pairs from biomedical journals. & Simplify scientific text. & 1 \\ 
    COVID-19 Open Research Dataset (CORD19) \cite{RN428} & >140k articles on COVID-19, SARS, MERS (72k full-text). & COVID-19 literature retrieval \& QA. & 1 \\ 
    COYO-700M (COYO) \cite{RN1496} & 747M image-text pairs with metadata. & Support robust vision-language models. & 1 \\ 
    CREAK \cite{RN1512} & Human-authored true/false entity claims. & Fact-checking and commonsense reasoning. & 1 \\ 
    CrossCodeEval \cite{RN1769} & Multilingual code completion benchmarks in four langs. & Assess cross-language code completion generalization. & 1 \\ 
    CrossCodeLongEval \cite{RN1746} & 5k chunk + 5k function completions from 1500 repos. & Evaluate large-span code completion. & 1 \\ 
    CSQA2.0 \cite{RN1513} & Multiple-choice commonsense QA questions. & Evaluate advanced commonsense reasoning. & 1 \\ 
    Curated Golden Evaluation \cite{RN1571} & Standard queries with tickets and authoritative solutions. & Benchmark retrieval and answer accuracy. & 1 \\ 
    CuratedTrec (CT) \cite{RN430} & 867 open-domain factoid questions. & Benchmark factoid QA systems. & 1 \\ 
    Current Events \cite{RN141} & 910 multiple-choice questions from Aug–Nov 2023 U.S. news articles. & Test LLM’s ability to learn new facts via fine-tuning/RAG. & 1 \\ 
    CXR-PRO \cite{RN582} & 248,236 chest X-ray images with de-identified metadata. & Support thoracic disease detection models. & 1 \\ 
    CyberAttack Sensing and Information Extraction (CASIE) \cite{RN431} & 1,000 English news articles on cybersecurity events. & Extract cybersecurity event information. & 1 \\ 
    DailyDialog \cite{RN432} & 13,118 daily-life multi-turn dialogues. & Develop human-like conversational agents. & 1 \\ 
    Data Mining and Text Analytics Course Materials Corpus \cite{RN1744} & ~500 pages of course textbooks, transcripts, figures. & RAG-enabled Q\&A and knowledge retrieval for course. & 1 \\ 
    De-identified electronic health records \cite{RN1421} & 2,278 malnutrition-related clinical notes & Validate summarization and extraction & 1 \\ 
    Defects for Java version 1.2 (Defects4J (v1.2)) \cite{RN433} & 20,109 KLOC of Java code \& tests with real bugs. & Evaluate automated bug repair models. & 1 \\ 
    DialogSum \cite{RN1748} & ~13k multi-speaker dialogues with human summaries. & Evaluate conversational summarization. & 1 \\ 
    DigMinecraft \cite{RN1574} & Images and step-by-step task instructions & Minecraft planning retrieval & 1 \\ 
    Discrete Reasoning Over Paragraphs (DROP) \cite{RN1775} & 96k questions requiring numeric and logical reasoning. & Benchmark discrete reasoning in QA. & 1 \\ 
    Django \cite{RN434} & NL descriptions and Django implementation code. & Evaluate NL-to-code generation on Django framework. & 1 \\ 
    Doc2Dial (D2D) \cite{RN1784} & Document-grounded QA across four domains with long texts. & Benchmark passage retrieval in conversational QA. & 1 \\ 
    DomainRAG \cite{RN1468} & Multiple RAG sub-datasets (extractive, noisy, etc.). & Benchmark domain-specific retrieval-augmented generation. & 1 \\ 
    DoQA \cite{RN1789} & Conversational QA over cooking, travel, movie forums. & Domain-specific dialogue QA with unanswerables. & 1 \\ 
    Drug-Drug Interactions (DDI) \cite{RN1583} & 1,025 texts from Medline and DrugBank. & Identify and classify drug interactions. & 1 \\ 
    Dynamed \cite{RN581} & Clinically organized summaries on 3,200+ topics. & Point-of-care clinical reference tool. & 1 \\ 
    EHRAgent \cite{RN1797} & Four exemplar EHR cases + 700 patient ``experience'' records. & Complex reasoning over EHR-based patient scenarios. & 1 \\ 
    Emotion-Specific Dialogue \cite{RN1490} & Chinese dialogues annotated for five emotions. & Train emotion-conditioned dialogue agents. & 1 \\ 
    EN.MC \cite{RN1780} & 229 multiple-choice QAs on novel contexts. & Benchmark novel-based MCQA. & 1 \\ 
    En.QA \cite{RN1780} & 351 QAs on long novels (~150k words context). & Test QA over very long texts. & 1 \\ 
    Encyclopedic-VQA \cite{RN1556} & 221k image QA pairs linked to 16.7k entities. & Knowledge-based visual question answering. & 1 \\ 
    EntityQuestion (EQ) \cite{RN1452} & 17,300 QA pairs on 24 relation types & Assess entity-centric knowledge retrieval & 1 \\ 
    European Association for the Study of the Liver Guidelines (EASL) \cite{RN1486} & HCV screening, diagnosis, and treatment guidelines. & Hepatology clinical decision support. & 1 \\ 
    Extreme Summarization (XSum) \cite{RN438} & 226,711 news articles for single-sentence summaries. & Support abstractive summarization models. & 1 \\ 
    Facebook Books \cite{RN1515} & User–book interactions data. & Research book recommendation systems. & 1 \\ 
    Fact Extraction and VERification Over Unstructured and Structured information (FEVEROUS) \cite{RN440} & 87,026 claims with text and table evidence. & Automate claim verification using text/tables. & 1 \\ 
    FAct Verification from Information-seeking Questions (FaVIQAmbig) \cite{RN441} & 188,000 true/false claims from info-seeking queries. & Generate and assess factual QA claims. & 1 \\ 
    FactKG \cite{RN1773} & Claims aligned with knowledge-graph triples. & Assess verification over structured KG. & 1 \\ 
    Factual Recall Questions \cite{RN1759} & 30 metadata-style queries (author, decision year, citation, etc.). & Assess factual recall accuracy in legal RAG. & 1 \\ 
    FACTUALITYPROMPTS \cite{RN1538} & Prompts targeting factual accuracy and entity hallucinations & Evaluate factual consistency in generation & 1 \\ 
    False Premise Questions \cite{RN1759} & 22 queries embedding legally incorrect assumptions. & Probe AI’s handling of contrafactual legal prompts. & 1 \\ 
    Fermi \cite{RN1460} & Estimation ``Fermi problems.'' & Reason about numeric magnitudes and estimates. & 1 \\ 
    Fifty-Four Question-Answer Pairs for Few-Shot Learning \cite{RN605} & 54 hepatologist-crafted QA examples. & Evaluate few-shot learning in clinical scenarios. & 1 \\ 
    FinanceBench \cite{RN1450} & 80 docs, 141 finance QA questions & Open-book financial QA & 1 \\ 
    Financial News \cite{RN1741} & 79k Chinese news articles with ChatGPT summaries. & Improve summarization and market context knowledge. & 1 \\ 
    Financial Reports \cite{RN1741} & 120k equity research reports with same-day price data. & Teach LLMs technical analysis and trend prediction. & 1 \\ 
    Financial Reports CoT \cite{RN1741} & 200 CoT annotations on financial report predictions. & Teach rationale-rich stock movement predictions. & 1 \\ 
    FLAN \cite{RN1543} & Natural language instructions for zero-shot learning & Boost zero-shot performance and generalization & 1 \\ 
    FloodBrain ablation study dataset \cite{RN1747} & 26 paired human and FloodBrain flood reports & Evaluate pipeline component impact. & 1 \\ 
    FloodBrain evaluation dataset \cite{RN1747} & 10 human vs 10 FloodBrain-generated flood reports. & Compare generated vs human summaries. & 1 \\ 
    FreebaseQA \cite{RN1778} & 28k trivia-style QAs mapped to Freebase entities. & KB-grounded question answering. & 1 \\ 
    FreshQA \cite{RN1517} & 600 questions with rapidly changing answers. & Test QA on dynamic answers needing external search. & 1 \\ 
    Gaokao-MM \cite{RN1469} & 646 MCQs across 8 subjects with 897 images. & Test multimodal perception and reasoning. & 1 \\ 
    Gender-Specific Dialogue \cite{RN1703} & Chinese dialogues labeled by speaker gender. & Model gendered linguistic features. & 1 \\ 
    General Legal Research \cite{RN1759} & 80 open-ended legal research questions (common-law, bar exams, doctrine). & Benchmark legal-AI retrieval for practicing attorneys. & 1 \\ 
    GIT \cite{RN1582} & Biomedical triple-extraction dataset for non-drug therapies. & Support biomedical relation extraction models. & 1 \\ 
    GIT Relation Extraction (GITRE) \cite{RN1582} & Sentences with head/tail entities and relations. & Predict relationships between biomedical entities. & 1 \\ 
    GPT-Generated Answer Evaluation Corpus \cite{RN1711} & 100 answers with TA and automated correctness labels. & Quantify model factual accuracy metrics. & 1 \\ 
    GraphQA \cite{RN1407} & ntegrates ExplaGraphs, SceneGraphs, WebQSP into QA. & Graph-based QA benchmark. & 1 \\ 
    GSM-HARD \cite{RN1413} & GSM8K variant with larger numeric values & Test arithmetic robustness & 1 \\ 
    GSM8K  \cite{RN1412} & 8.5k grade-school math word problems & Benchmark multi-step math reasoning & 1 \\ 
    HANS \cite{RN1536} & Heuristic-bias evaluation for NLI. & Test NLI heuristic vulnerability. & 1 \\ 
    Harry Potter Series (Books3 subset) \cite{RN1761} & Full text of seven books (~1 M words). & Study model memorization and extraction from training. & 1 \\ 
    Harvard Law Case Corpus \cite{RN1576} & Extensive collection of Harvard Law case texts. & Pretrain/fine-tune legal language models. & 1 \\ 
    Harvard-FairVLMed \cite{RN1783} & Multimodal fundus images with associated textual data. & Fairness evaluation in ophthalmic vision-language. & 1 \\ 
    HealthcareMagic-101 \cite{RN1546} & 200k doctor-patient medical dialogues & Model sensitive medical conversational contexts & 1 \\ 
    Hearthstone \cite{RN1765} & Game-card logic code paired with card names. & Benchmark NL-to-code on game logic generation. & 1 \\ 
    Historical Issue Tickets \cite{RN1571} & Customer service tickets parsed into hierarchical trees. & Improve retrieval/QA over support tickets. & 1 \\ 
    Hospital Neurology Discharge Summaries \cite{RN1712} & 100 anonymized neurology discharge summaries. & Personalize advice and track recovery via memory. & 1 \\ 
    Human-Edited Counterfactuals Subset of IMDb \cite{RN340} & 1.7K movie reviews manually sentiment-inverted. & Augment data via sentiment counterfactuals. & 1 \\ 
    Human-Generated Responses \cite{RN1435} & Free-text pre-op instructions by junior doctors & Baseline pre-op instruction generation & 1 \\ 
    HumanEval \cite{RN1408} & 164 Python programming tasks with unit tests. & Evaluate code generation correctness. & 1 \\ 
    HumanEval+ \cite{RN1409} & 164 tasks with 80× more test cases & Robustness evaluation for code generation & 1 \\ 
    HybriDialogue (HDial) \cite{RN1792} & QA on hybrid pages (text + tables) in conversation. & Mixed-modal conversational reasoning. & 1 \\ 
    IMDB (Internet Movie Database) \cite{RN577} & Subsets of movie reviews and associated metadata. & Sentiment analysis and recommendation tasks. & 1 \\ 
    InferredBugs \cite{RN348} & ~6,200 repos; 8,280 bug-fix patches. & Suppoty models on static-analysis bug fixes. & 1 \\ 
    Infineon Developer Community Forum Questions \cite{RN1578} & Technical Q\&A with expert answers. & Benchmark chatbot against forum solutions. & 1 \\ 
    Infineon Product Documents \cite{RN1577} & Datasheets and product guides. & Retrieval for technical RAG systems. & 1 \\ 
    InfoSeek \cite{RN1557} & 1.3M image-QA triplets for ~11k entities. & Assess external knowledge integration in VQA. & 1 \\ 
    INSCIT \cite{RN1788} & Under-specified Wikipedia QA requiring clarification. & Test clarification question generation. & 1 \\ 
    IU-Xray \cite{RN1782} & Chest X-rays paired with detailed diagnostic reports. & Support medical image-reporting systems. & 1 \\ 
    Joint Research Centre Acquis (JRCAcquis) \cite{RN447} & ~8,000 legal docs per language, 20+ EU languages. & Multilingual legal parallel corpus. & 1 \\ 
    Jurisdiction or Time-Specific Research \cite{RN1759} & 70 questions on jurisdictional splits or overturned precedents. & Test RAG on time-sensitive legal rule retrieval. & 1 \\ 
    Knowledge Intensive Language Tasks (KILT) \cite{RN1494} & 11 datasets for fact checking, QA, entity linking. & Unified evaluation of knowledge-intensive tasks. & 1 \\ 
    Labeled EDGAR (LEDGAR) \cite{RN1427} & 846K contract provisions with 12.6K refined labels. & Contract clause classification. & 1 \\ 
    Lambada \cite{RN1532} & Cloze tasks requiring broad discourse context. & Test long-range dependency in LMs. & 1 \\ 
    Language Model Personalization (LaMP) \cite{RN1485} & Seven classification and generation tasks. & Benchmark personalized model outputs. & 1 \\ 
    Lecture-Material \cite{RN1711} & Lecture notes, slides, exercise sheets corpus. & RAG retrieval for course-related queries. & 1 \\ 
    LegalBench Collection \cite{RN1568} & 50 manual legal QA pairs. & Small-scale legal QA benchmarking. & 1 \\ 
    LightQA \cite{RN358} & QA from role-playing dialogues with final utterance. & Evaluate factual QA in game dialogue contexts. & 1 \\ 
    LightWild \cite{RN452} & 462K utterances across 41K RPG episodes. & Support dialogue agents in fantasy settings. & 1 \\ 
    LiveQA \cite{RN1692} & Real medical questions with long-form answers. & Evaluate clinical long-answer generation. & 1 \\ 
    LLaVA-Instruct \cite{RN1558} & 158k image–instruction training pairs. & Visual instruction tuning for MLLMs. & 1 \\ 
    Lumos-QG-Generated QA Dataset (9 000 Pairs) \cite{RN1744} & 9,000 auto-generated QA pairs from course materials. & Expand knowledge base for Alexa skill and evaluation. & 1 \\ 
    lyft\_2021 \cite{RN1527} & Lyft 2021 document used for chunking benchmark queries. & Benchmark document-chunking techniques. & 1 \\ 
    Massive Multi-discipline Multimodal Understanding (MMMU) \cite{RN1560} & 11.5k college-level multimodal exam questions. & Expert-level multimodal reasoning evaluation. & 1 \\ 
    Math Nation Queries \cite{RN134} & 51 factual/conceptual math questions from forum. & Benchmark math QA from student discussions. & 1 \\ 
    MathVista \cite{RN1470} & 6,141 math problems with diagrams, charts, plots. & Evaluate multimodal math reasoning. & 1 \\ 
    Medical Transcription Samples (MTsample) \cite{RN1586} & Transcriptions across 40+ clinical specialties. & Research clinical text classification patterns. & 1 \\ 
    MedicationQA \cite{RN1693} & Long-form QA focused on medication queries. & Test medication-related answer accuracy. & 1 \\ 
    MedInstruct \cite{RN1696} & Biomedical instructions: QA, summarization, MCQs. & Fine-tune models on diverse clinical tasks. & 1 \\ 
    MedMCQA \cite{RN1694} & Multiple-choice biomedical questions & Benchmark biomedical QA systems. & 1 \\ 
    MedQA \cite{RN1695} & Multiple-choice medical exam questions & Evaluate medical QA models. & 1 \\ 
    MetaQA \cite{RN1474} & 400k questions covering single- and multi-hop reasoning. & Test end-to-end KG QA systems. & 1 \\ 
    Microsoft COCO (MSCOCO) \cite{RN453} & 328K images, 2.5M labeled object instances. & Scene understanding and object detection. & 1 \\ 
    Microsoft Research Paraphrase Corpus (MSRPC) \cite{RN1483} & 2.2k train, 550 val, 1.1k test paraphrase pairs. & Evaluate paraphrase detection. & 1 \\ 
    Microsoft Research Video Description Corpus (MSVD) \cite{RN1505} & 1,970 YouTube clips with ~80k English descriptions. & Benchmark video captioning models. & 1 \\ 
    Microsoft Research Video to Text (MSRVTT) \cite{RN1506} & 10,000 videos with ~200k captions. & Video captioning evaluation across domains. & 1 \\ 
    MIMIC-CXR \cite{RN1781} & Large public CXR images with radiology reports. & Develop chest X-ray interpretation models. & 1 \\ 
    Minecraft Wiki \cite{RN1573} & Thousands of community-curated Minecraft articles & Retrieval for planning tasks & 1 \\ 
    Mintaka \cite{RN1473} & Knowledge graph QA with complex, diverse questions. & Knowledge graph QA benchmark. & 1 \\ 
    MMBench (MMB) \cite{RN1561} & 3k multiple-choice questions covering 20 abilities. & Benchmark fine-grained multimodal capabilities. & 1 \\ 
    Mol-Instructions \cite{RN1687} & Off-the-shelf biomedical instruction tasks. & Instruction-tuning biomedical models. & 1 \\ 
    MongoDB-Logs (Chat \& Cost) \cite{RN1711} & Conversation logs and token-cost data. & The logs underpin post-hoc accuracy checks, cost calculations and support future optimisation of the chatbot service. & 1 \\ 
    MongoDB-QA (Question Answer Pairs) \cite{RN1711} & 170 validated course QA pairs. & It is sampled by the QAGeneration-Chain to generate quick practice exercises for students. & 1 \\ 
    Mostly Basic Programming Problems (MBPP) \cite{RN1410} & 974 beginner Python problems with tests & Evaluate beginner-level code models & 1 \\ 
    Mostly Basic Programming Problems+ (MBPP+) \cite{RN1409} & MBPP tasks with added test cases & Enhanced MBPP evaluation coverage & 1 \\ 
    MovieLens100K \cite{RN1514} & 100k movie ratings by various users. & Benchmark recommendation algorithms. & 1 \\ 
    MS-CXR \cite{RN457} & 1,153 chest X-rays with paired radiology reports. & Evaluation CXR interpretation and report models. & 1 \\ 
    Multi-Domain Wizard-of-Oz version 2.1 (MultiWOZ 2.1) \cite{RN504} & 10,438 dialogs across seven domains with slots. & Develop and benchmark multi-domain dialogue. & 1 \\ 
    Multi-Genre Natural Language Inference (MNLI) \cite{RN458} & 433k sentence pairs labeled entailment/contradiction/neutrality. & Evaluate natural language inference models. & 1 \\ 
    Multi-programming Language Commit Message (MCMD) \cite{RN459} & 2.25M commit messages across five programming languages. & Evaluate semantic code search capabilities. & 1 \\ 
    Multi-Sentence Reading Comprehension (MultiRC) \cite{RN460} & 800 paragraphs with 6,000 multi-sentence questions. & Evaluation comprehension over multi-sentence contexts. & 1 \\ 
    Multimodal Evaluation (MME) \cite{RN1559} & 14 tasks in cognition and perception categories. & Standardized benchmark for multimodal LLMs. & 1 \\ 
    Natural Language to Bash (NL2Bash) \cite{RN466} & 9,000+ English descriptions paired with Bash commands. & Translate natural language to shell commands. & 1 \\ 
    Natural Language to Command Line (NLC2CMD) \cite{RN467} & 100 NL-to-command evaluation examples. & Build NL-to-command translation systems. & 1 \\ 
    New York Times (NYT) \cite{RN471} & 1.8M articles published between 1987–2007. & News summarization. & 1 \\ 
    NewsQA \cite{RN472} & 119k QA pairs from 12.7k CNN news articles. & Human-generated question-answer pairs developed from news articles from CNN & 1 \\ 
    NoCaps \cite{RN1495} & ~15k images of novel objects without MSCOCO overlap. & Evaluate novel-object captioning. & 1 \\ 
    North American HCV Guidelines \cite{RN1487} & AASLD-IDSA supplemental HCV practice guidelines. & Supplementary HCV clinical reference. & 1 \\ 
    Online Sources Nursing Knowledge JSON \cite{RN1712} & Scraped nursing instructions and academic papers JSON. & Supply RAG pipeline with clinical knowledge. & 1 \\ 
    OpenQA-NQ (subset of Natural Questions) \cite{RN473} & 13M evidence blocks from Wikipedia for QA retrieval. & Open-retrieval question answering. & 1 \\ 
    OpenStax Prealgebra Textbook \cite{RN506} & Textbook sections on prealgebra & The content from the math textbook is used to generate responses to real student questions. & 1 \\ 
    OpenStreetMap Planet dump \cite{RN1772} & Global vector map data: roads, buildings, POIs. & Enrich flood maps with geographic data. & 1 \\ 
    Osaka Personal Activity Trajectory \cite{RN1753} & 2,102 daily check-in trajectories, 537 synthetic samples. & Evaluate mobility framework’s city generalization. & 1 \\ 
    ParaSCI-ACL \cite{RN1484} & 28,883 scientific paraphrase training examples. & Scientific-domain paraphrase generation. & 1 \\ 
    Patient Inquiry Dataset \cite{RN1712} & Timestamped patient questions during system testing. & Evaluate conversational performance and short-term memory. & 1 \\ 
    Patient Symptom Record Dataset \cite{RN1712} & Daily self-reported vital signs and symptom notes. & Monitor condition changes and trigger alerts. & 1 \\ 
    PDFTriage  (PDFT) & Questions on PDF document structures. & Benchmark document-structure QA tasks. & 1 \\ 
    PMC Full-text \cite{RN1698} & Full-text articles from PubMed Central. & Enable retrieval for biomedical question answering. & 1 \\ 
    Polling-based Object Probing Evaluation (POPE) \cite{RN1562} & Binary yes/no questions from ground truth objects/negatives. & Assess object hallucination in V-L models. & 1 \\ 
    Pre-training Corpus \cite{RN1545} & 330 B tokens from 15 high-quality sources. & Pretrain RETRO and GPT language models. & 1 \\ 
    Probably-Asked Questions (PAQ) \cite{RN475} & 65 M auto-generated QA pairs & Semi-structured KB QA knowledge base. & 1 \\ 
    PTB-XL \cite{RN1564} & 21,837 12-lead ECG records with cardiologist annotations. & Arrhythmia diagnosis and zero-shot eval. & 1 \\ 
    PTB-XL+ \cite{RN1566} & Adds algorithm-extracted ECG features for each record. & Detailed ECG feature analysis for diagnosis. & 1 \\ 
    PubMed Abstract \cite{RN1697} & Corpus of PubMed abstracts. & Provide domain evidence for QA retrieval. & 1 \\ 
    PwC Reading-Comprehension Corpus \cite{RN1777} & ~241k passage-question-answer triples. & Research on large-context compression. & 1 \\ 
    Python Code Summarization Dataset (PCSD) \cite{RN477} & 150 k function–docstring pairs & Code summarization. & 1 \\ 
    PyTorrent \cite{RN1764} & ~2M Python methods from PyPI/Anaconda packages. & Code exemplar retrieval for Python generation. & 1 \\ 
    QReCC \cite{RN1786} & Open-domain conversational QA over web docs (avg 5K words). & Zero-shot conversational retrieval and QA. & 1 \\ 
    QuAIL \cite{RN1776} & ~15k multiple-choice questions across varied texts. & Evaluate adaptive QA across question types. & 1 \\ 
    QuALITY \cite{RN1509} & MCQs from stories/articles (multiple-choice). & Narrative comprehension evaluation. & 1 \\ 
    QuaRTz \cite{RN1465} & 3,864 MCQs on qualitative relationships. & Semantic and linguistic reasoning in QA. & 1 \\ 
    Question Answering in Context (QuAC) \cite{RN1785} & Multi-turn dialogues over Wikipedia with answerable turns. & Conversational QA with linked long contexts. & 1 \\ 
    Quora Question Pairs 140K (QQP) \cite{RN1482} & 134k train, 5k val, 5k test paraphrase pairs. & Paraphrase detection and generation. & 1 \\ 
    Quora Question Pairs 50K (QQP) \cite{RN1481} & 50k paraphrase question pairs. & Paraphrase detection and generation. & 1 \\ 
    RACE \cite{RN1534} & Exams-derived reading comprehension dataset. & Benchmark multi-paragraph comprehension. & 1 \\ 
    RAG Comparison (Derived from the SPOKE KG) \cite{RN1569} & Biomedical questions from SPOKE KG entity associations. & Compare RAG: KG, Cypher, full-text methods. & 1 \\ 
    RAG-Fusion Query Set \cite{RN1492} & Dynamically generated multi-query sets. & Enhance retrieval via rank fusion. & 1 \\ 
    RAGTruth \cite{RN1579} & ~18,000 LLM-generated responses with quality labels. & Benchmark hallucination detection in RAG. & 1 \\ 
    Reading Comprehension with Commonsense Reasoning Dataset  (ReCoRD) \cite{RN478} & 70 k passages, 120 k queries & Commonsense reading comprehension & 1 \\ 
    REALTOXICITYPROMPTS \cite{RN1539} & Prompts engineered to elicit toxic language. & Evaluate worst-case toxicity in outputs. & 1 \\ 
    Reddit Webis-TLDR-17 \cite{RN1548} & Reddit posts paired with short summaries & Test summarization with varied tones & 1 \\ 
    ReliefWeb flood reports \cite{RN1747} & Human-authored situational flood event reports. & Benchmark report factual accuracy. & 1 \\ 
    Research Dataset \cite{RN1741} & 42k finance texts merging sentiment, numeric, headline tasks. & Pretrain/fine-tune LLMs on financial language. & 1 \\ 
    Retrieval-Augmented Generation Benchmark (RGB ) \cite{RN622} & 1,000 English \& Chinese QA & Evaluate retrieval-augmented generation & 1 \\ 
    RiddleSense \cite{RN1814} & ~5,000 riddles with answer options requiring creative reasoning. & Challenge models on linguistic creativity and commonsense. & 1 \\ 
    Roles Across Multiple Sentences (RAMS) \cite{RN481} & 3,993 docs, 9,124 event annotations & Multi-sentence semantic role labeling & 1 \\ 
    RTLLM \cite{RN1521} & RTL generation benchmark tasks. & Evaluate LLM-based RTL design generation. & 1 \\ 
    SamSum \cite{RN1779} & 16k messenger-style dialogues with abstractive summaries. & Train dialogue summarization systems. & 1 \\ 
    SBU Captions (SBU) \cite{RN1497} & ~1M Flickr-based image–caption pairs. & Large-scale image captioning research. & 1 \\ 
    SceneGraphs (from GQA) \cite{RN1471} & 100k scene graphs of images for visual reasoning. & Support spatial and visual inference tasks. & 1 \\ 
    Scoliosis Research Society (SRS) \cite{RN508} & Educational, research, patient resources & Support spinal deformity care & 1 \\ 
    SearchQA \cite{RN482} & 140 k QA pairs, 6.9 M snippets & QA simulating real web search & 1 \\ 
    Self-Instruct \cite{RN1541} & LM-generated instruction examples & Support models on diverse self-generated directives & 1 \\ 
    Sentiment-Specific Dialogue \cite{RN994} & English dialogues labeled by sentiment. & Generate sentiment-controlled responses. & 1 \\ 
    ServiceNow Internal Data \cite{RN1502} & Annotated queries with structured workflow JSON. & Translate NL requests into workflows. & 1 \\ 
    SocialIQA (SIQA) \cite{RN1456} & 38,000 social-context multiple-choice QA pairs. & Test commonsense reasoning in social contexts. & 1 \\ 
    SODA \cite{RN1540} & High-quality social dialogue examples & Enhance conversational fine-tuning & 1 \\ 
    SQA \cite{RN1791} & Conversational QA over single Wikipedia tables & Compositional multi-column table QA. & 1 \\ 
    SQuAD v2 \cite{RN1425} & 150k QAs plus 50k unanswerable questions on Wikipedia. & QA with answer/no-answer classification. & 1 \\ 
    Stanford Sentiment Treebank (SST2) \cite{RN1549} & 215k phrases labeled for fine-grained sentiment. & Benchmark sentiment classification & 1 \\ 
    StockQA \cite{RN1741} & 21k Chinese QA pairs from real stock-price sequences. & Train time-series reasoning for investor queries. & 1 \\ 
    TACRED \cite{RN1520} & Adapted TACRED for zero/few-shot slot filling (41 types). & Benchmark relation extraction and slot filling. & 1 \\ 
    TAM Questionnaire Response Set \cite{RN1711} & 30 students’ Likert-scale survey responses. & Evaluate user acceptance via factor/regression. & 1 \\ 
    TFix \cite{RN486} & 100 k code error-fix pairs & Evaluate code repair models & 1 \\ 
    The human cost of disasters (2000-2019) \cite{RN1770} & Global disaster human-impact records 2000–2019. & Analyze flood impacts for planning. & 1 \\ 
    The Pile \cite{RN487} & 825 GiB text from 22 sources & Pretrain diverse language models & 1 \\ 
    The Stack \cite{RN1768} & 3 TB public source code from GitHub. & Pretrain and fine-tune code language models. & 1 \\ 
    Tokyo Personal Activity Trajectory \cite{RN1753} & 100 users’ time-ordered GPS check-ins (2019–2022). & Model realistic human mobility patterns. & 1 \\ 
    ToolQA \cite{RN1518} & Personal-agenda questions assessing external tool use. & Measure LLM integration of external tools in QA. & 1 \\ 
    TopiOCQA (TCQA) \cite{RN1787} & QA over full Wikipedia with topic shifts. & Evaluate topic-transition conversational QA. & 1 \\ 
    TREC-COVID \cite{RN1426} & Dynamic COVID-19 docs with topics and relevance labels. & Pandemic literature retrieval evaluation. & 1 \\ 
    True/False dataset \cite{RN1569} & True/false statements on gene-disease and drug-disease. & Benchmark biomedical assertion verification. & 1 \\ 
    UltraDomain - Agriculture \cite{RN1794} & 2,017,886 tokens from 12 college-agriculture texts & Evaluate RAG’s sense-making in agriculture domain & 1 \\ 
    UltraDomain - CS \cite{RN1794} & 2,306,535 tokens from 10 computer-science texts & Test RAG on technical computer-science content & 1 \\ 
    UltraDomain - Legal \cite{RN1794} & 5,081,069 tokens from 94 legal textbook documents & Benchmark RAG on complex legal language and reasoning & 1 \\ 
    UltraDomain - Mixed \cite{RN1794} & 619,009 tokens across 61 humanities texts & Challenge RAG with heterogeneous humanities content & 1 \\ 
    Unnatural Instructions \cite{RN1542} & Minimally human-curated challenging instructions & Augment instruction tuning diversity & 1 \\ 
    UpToDate & Clinical decision support content by Wolters Kluwer. & Point-of-care medical reference. & 1 \\ 
    VATEX \cite{RN1504} & 25,991 train, 9k val/test English video captions. & Multilingual and multi-modal captioning. & 1 \\ 
    VerilogEval \cite{RN1522} & Verilog code generation tasks. & Assess LLM Verilog functional correctness. & 1 \\ 
    VerilogEval-syntax \cite{RN1523} & 200+ clustered Verilog syntax error examples. & Test syntax-error correction in Verilog. & 1 \\ 
    Visual Question Answering (VQA) \cite{RN491} & 254,721 images with ~760k questions and ~10M answers. & Visual QA tasks combining vision and language. & 1 \\ 
    W3C-Email & Emails similar to GPT-Neo’s training distribution & Study retrieval-augmented memorization effects & 1 \\ 
    Web Search & - & - & 1 \\ 
    WebQA \cite{RN492} & 34,200 train, 5,000 val, 7,500 test QA pairs; ~390k images. & Multimodal web-based QA benchmarking. & 1 \\ 
    Weibo \cite{RN1475} & 4.4M post-response pairs from Sina Weibo. & Support short-text conversation models. & 1 \\ 
    WikiPassageQA \cite{RN1461} & 4,165 QA with long answer passages. & Reading comprehension with long answers. & 1 \\ 
    Wikipedia (October 2017) \cite{RN445} & Snapshot of English Wikipedia articles. & Historic Wikipedia text for NLP. & 1 \\ 
    Wikipedia Evaluation (WikiEval) \cite{RN511} & 50 Wikipedia pages covering diverse topics. & Evaluate retrieval-augmented systems. & 1 \\ 
    Wikipedia Passages \cite{RN497} & 6M+ articles, ~3.8B words across languages (as of 2021). & Large-scale text corpus for NLP. & 1 \\ 
    WinoGrande \cite{RN1535} & Pronoun-resolution tasks in complex contexts. & Assess coreference resolution capability. & 1 \\ 
    WitQA \cite{RN1451} & 14k factual QA pairs on 32 relation types & Evaluate factual QA across relations & 1 \\ 
    Wizard of the Internet (WizInt) \cite{RN349} & 9,633 dialogues, 93,665 utterances, 29,500 URLs. & Dialogue with live internet search. & 1 \\ 
    WNED \cite{RN1467} & 320 documents with 6,821 linkable mentions. & Evaluate entity linking systems. & 1 \\ 
    Word-in-Context (WiC) \cite{RN1537} & Word-in-context disambiguation pairs. & Evaluate word sense disambiguation. & 1 \\ 
    Worker and AI Collaboration for Natural Language Inference (WaNLI) \cite{RN500} & 107,885 NLI examples combining human and GPT-3 data. & Natural language inference with AI mix. & 1 \\ 
    Yelp Reviews \cite{RN502} & 1.1M+ reviews, 42k businesses, 400k tips, check-ins. & Recommendation and sentiment analysis. & 1 \\ 
    Yelp. 2021 \cite{RN1580} & Business attributes and reviews with detailed schema. & Data-to-text generation and hallucination tests. & 1 \\ 
    ZINC-15 \cite{RN1449} & 1.54B filtered SMILES strings & Virtual screening compound datasets & 1 \\ 

\end{longtable}
\end{footnotesize}
\end{landscape}
\twocolumn

\section{Acknowledgment}
This research is supported by the Advanced Research and Engineering Centre (ARC) in Northern Ireland, funded by PwC and Invest NI. The views expressed are those of the authors and do not necessarily represent those of ARC or the funding organisations.

The authors appreciate the use of the Kelvin\textsuperscript{2} High Performance Computing cluster at Queen’s University Belfast for computational work.

\bibliographystyle{IEEEtran}
\bibliography{bib/references}

\newpage

\begin{IEEEbiography}[{\includegraphics[width=1in,height=1.25in,clip,keepaspectratio]{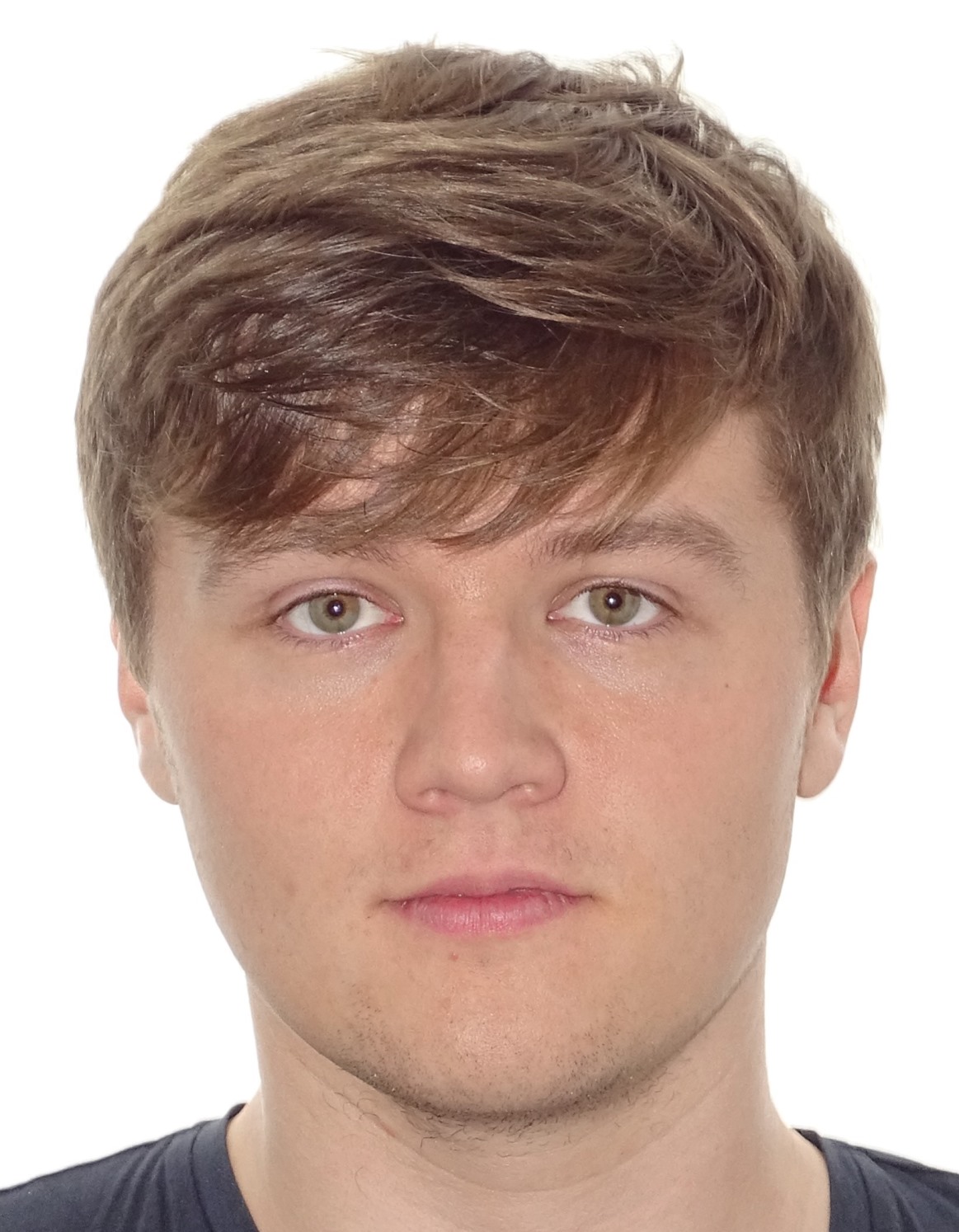}}]{Andrew Brown} received the BSc degree in Computer Science (First Class Honours) from Queen's University Belfast, UK, in 2022, and is currently pursuing the PhD degree in Computer Science at Queen's University Belfast. His research focuses on natural language processing for document understanding and business information extraction. He has served as a Demonstrator in video analytics and machine learning, cloud computing, and AI for health at Queen's University Belfast (2021--2024), and previously worked as a Junior Software Engineer with Congruity360 (2022--2023). He received the Associate Fellowship of the Higher Education Academy in 2021. His research interests include information extraction, applied machine learning, and AI systems for documents.
\end{IEEEbiography}

\begin{IEEEbiography}[{\includegraphics[width=1in,height=1.25in,clip,keepaspectratio]{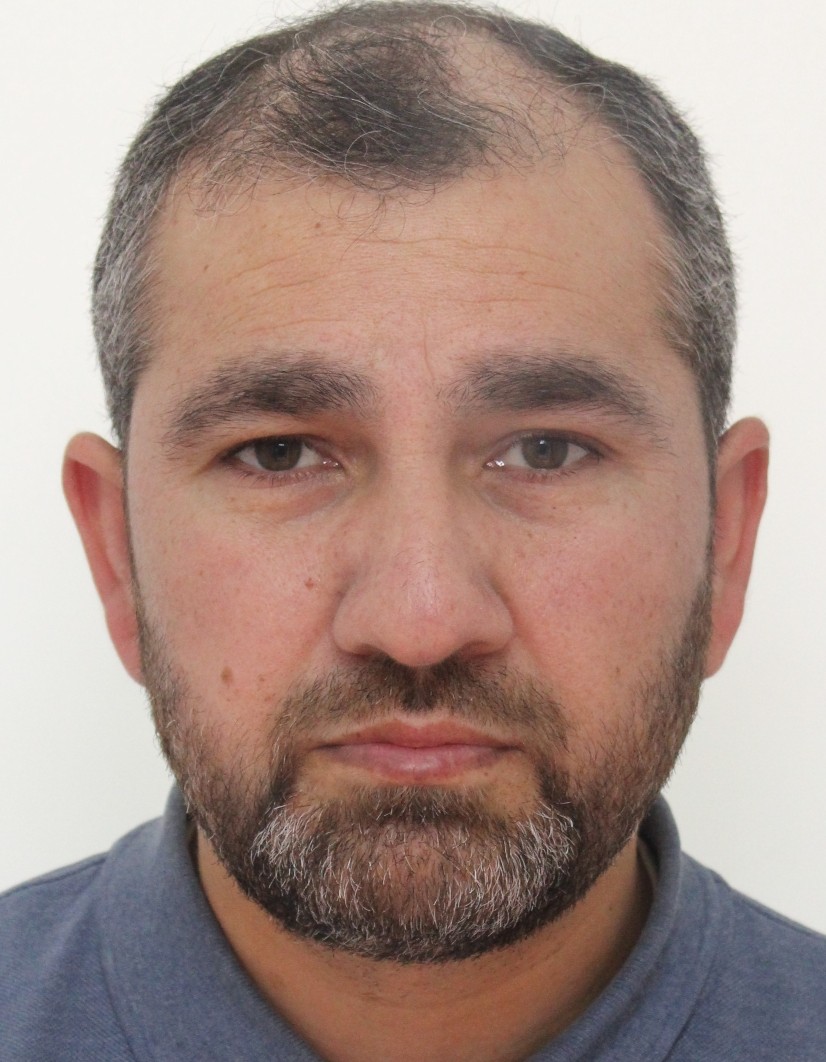}}]{Muhammad Roman} received the Ph.D. degree in Computer Science from Kohat University of Science and Technology (KUST), Pakistan, in 2021, specializing in Natural Language Processing, Information Retrieval, and Large Language Models. He has over 16 years of experience in AI research and development, with expertise in Retrieval-Augmented Generation, multimodal AI, and AI-driven orchestration. His current work focuses on LLM-based multi-agent systems for energy flexibility services, digital product passports, dataspaces, cross-sector data sharing, and complete lifecycle analysis data, including renewable energy integration, energy attribute certificates, and carbon footprint tracking. He has authored multiple journal publications, and his research interests include LLMs, document AI, compliance automation, and AI-native networking.
\end{IEEEbiography}

\begin{IEEEbiography}[{\includegraphics[width=1in,height=1.25in,clip,keepaspectratio]{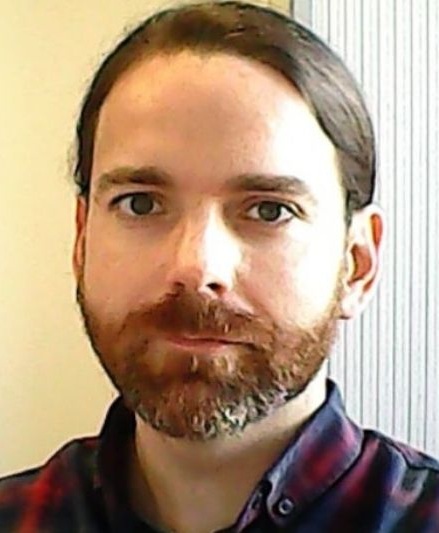}}]
{Barry Devereux} is a Senior Lecturer in the School of Electronics, Electrical Engineering and Computer Science at Queen's University Belfast. His research spans computational cognitive neuroscience and natural language processing, with interests in semantics, LLM analysis and interpretability, text mining, and biomedical and clinical text data. He has published in venues such as Computational Linguistics, COLING, EMNLP, and Cognitive Science, including work on the representation of compound semantics in LLMs, retrieval-augmented generation with knowledge graphs, and modelling human neuroimaging data in vision and language processing. He is programme director of the QUB MSc programme in Artificial Intelligence and serves as an Area Chair for the ACL Rolling Review.
\end{IEEEbiography}

\end{document}